\documentclass[a4paper,12pt]{article}

\usepackage{emptypage} 

\usepackage{graphicx}  

\usepackage{imakeidx}
\makeindex

\usepackage{amsmath}
\usepackage{mathtools}
\usepackage{amssymb}
\usepackage{amsthm}
\usepackage[makeroom]{cancel}

\numberwithin{equation}{section} 

\usepackage{titlesec}

\usepackage[most]{tcolorbox}

\usepackage{float}
\usepackage{caption}
\usepackage{marginnote}
\usepackage[top=2.5cm, bottom=2.5cm, outer=3cm, inner=3cm, heightrounded, marginparwidth=1.5cm, marginparsep=0.5cm]{geometry}
 
\usepackage{hyperref}
\hypersetup{
  colorlinks   = true,    
  urlcolor     = blue,    
  linkcolor    = blue,    
  citecolor    = aquamarine      
}

\newcommand{\beq}{\begin{equation}}
\newcommand{\eeq}{\end{equation}}

\newtheorem{theorem}{Exercise}[section]
\newtheorem{note}{Box}[section]
\theoremstyle{remark}

\usepackage{etoolbox} 
\patchcmd{\thebibliography}{\section*{\refname}}{}{}{}

\usepackage{tikz}
\usetikzlibrary{matrix,backgrounds,calc,decorations.pathreplacing}
\tikzstyle{operator} = [draw,fill=white,minimum size=1.5em] 
\tikzstyle{phase} = [draw,fill,shape=circle,minimum size=8pt,inner sep=0pt] 
\newcommand{\ket}[1]{\ensuremath{\left|#1\right\rangle}}

\begin{document}

\thispagestyle{empty} 
 
\begin{center} 
\Large\textbf{{A Short Introduction to \\
Quantum Computing for Physicists }}\\
\end{center}
\begin{center} 
Oswaldo Zapata
\end{center}

\begin{abstract}
\noindent
These notes provide an introduction to standard topics on quantum computation and communication for those who already have a basic knowledge of quantum mechanics. The main target audience are professional physicists as well as advanced students of physics; however, engineers and computer scientists may also benefit from them.
\end{abstract}

\tableofcontents
\newpage

\section{Introduction}

The goal of the present notes is to introduce the theoretical framework a trained physicist needs to get into quantum computing. Thus, if you are a physicist and you want to learn the basics of quantum computing, these notes are for you. In a matter of hours (maybe dedicating an entire weekend), you will be able to learn all the basics of quantum computer science. 

If, as I suppose, you are a physicist, then at some point in your career you took a proper course on quantum mechanics. Of course, I do not assume that you remember everything you studied then, however, I do assume that you already went through all the standard topics as found in the books by Sakurai or Cohen Tannoudji \emph{et al}. This will allow me to focus on aspects of quantum computing that I think are new to you as a physicist. That said, if you think that you forgot most of what you learned about quantum mechanics, you should not worry. Sincerely speaking, the use of quantum mechanics in quantum computing is relatively simple. Moreover, to help you, in general I recall the main physical and mathematical concepts and I provide explicit calculations so you can easily follow what I am explaining. 

Quantum computing is usually described as lying at the intersection of quantum mechanics, mathematics and computer science. As I said, I assume that you studied quantum mechanics. Now, concerning mathematics, I am afraid that most physicists are not familiar with the way computer scientists learn the subject. Here I am not referring, of course, to the mathematics used in quantum mechanics, such as linear algebra, but to subjects like formal logic, models of computation or complexity theory. Since I am not an expert in the field, I will simply sketch the main ideas without entering too many details. The interested reader may look at the appropriate literature. Concerning the most basic notions of computer science, such as Boolean algebra and circuits, I assume that you are barely familiar with them (maybe at the level of the first few lines of a Wikipedia article). 

The notes are organized as follows: In Section 2, I introduce the quantum systems relevant to quantum computing and review the mathematical formalism necessary to understand them. In Section 3, I describe how these quantum systems can be manipulated and measured. In Section 4, I review some clever ways physicists and computer scientists have found, at least theoretically, to modify the quantum systems in order to compute certain tasks more efficiently than classical computational methods. In Section 5, I explain how the destructive effect of the environment can be reduced so it does not destroy the quantum nature of the system. 

A short comment on the organization of these notes. While Sections 2 and 3 must be read one after the other, Sections 4 and 5 are rather independent of each other. So, after reading Sections 2 and 3, read Sections 4 and 5 in the order that suits you.

The Boxes you find within the main text contain additional material that I consider supplementary. Some of them review topics that I assume you already know and some others expand the main text. My recommendation is that while reading these notes, you give a quick glimpse at the Boxes to see what they are about and, depending on your knowledge, read or skip them. If you decide to skip them, you can always come back to them at a later time. 

Concerning the Exercises, I have added them to help you understand and become familiar with the subject, not to make you smarter. So, try to do them; they are relatively easy. 

I wish to thank my physics friends for reading the notes, suggesting many improvements and, crucially, testing that you can indeed learn from them. I hope they will be helpful to you as well. 

I am planning to continue adding new material to these notes; thus, if you have any feedback (maybe you find a typo, you think that I say something that is not completely correct, I ignored a subject or its presentation can be improved, or any other reason you may have), I will sincerely appreciate it if you send me an email to \href{mailto:zapata.oswaldo@gmail.com}{zapata.oswaldo@gmail.com}.

\vspace{25pt}
\hrule
\vspace{25pt}

Before moving on to the technical details, let me give a
very brief overview of the history of the subject. This will allow you to see the content of these notes in perspective. 

The first people who thought about the possibility and the necessity of building quantum computers were Yuri Manin
(1980) and Richard Feynman (1982). Feynman's vision
was more elaborate, and he considered the advantage
of a quantum computer over a classical one for
simulating complex quantum systems such as molecules. The next
important development was the invention by David Deutsch (1985) of
the first quantum algorithm with a computational
advantage over classical models of computation.
Almost a decade later, there was the discovery by Peter Schor (1985) that quantum
computers may be more efficient at solving the prime
factorization problem, a scheme widely used to secure the transmission of data. A couple of years later Lov Grover (1996) created and proved that his quantum algorithm for finding
an element in a large set of data
was more efficient than any possible classical algorithm.
The last breakthrough we want to mention is the
discovery, also by Peter Shor (1995), that quantum information can indeed be protected
against the pernicious effects of the environment.

Look at the Bibliography or popular science literature for more on the history of quantum computing.  

\section{Quantum Bits}

A \index{Computer}\emph{computer} is a physical device that, when supplied with the correct set of data, generally known as the \emph{input}, provides another set of data, the \emph{output}. From this general definition it follows that despite our familiarity with modern personal computers, a computer is not necessarily an electronic device. Actually, the first computer conceived and built under the supervision of Charles Babbage in the 19th century was a purely mechanical device with no electronics in it. 

If you think for a
moment about this wide-ranging definition, you will quickly realize that there are infinite
different ways we can write (encode) the initial message we want to communicate to the computer.
Ultimately, the way we should encode it will depend
on the language spoken by the device, that is,
the system of words and rules used by the computer to operate. As with human language, the basic elements of the language of the computer are the words and characters used to construct it.

To make the transition from classical to quantum information processing as smooth as possible, we will start reviewing the basics of classical information theory. Then, we will concentrate on the quantum case.

\subsection{Classical Bits}

As you certainly already know, the language spoken by ordinary computers is the \index{Binary system}\emph{binary system}. The latter assumes that every piece of
information, for example, a number, a letter or a color, has a unique expression as a finite sequence of zeros and ones. In the binary system, the number $39$ is written $100111$. Sometimes, by convention, the sequence $01000001$ is assigned to the letter A and $11111111~00000000~00000000$ to the color red.

These sequences of zeros and ones are called \index{Bit string}\emph{bit strings} and are somehow equivalent to the words used by humans. Each individual digit of a binary string
is called a \index{Bit}\emph{bit} (from \emph{bi}nary dig\emph{it}) and is the most basic piece of classical information. This is the analog of the letters used in alphabetic languages. The number of bits in a bit string is known as the \emph{size} of the string.

Here we will only be interested in the binary system applied to numbers. If you are given a positive integer number $N$ in the usual decimal system, the corresponding binary string will be given by the following formula,
\beq\label{bin-dec}
N=2^{n-1}b_1+2^{n-2}b_2+\ldots +2^0 b_n \longleftrightarrow b_1\,b_2\, \ldots \, b_n  \,.
\eeq
For example,
\begin{equation*}
  39=2^{5}b_1+2^{4}b_2+ 2^{3}b_3+2^{2}b_4+2^{1}b_5+2^0 b_6= 2^{5}1+2^{4}0+ 2^{3}0+2^{2}1+2^{1}1+2^0 1  \,,  
\end{equation*}
thus,
\begin{equation*}
39 \longleftrightarrow 100111 \,.
\end{equation*}
\begin{theorem}
\textup{Write the bit string equivalent to every natural number from $1$ to $20$.}
\end{theorem}
\begin{theorem}
\textup{Express $56$ and $83$ in binary notation.}
\end{theorem}

\subsection{Single Qubits}
The words a quantum computer understands, that is, the carriers of information, are called
\emph{quantum bits} or \index{Qubit}\emph{qubits}, for short. The simplest piece of quantum information is the \index{Single qubit, 1 qubit}\emph{single qubit} or 1 \emph{qubit}. It is a two-level quantum system described by a complex two-dimensional unit state vector 
\beq
|\psi\rangle = a|\varphi_1\rangle+b|\varphi_2\rangle \,,
\eeq
where $a$ and $b$ are complex numbers, $a,b\in \mathbb{C}^2$, and the vectors $|\varphi_1\rangle$ and $|\varphi_2\rangle$ are two arbitrary orthonormal vectors
spanning the Hilbert space $\mathcal{H}\cong \mathbb{C}^2$ where the qubit $|\psi\rangle$ lives. The real number $|a|^2$ is the probability of measuring the system in the state $|\varphi_1\rangle$ and $|b|^2$ the probability of measuring it in $|\varphi_2\rangle$. Of course, since the only possible outcomes of a measurement are $|\varphi_1\rangle$ and $|\varphi_2\rangle$, it follows that $|a|^2+|b|^2=1$. I remind you that the basis vectors $|\varphi_1\rangle$ and $|\varphi_2\rangle$ are chosen to be orthonormal, that is, $\langle \varphi_r|\varphi_s\rangle=\delta_{rs}$, where $r,s=1,2$, because we want the two states to be perfectly distinguishable. The symbol $\langle~|~\rangle$, of course, indicates the inner product on the Hilbert space 
$\mathcal{H}$.

\begin{theorem}
\textup{How is the inner product on a Hilbert space usually defined?}
\end{theorem}

If you are the sort of person that prefers to have a physical picture in mind, you may think of a qubit as an electron with two possible spins, a spin up $|\uparrow\,\rangle$ and a spin down $|\downarrow\,\rangle$, a photon with a vertical $|\uparrow\,\rangle\,$ and a horizontal $|\rightarrow\,\rangle$ polarization, or an atom with two energy level states $|E_0\rangle$ and $|E_1\rangle$.  We will not use explicitly any of these physical representations; however, at times it can be handy to have these pictures in mind. This is somehow analogous to the correspondence made in classical circuit
theory between the binary values 0 and 1 and a zero or non-zero voltage, respectively, along a piece of wire. In both cases, classical and quantum, a purely theoretical discussion can be carried out without paying attention to any of these real implementations. This is the approach we will take in these notes. 

Even though you already studied most of the quantum mechanics used in quantum computing, there are various conventions and original points of view that are worth following. To begin, we will express the state vector of a single qubit as follows,
\beq\label{1qubit}
|q\rangle = \alpha_0|0\rangle+\alpha_1|1\rangle \,.
\eeq
The notation $|q\rangle$ is unconventional. In fact, as usual in quantum mechanics, most authors use $|\psi\rangle$. However, we follow the standard convention employed in quantum computing and denote the orthonormal basis vectors by $|0\rangle$ and $|1\rangle$ to emphasize the similitude with the classical binary system. The set $\{|0\rangle,|1\rangle\}$ is known as the \index{Computational basis} \emph{computational basis}. If a state vector, say $|i\rangle$, can only take the values $|0\rangle$
or $|1\rangle$, it is usual to
simplify the notation by writing $i\in \{0,1\}$ or $i=0,1$ instead of
$|i\rangle\in \{|0\rangle,|1\rangle\}$. Notice that in our notation, if $i,j=0,1$, then $\langle i|j \rangle =\delta_{ij}$. We will use $\mathcal{H}_q\cong \mathbb{C}^2$ to refer to the Hilbert space of a single qubit.

Another useful set of orthonormal
vectors in the Hilbert space of a single qubit $\mathcal{H}_q$ is the so called \index{Hadamard basis}\emph{Hadamard basis} $\{|+\rangle,|-\rangle\}$. The latter is given in terms of the computational basis vectors by
\beq
|+\rangle=\frac{1}{\sqrt{2}}|0\rangle+\frac{1}{\sqrt{2}}|1\rangle \,,\qquad
|-\rangle=\frac{1}{\sqrt{2}}|0\rangle-\frac{1}{\sqrt{2}}|1\rangle \,.
\eeq 
The converse relations are
\beq
|0\rangle=\frac{1}{\sqrt{2}}|+\rangle+\frac{1}{\sqrt{2}}|-\rangle \,,\qquad
|1\rangle=\frac{1}{\sqrt{2}}|+\rangle-\frac{1}{\sqrt{2}}|-\rangle \,.
\eeq 
The state vector $|q\rangle$ of a single qubit can then be rewritten as $|q\rangle_H=\alpha_+|+\rangle+\alpha_-|-\rangle$, where
\beq
\alpha_+=\frac{\alpha_0+\alpha_1}{\sqrt{2}} \,, \qquad \alpha_-=\frac{\alpha_0-\alpha_1}{\sqrt{2}} \,.
\eeq
According to definition \eqref{1qubit}, the state vector of a single qubit is a function of the two complex numbers $\alpha_0$ and $\alpha_1$. That is, we can write more explicitly
\beq
| q(\alpha_0,\alpha_1) \rangle =\alpha_0|0\rangle+\alpha_1|1\rangle \,.
\eeq
Now, since two complex numbers are equivalent to four real numbers and the normalization condition imposes that $|\alpha_0|^2+|\alpha_1|^2=1$, these four numbers reduce to three. Additionally, since
two state vectors that differ by a global phase, in fact represent the same physical system,
the three real numbers finally reduce to two. The
new variables, that we denote $\theta$ and $\phi$, with $\theta\in [0,\pi]$ and $\phi\in [0,2\pi)$, can be chosen so that
\beq
|\alpha_0|=\cos(\theta/2) \,, \qquad |\alpha_1|=\sin(\theta/2) \,.
\eeq
Note that $|\alpha_0|^2+|\alpha_1|^2$ is still equal to 1. 
\begin{theorem}
\textup{Complete the missing steps.}
\end{theorem}
The general expression of the single-qubit state vector in these new variables is
\beq\label{1qubitbloch}
| q(\theta,\phi) \rangle =\cos(\theta/2)|0\rangle + e^{i\phi}\sin(\theta/2)|1\rangle \,. 
\eeq
In particular,
\beq
|q(0,\phi)\rangle=|0\rangle \,,\qquad 
|q(\pi,\phi)\rangle=|1\rangle \,.
\eeq
We also have,
\beq
|q(\pi/2,0)\rangle=\frac{1}{\sqrt{2}}|0\rangle+\frac{1}{\sqrt{2}}|1\rangle = |+\rangle\,,
\eeq
and 
\beq
|q(\pi/2,\pi)\rangle=\frac{1}{\sqrt{2}}|0\rangle-\frac{1}{\sqrt{2}}|1\rangle = |-\rangle\,.
\eeq 
This parametrization of the state vector of a single qubit has a useful visual representation. Suppose that the variables $\theta$ and $\phi$ are the usual spherical coordinates. Then, the state vector of a qubit will be represented by a point --- or arrow --- on the unit sphere. For example, the north pole corresponds to the basis state vector $|0\rangle$ and the south pole to $|1\rangle$. This unit sphere is called the \index{Bloch sphere} \emph{Bloch sphere}. 
\begin{figure}[H]
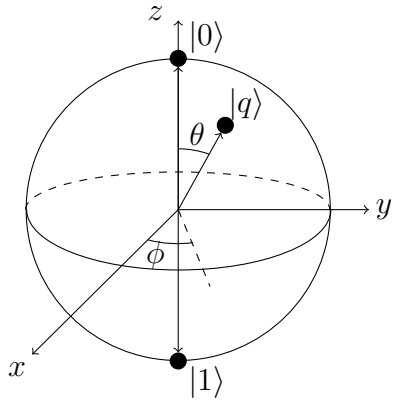

  \centering
\include{images/blochsphere}
 \caption{The Bloch sphere.}
 \label{fig:blochsphere}
\end{figure}
\begin{theorem}
\textup{What is the position of the Hadamard basis vectors $|+\rangle$ and $|-\rangle$ in the Bloch sphere?}
\end{theorem}
\begin{theorem}
\textup{Show that orthogonal states are anti-parallel in the Bloch sphere.}
\end{theorem}

\subsection{Multiple Qubits} 

If a single qubit is a quantum system whose state vector lives in a two-dimensional complex Hilbert space, $|q\rangle=|q_1\rangle\in \mathcal{H}_{q_1}\cong\mathbb{C}^2$,  
a \emph{2 qubit} is a quantum system whose state vector lives in a Hilbert space which is the tensor product of the Hilbert spaces of two single qubits, $|q_2\rangle\in \mathcal{H}_{q_2}=\mathcal{H}_{q_1} \otimes \mathcal{H}_{q'_1} \cong \mathbb{C}^{2^2}$. 

In order to have a clean notation for higher qubits, we will rewrite the state vector of a single qubit as follows,
\beq
|q_1\rangle=\alpha_0|0\rangle+\alpha_1|1\rangle=\sum_{i=0,1} \alpha_{i}|i\rangle \,\,.
\eeq
Following the same notation, the state vector of a 2 qubit is simply
\beq
|q_2\rangle
=\alpha_{00}|0\,0\rangle+\alpha_{01}|0\,1\rangle+\alpha_{10}|1\,0\rangle+\alpha_{11}|1\,1\rangle =\sum_{i,j} \alpha_{ij}|ij\rangle \,,
\eeq
where $i,j=0,1$. From now on, to avoid cluttering the formulas, we will assume that --- unless otherwise indicated --- the indices $i, j, k$ under the summation symbol take the values $0$ and $1$.
By convention, the first element in the ket $|i \, j\rangle$ represents a computational
basis vector of $\mathcal{H}_{q_1}$ and the second element a basis vector of $\mathcal{H}_{q'_1}$. Thus, $\langle i\,j | k\, l\rangle=\langle i | k \rangle\langle j | l \rangle=\delta_{ik}\delta_{jl}$. 
The mutually orthonormal states $|00\rangle, |01\rangle, |10\rangle$ and $|11\rangle$ form the \index{Computational basis} \emph{computational basis} of $\mathcal{H}_{q_2}$. 

\begin{theorem}
\textup{Do you remember how the inner product on $\mathcal{H}_{q_2}$ is given in terms of the inner products on the individual Hilbert spaces $\mathcal{H}_{q_1}$ and $\mathcal{H}_{q'_1}$?}
\end{theorem}

\noindent \textbf{Note:} If you had problems understanding the beginning of this section, I recommend you to read the following Box. It summarizes the main mathematical concepts and conventions we will use to describe multiple qubits. If you understood everything, then you can confidently skip it. 

\vspace{10pt}
\begin{tcolorbox}[breakable, enhanced]\vspace{5pt}
\begin{note}\label{box:tensorproductspaces}
$\mathrm{\mathbf{Tensor~ product~ spaces.}}$
\end{note} 
A 2 qubit, simply put, is the
composite system of two single qubits. Here we must remember that, since the single qubits can interact between them, the complete description
of the whole  2 qubit system may contain information that is not available at the level of the individual qubits.

~~The Hilbert space $\mathcal{H}_{q_2}$ of the composite system is
given by the \index{Tensor product} \emph{tensor product} of the two individual
Hilbert spaces,
\beq
\mathcal{H}_{q_2}=\mathcal{H}_{q}\otimes \mathcal{H}_{q'} \,.
\eeq
This means the following: given the single qubits $|q\rangle, |\tilde{q}\rangle\in \mathcal{H}_{q}$ and $|q'\rangle, |\tilde{q}'\rangle\in \mathcal{H}_{q'}$, the \emph{tensor product of two vectors} is a map
\beq
\otimes\colon \mathcal{H}_{q}\otimes \mathcal{H}_{q'} \to \mathcal{H}_{q}\otimes\mathcal{H}_{q'} \,,
\eeq
which satisfies
\beq
c(|q\rangle\otimes |q'\rangle)=(c|q\rangle)\otimes |q'\rangle=
|q\rangle\otimes(c|q'\rangle) \,,
\eeq
for every complex constant $c$, and 
\begin{align}
|q\rangle \otimes (|q'\rangle+|\tilde{q}'\rangle) &= |q\rangle\otimes |q'\rangle + |q\rangle\otimes|\tilde{q}'\rangle \,,\\
(|q\rangle +|\tilde{q}\rangle )\otimes (|q'\rangle)&= |q\rangle\otimes |q'\rangle + |\tilde{q}\rangle\otimes|q'\rangle \,.
\end{align}
We can use this definition of the tensor product between vectors to define the tensor product between entire Hilbert spaces. 

~~If $\{|0\rangle,|1\rangle \}$ is a basis for $\mathcal{H}_q$ and $\{|0'\rangle,|1'\rangle \}$, is a basis for $\mathcal{H}_{q'}$, the tensor product of these basis vectors, that is, $|0\rangle\otimes |0'\rangle$, $|0\rangle\otimes |1'\rangle$, $|1\rangle\otimes |0'\rangle$ and $|1\rangle\otimes |1'\rangle$ are basis vectors for $\mathcal{H}_{q_2}= \mathcal{H}_q \otimes \mathcal{H}_{q'}$. In other words, every element $|q_2\rangle$ in $\mathcal{H}_{q_2}$ has a unique expression of the form
\beq
|q_2\rangle=\sum_{i,j'}\alpha_{ij'}|i\rangle\otimes|j'\rangle \,,
\eeq
where the coefficients $\alpha_{ij'}$ are complex numbers.
Often, to lighten the notation, one drops the symbol $\otimes$ between the vectors. Additionally, one simply writes $|j\rangle$ instead of $|j'\rangle$ because it is clear that the second basis vector is in $\mathcal{H}_{q'}$. Thus, 
\beq
\mathcal{H}_{q_2}\ni |q_2\rangle=\sum_{i,j}\alpha_{ij}|i\rangle|j\rangle \in \mathcal{H}_{q} \otimes \mathcal{H}_{q'} \,.
\eeq 
A further simplification is to write $|i\,j\rangle$ instead of $|i\rangle|j\rangle$:
\beq
|q_2\rangle=\sum_{i,j}\alpha_{ij}|i\,j\rangle \,.
\eeq 
The inner product on the Hilbert space $\mathcal{H}_{q_2}$ is related to the inner products on $\mathcal{H}_{q}$ and $\mathcal{H}_{q'}$ by the following formula,
\begin{align}
\langle q_2|q'_2\rangle&=\Big(\sum_{i,j}\alpha_{ij} |i\,j\rangle , \sum_{k,l}\alpha'_{kl} |k\,l\rangle\Big)=\sum_{i,j,k,l} \alpha_{ij}^*\alpha'_{kl} \langle i\,j| k\,l\rangle \nonumber \\
&=\sum_{i,j,k,l}\alpha_{ij}^*\alpha'_{kl} \langle i| k\rangle \langle j| l\rangle =\sum_{i,j}\alpha_{ij}^*\alpha'_{ij} \,.
\end{align}
\begin{theorem}
\textup{How would you define the tensor product between $n$ single-qubit Hilbert spaces?}
\end{theorem}
\vspace{5pt}
\end{tcolorbox}
\vspace{10pt}

Remember that composite quantum systems, such as 2 qubits, can be \index{Entangled state}\emph{entangled}, namely, can be in a physical state whose corresponding vector cannot be written as the tensor product
of single qubits. In other words, an entangled state is not a \index{Product state}\emph{product state}. What we mean by this is the following: if we multiply two
single qubits,
\begin{align}\label{entangled2q}
|q\rangle|q'\rangle&=\big(\alpha_0|0\rangle+\alpha_1|1\rangle\big)\big(\alpha_{0'}|0'\rangle+\alpha_{1'}|1'\rangle\big)\nonumber \\[5pt]
&=\alpha_0\alpha'_0|0\,0\rangle+\alpha_0\alpha'_1|0\,1\rangle+\alpha_1\alpha'_0|1\,0\rangle+\alpha_1\alpha'_1|1\,1\rangle \nonumber \\[3pt]
&=\sum_{i,j} \alpha_i\alpha'_j |i\,j\rangle \,,
\end{align}
the entangled states in $\mathcal{H}_{q_2}$ are those for which $\alpha_{ij}\neq \alpha_i\alpha'_j$.

Entangled states are a purely quantum phenomenon. They generally result from the interaction of two or more quantum systems.

\begin{theorem}
\textup{Convince yourself that $1/\sqrt{2}(|0\,0\rangle+|1\,1\rangle)$ is an entangled state.}
\end{theorem} 

For 3 qubits, the definition is similar: $ |q_3\rangle\in \mathcal{H}_{q_3}=\mathcal{H}_{q'_1}\otimes\mathcal{H}_{q''_1}\otimes \mathcal{H}_{q'''_1}\cong \mathbb{C}^{2^3}$. In the computational basis $\{|i\,j\,k\rangle\}$ of $\mathcal{H}_{q_3}$,
\beq
|q_3\rangle =\sum_{i,j,k} \alpha_{ijk}|i\,j\,k\rangle \,.
\eeq
\begin{theorem}
\textup{What condition is satisfied by the
entangled states in $\mathcal{H}_{q_3}$?}
\end{theorem}
\begin{theorem}
\textup{Does the 3-qubit state vector $1/\sqrt{2}(|0\,0\,0\rangle+|1\,1\,1\rangle)$, known as the \index{GHZ state} \emph{GHZ state}, represents an entangled system?}
\end{theorem} 

The generalization to $n$ qubits is straightforward. A \emph{multiple qubit} or \index{Multiple or $n$ qubit} $n$ \emph{qubit}, for $n\geq 2$, is a quantum system whose state vector $|q_n\rangle \in \mathcal{H}_{q_n}=\mathcal{H}_{q'_1}\otimes\ldots\otimes \mathcal{H}_{q'^n_1}\cong \mathbb{C}^{2^n}$. We will often use the notation $|Q\rangle=|q_n\rangle$ and $\mathcal{H}_Q=\mathcal{H}_{q_n}$. In the \index{Computational basis}\emph{computational basis} $\{|i_1\ldots i_n\rangle\}$ of $\mathcal{H}_Q$, the multiple qubit state vector $|Q\rangle$ is given by the linear combination
\beq
|Q\rangle =\sum_{i_1, \ldots, i_n}\alpha_{i_1 \ldots i_n}|i_1\ldots i_n\rangle \,,
\eeq
where the coefficients $\alpha_{i_1 \ldots i_n}$ are complex numbers.
\begin{theorem}
\textup{What is the condition satisfied by the
entangled states in $\mathcal{H}_Q$?}
\end{theorem}
To simplify the notation further, usually the bit
string $i_1 \ldots i_n$ appearing in the state vector $|i_1\ldots i_n\rangle$ is expressed in decimal notation using \eqref{bin-dec},
\beq\label{qubitindecnotation}
|Q\rangle =\sum_{x=0}^{2^n-1} \alpha_x |x \rangle \,.
\eeq
For example, a 2 qubit can alternatively be written in binary or decimal notation
\begin{align}
|q_2\rangle&=\alpha_{00}|0\,0\rangle+\alpha_{01}|0\,1\rangle+\alpha_{10}|1\,0\rangle+\alpha_{11}|1\,1\rangle\nonumber\\[5pt]
&=\alpha_{0}|0\rangle+\alpha_{1}|1\rangle+\alpha_{2}|2\rangle+\alpha_{3}|3\rangle \,.
\end{align}
Even though the first two terms in the last line look exactly the same as the definition \eqref{1qubit} of a single qubit state vector, there is no risk of confusion because the context will always clearly indicate the one we will be dealing with.

\section{Quantum Circuits}
Before we start building a computer, we need to decide in
advance what sort of tasks it will perform and find
the most efficient way of achieving them. Later on we
will have time to come back to the concept of efficiency
in computer science. However, let us give you an
intuitive idea. Suppose we have to automatically generate and
tabulate the values of a given polynomial function
between two real numbers. To do this, we can use Babbage's ``Difference Engine," a heavy, slow and
expensive mechanical device. In principle, there is nothing wrong
with it. However, I think we all agree that today this is not the
most efficient way of performing our tasks. That is, it is
not enough to come up with clever theoretical ideas;
these ideas must be transformable into practical devices
that can process information efficiently. This interplay
between theoretical and practical aspects is key in
computer science. It was the invention of the
transistor in 1947 that consolidated the classical
circuit model of computation and gave rise to modern computers.
We start this section with a brief overview of
digital circuits to better understand how quantum
computing relies on, but also goes beyond this classical model.

\subsection{Classical Circuit Gates}

As we said, an ordinary digital computer
understands the binary language of zeros and ones. We provide our computer with a string of zeros and ones (the input), it processes them and at the end it delivers a new string of zeros and ones (the output). This process, which can be mechanical, electric, or of any other physical nature, is in general
expressed mathematically by a function $f$ from the space of bit strings of size $l$ to the space of bit strings of size $m$, $f\colon \{0,1\}^l\to \{0,1\}^m$. These functions are called\index{Boolean function}\emph{(vector-valued) Boolean functions}. Here we are interested in these functions, that is, in the way the device processes information.

Computer science
is a subject that, at least as we approach it here,
is at its core in part theoretical and in part practical. Let us say we have a
Boolean function $f\colon \{0,1\}^l\to \{0,1\}^m$ and we want to build a device that performs the same operation as $f$. How should we proceed? Theoretical computer scientists have arrived at the conclusion that any binary function $f$, no matter how difficult it is, can always be reconstructed by using a combination  of functions that are actually easier to materialize in the real world. These more elementary functions are called \index{Gate (classical logic)}\emph{elementary} or \emph{basic logic gates}. 
This is the essence of the \index{Circuit model of computation (classical)}\emph{classical circuit model of computation}. 

The \index{NOT gate (classical)}\emph{NOT gate} is one of these classical basic functions,
\beq
\mathrm{NOT} \colon \{0,1\} \to \{0,1\}\,, \qquad b \mapsto \mathrm{NOT} (b)=\bar{b}\,.
\eeq
The bar over the letter $b$ denotes the logic negation of the bit $b$. In simple words, if the input is 0, then the output is 1, and
vice versa. We can also represent the action of the NOT gate on a bit as follows,
\beq
0\xmapsto{~\mathrm{NOT}~}1 \,, \qquad 1\xmapsto{~\mathrm{NOT}~}0 \,.
\eeq
The next basic gate is the \emph{OR gate}, 
\beq
\mathrm{OR}\colon \{0,1\}^2 \to \{0,1\} \,, \qquad b_1b_2\mapsto \mathrm{OR}(b_1b_2) \,,
\eeq
given explicitly by,
\beq
0\,0\xmapsto{~\mathrm{OR}~}0 \,, \qquad 0\,1\xmapsto{~\mathrm{OR}~}1 \,, \qquad 1\,0\xmapsto{~\mathrm{OR}~}1 \,, \qquad 1\,1\xmapsto{~\mathrm{OR}~}1 \,.
\eeq
Note that, in contrast to the NOT gate, the input of an OR gate is a string of size $2$. So, we call it a $2$-bit gate.
The last basic gate on our list is the \emph{AND gate},
\beq
\mathrm{AND}\colon \{0,1\}^2 \to \{0,1\} \,,\qquad b_1b_2\mapsto \mathrm{AND}(b_1b_2) \,,
\eeq
which transforms
\beq
0\,0\xmapsto{~\mathrm{AND}~}0 \,, \qquad 0\,1\xmapsto{~\mathrm{AND}~}0 \,, \qquad 1\,0\xmapsto{~\mathrm{AND}~}0 \,, \qquad 1\,1\xmapsto{~\mathrm{AND}~}1 \,.
\eeq
The result we referred above establishes that
any Boolean function $f\colon \{0,1\}^l \to \{0,1\}^m$ can be expressed as a composition of these elementary gates. It is then said that the gates NOT, OR and AND form a \index{Universal set of gates (classical)}\emph{universal set of (classical) (logic) gates}.

Just as every component of an electric circuit
has a visual representation, the three electronic
gates just mentioned have also a corresponding \index{Circuit diagram (classical)}\emph{circuit diagram},
\begin{figure}[H]
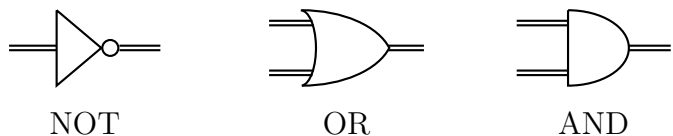

  \centering
  \include{images/classicalgates}
 \caption{Three classical basic electronic gates.}
 \label{fig:classicalgates}
\end{figure}
\noindent By convention, the inputs enter from the left of the gate and the outputs exist from the right. The double lines
represent the wires through which the data, namely,
the bit strings, flow to go from one gate to the next.
A classical circuit, which, as we said, can always be made using only
the NOT, OR and AND gates, will consequently have an associated
visual representation, in general a convoluted circuit diagram,
showing every single element necessary to build it and the relative position between them.

\subsection{Single-Qubit Gates}

As well as every Boolean function can be thought of
as a concatenation of elementary logic gates, we
will see that any unitary transformation on a qubit
can be decomposed into a sequence of elementary
quantum gates. 

As you know, according to quantum mechanics, the evolution of a quantum system is given by the action of a unitary operator on the state vector that describes the system at some moment in time. That is, if our quantum system is an $n$ qubit, it will evolve from its initial state $|Q_0\rangle$ to its final state $|Q_f\rangle$ according to $|Q_0\rangle\xmapsto{U}|Q_f\rangle = U|Q_0\rangle$. In quantum computing, unitary transformations acting on qubit state vectors, especially when the number of qubits is small, are also called \index{Gate (quantum)}\emph{(quantum logic) gates} or \index{Unitary}\emph{unitaries}. In this subsection we will only deal with unitaries on single qubits.
\begin{figure}[H]
 \centering
\input{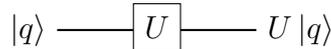}
 \caption{Circuit diagram of a single-qubit gate.}
 \label{fig:singlequbitgate}
\end{figure}
\noindent As for classical circuits, the qubits move from left to right. However, notice that we use single lines to represent the quantum communication channels (to distinguish them from the double lines we used above for classical wires).

\begin{theorem}
\textup{Why quantum transformations must be unitary, $U^{-1}=U^{\dagger}$?}
\end{theorem}

Because the Hilbert space of a single qubit is a 2-dimensional vector space, it is usual to express the computational basis vectors in
column vector notation, 
\beq\label{q1computationalbasis}
|0\rangle=
\begin{bmatrix}
1 \\
0
\end{bmatrix} \,, \qquad 
|1\rangle=
\begin{bmatrix}
0 \\
1
\end{bmatrix} \,.
\eeq 
\begin{theorem}
\textup{Show that the matrices assigned to the computational basis vectors are indeed consistent with the orthonormality condition we imposed on them.}
\end{theorem}
\noindent With this choice, the state vector of the single qubit \eqref{1qubit} has the column vector form
\beq
|q\rangle=
\alpha_0\begin{bmatrix}
1 \\
0
\end{bmatrix} 
+\alpha_1
\begin{bmatrix}
0 \\
1
\end{bmatrix} 
=
\begin{bmatrix}
\alpha_0 \\
\alpha_1
\end{bmatrix} 
\,.
\eeq
Correspondingly, its evolution will be determined by a \emph{single-qubit gate}\index{Single-qubit gate} represented by a $2\times 2$ matrix
\beq\label{2x2matrix}
U=
\begin{bmatrix}
U_{00} & U_{01} \\
U_{10} & U_{11}
\end{bmatrix}
\,.
\eeq 
Then, when the single qubit $|q\rangle$ enters the gate $U$, on the other side of the gate exists a state 
\beq
U|q\rangle=
\begin{bmatrix}
U_{00}\alpha_0 +U_{01}\alpha_1 \\
U_{10}\alpha_0 + U_{11}\alpha_1
\end{bmatrix}
\,.
\eeq 

\begin{theorem}
\textup{Show that in index notation
\beq
U|0\rangle=\sum_i U_{i0}|i\rangle \,, \qquad 
U|1\rangle=\sum_i U_{i1}|i\rangle \,,
\eeq
and thus, more generally,
\beq
U|q\rangle=\sum_{i,j} \alpha_j U_{ij} |i\rangle \,.
\eeq}
\end{theorem}
\noindent If we are not given the explicit matrix representation of the single-qubit gate as in \eqref{2x2matrix}, but only its action on the computational basis vectors, the single-qubit gate is abstractly given by the ket-bra expression
\beq
U=\sum_{i,j} U_{ij}|i\rangle\langle j| \,.
\eeq
From here, we can find the matrix by using the following formula:
\beq\label{2x2matrixinnerproducts}
U=
\begin{bmatrix}
\langle 0|U|0\rangle & \langle 0|U|1\rangle \\
\langle 1|U|0\rangle & \langle 1|U|1\rangle
\end{bmatrix}
\,.
\eeq 
That is, the elements of a $2\times 2$ matrix associated to a single-qubit gate are given by
\beq
U_{ij}=\langle i |U|j\rangle \,.
\eeq

If a single qubit enters two gates, first $U_1$ and then $U_2$, quantum mechanics tells us that the outgoing qubit will be $U_2(U_1|q\rangle)$.
\begin{figure}[H]
  \centering
\input{images/2singlequbitgates}
 \caption{Two consecutive single-qubit gates.}
 \label{fig:2singlequbitgates}
\end{figure}
\begin{theorem}
\textup{Show that
\beq
U_2U_1|q\rangle=\sum_{i,j,k} \alpha_j U_{2,ik}U_{1,kj}|i\rangle \,,
\eeq
where $U_1$ and $U_2$ are two arbitrary single-qubit gates. Generalize this formula to $N$ consecutive gates.}
\end{theorem}
A set of unitary transformations that play a key role in quantum computation and communication are the \index{Pauli matrices}\emph{Pauli matrices} (the same Pauli matrices you certainly encountered when you studied the spin of the electron):
\beq
\sigma _X=X=
\begin{bmatrix}
0 & 1 \\
1 & 0
\end{bmatrix} \,,\qquad 
\sigma_Y=Y=
\begin{bmatrix}
0 & -i \\
i & 0
\end{bmatrix} \,, \qquad 
\sigma_Z=Z=
\begin{bmatrix}
1 & 0 \\
0 & -1
\end{bmatrix}
\,.
\eeq
Most of the time we will refer to them as the $X, Y, Z$ \emph{gates} because this
is how they are actually called in quantum
computing. However, as we will see, the $\sigma$ notation is sometimes useful. 

\noindent Among the many properties of the Pauli matrices, I start by reminding you they are Hermitian, $\sigma_a^{\dagger}=\sigma_a$. In our notation $a=X,Y,Z$. From the physical point of view this is important because it is telling us that the Pauli matrices are observables. 
\begin{theorem} 
\textup{Show that every Pauli matrix $\sigma_a$ is its own inverse, that is, $(\sigma_a)^2=I$, where $I$ is the identity matrix. Verify that, however, the product of two different Pauli matrices satisfy $\sigma_a\sigma_b=-\sigma_b\sigma_a$.}
\end{theorem} 
\begin{theorem} 
\textup{Prove that any complex $2\times 2$ matrix can be uniquely written as a linear combination of the
Pauli matrices and the identity.}
\end{theorem}
\noindent If we apply the Pauli matrices on the computational basis vectors, we get
\begin{equation*}
X|0\rangle=
\begin{bmatrix}
0 & 1 \\
1 & 0
\end{bmatrix}
\begin{bmatrix}
1 \\
0
\end{bmatrix}=
\begin{bmatrix}
0 \\
1
\end{bmatrix}
=|1\rangle
\,, \qquad 
X|1\rangle=
\begin{bmatrix}
0 & 1 \\
1 & 0
\end{bmatrix}
\begin{bmatrix}
0 \\
1
\end{bmatrix}=
\begin{bmatrix}
1 \\
0
\end{bmatrix}
=|0\rangle \,,
\end{equation*}

\begin{equation*}
Y|0\rangle=
\begin{bmatrix}
0 & -i \\
i & 0
\end{bmatrix}
\begin{bmatrix}
1 \\
0
\end{bmatrix}=
\begin{bmatrix}
0 \\
i
\end{bmatrix}
=i|1\rangle
\,, \qquad 
Y|1\rangle=
\begin{bmatrix}
0 & -i \\
i & 0
\end{bmatrix}
\begin{bmatrix}
0 \\
1
\end{bmatrix}=
\begin{bmatrix}
-i \\
0
\end{bmatrix} 
=-i|0\rangle \,,
\end{equation*}

\begin{equation*}
Z|0\rangle=
\begin{bmatrix}
1 & 0 \\
0 & -1
\end{bmatrix}
\begin{bmatrix}
1 \\
0
\end{bmatrix}=
\begin{bmatrix}
1 \\
0
\end{bmatrix}
=|0\rangle
\,, \qquad 
Z|1\rangle=
\begin{bmatrix}
1 & 0 \\
0 & -1
\end{bmatrix}
\begin{bmatrix}
0 \\
1
\end{bmatrix}=
\begin{bmatrix}
0 \\
-1
\end{bmatrix} 
=-|1\rangle
\,.
\end{equation*}
This set of relations established by the Pauli matrices between the computational basis vectors, allow us to define the abstract opetators
\begin{alignat}{3}
X|0\rangle &=|1\rangle \,, \qquad
\hspace{3pt}X|1\rangle &&=|0\rangle \,,\\[5pt]
Y|0\rangle &=i|1\rangle \,, \qquad Y|1\rangle &&=-i|0\rangle \,,\\[5pt]
Z|0\rangle &=|0\rangle \,, \qquad \hspace{4pt}Z|1\rangle&&=-|1\rangle \,.
\end{alignat}
\begin{theorem} 
\textup{Use the formula \eqref{2x2matrixinnerproducts} to check that these operators indeed have the Pauli matrices as representations.}
\end{theorem} 
\noindent Note that the Pauli operator $X$ flips the computational basis vectors, $X|i\rangle=|\bar{i}\rangle=|1-i\rangle$. So, its action is similar to the classical NOT gate, $\mathrm{NOT}(b)=\bar{b}=1-b$. This explains why in quantum computing the $X$ operator is called the \index{Bit flip gate} \index{NOT gate (quantum)}\emph{bit flit gate} and is usually denoted NOT. 
\begin{theorem} 
\textup{Compute $X, Y, Z$ on $|+\rangle$ and $|-\rangle$. Interpret your results.}
\end{theorem}
\begin{theorem} 
\textup{What is the geometric interpretation of the action of the Pauli matrices on vectors in the Bloch sphere \ref{fig:blochsphere}?}
\end{theorem} 

\noindent In ket-bra notation the Pauli operator $X$ takes the following form,
\beq
X=|1\rangle\langle0|+|0\rangle\langle1|\,.
\eeq
Or, in terms of the Hadamard basis vectors,
\beq
X=|+\rangle\langle+|+|-\rangle\langle-| \,.
\eeq
\begin{theorem} 
\textup{Find the ket-bra expressions for $Y$ and $Z$.}
\end{theorem}
\begin{theorem} 
\textup{Using the column vector representation of the computational basis vectors $|0\rangle$ and $|1\rangle$, show that, in fact, the ket-bra expressions above reproduce the Pauli matrices.}
\end{theorem}

Being Hermitian, the Pauli matrices can be used to define the following unitary operators,
\beq
R_x(\alpha)=e^{-iX\alpha/2} \,,\quad R_y(\beta)=e^{-iY\beta/2} \,, \quad R_z(\gamma)=e^{-iZ\gamma/2} \,.
\eeq
where $\alpha,\beta,\gamma\in [0,2\pi)$. 
They can be written more compactly as
\beq
R_a(\theta_a)=e^{-i\sigma_a\theta_a/2} \,.
\eeq
The operator $R_a(\theta_a)$ on a single qubit \eqref{1qubitbloch} acts as a rotation of $\theta_a$ radians about the $a$ axis. 
We can rewrite them using trigonometric functions,
\beq
R_a(\theta_a)=\cos(\theta_a/2)I-i\sin(\theta_a/2)\sigma_a \,,
\eeq
\begin{theorem}
\textup{Prove the previous identity.}
\end{theorem}
\begin{theorem}
\textup{Suppose that $\mathbf{\hat{n}}=n_x\boldsymbol{\hat{\i}}+n_y\boldsymbol{\hat{\j}}+n_z\mathbf{\hat{k}}$ is a unit normal vector on the Bloch sphere and  $\boldsymbol{\sigma}=\sigma_x\boldsymbol{\hat{\i}}+\sigma_y\boldsymbol{\hat{\j}}+\sigma_z\mathbf{\hat{k}}$. Show that a rotation of an angle $\theta_{\mathbf{\hat{n}}}$ about the axis defined by $\mathbf{\hat{n}}$ is given by
\beq
R_{\mathbf{\hat{n}}}(\theta_{\mathbf{\hat{n}}})=e^{-i\mathbf{\hat{n}}\cdot \boldsymbol{\sigma} \theta_{\mathbf{\hat{n}}}/2}=\cos(\theta_{\mathbf{\hat{n}}}/2)I-i\sin(\theta_{\mathbf{\hat{n}}}/2)\mathbf{\hat{n}}\cdot \boldsymbol{\sigma}\,.
\eeq}
\end{theorem}
Another single-qubit gate which is extensively used
in quantum computing is the \index{Hadamard gate} \emph{Hadamard gate}, defined by its action on the computational basis vectors as follows
\beq
H|0\rangle=\frac{1}{\sqrt{2}}|0\rangle+\frac{1}{\sqrt{2}}|1\rangle \,,\qquad 
H|1\rangle=\frac{1}{\sqrt{2}}|0\rangle-\frac{1}{\sqrt{2}}|1\rangle \,,
\eeq
that is,
\beq
H|0\rangle=|+\rangle \,, \qquad H|1\rangle=|-\rangle \,.
\eeq
Thus, if a single qubit enters a Hadamard gate, the outgoing state will be
\beq
H|q\rangle=H(\alpha_0|0\rangle+\alpha_1|1\rangle)=\alpha_0H|0\rangle+\alpha_1H|1\rangle=\alpha_0|+\rangle+\alpha_1|-\rangle \,.
\eeq
The Hadamard gate, then, takes a state vector in the computational basis and shift it to the Hadamard basis. The converse is also true because
\begin{align}
H|+\rangle &=\frac{1}{\sqrt{2}}H|0\rangle+\frac{1}{\sqrt{2}}H|1\rangle=\frac{1}{\sqrt{2}}|+\rangle+\frac{1}{\sqrt{2}}|-\rangle=|0\rangle \,, \\ 
H|-\rangle &=\frac{1}{\sqrt{2}}H|0\rangle-\frac{1}{\sqrt{2}}H|1\rangle=\frac{1}{\sqrt{2}}|+\rangle-\frac{1}{\sqrt{2}}|-\rangle=|1\rangle \,,
\end{align}
so, 
\beq
H|q\rangle_H=H(\alpha_+|+\rangle+\alpha_-|-\rangle)=\alpha_+H|+\rangle+\alpha_-H|-\rangle
=\alpha_+|0\rangle+\alpha_-|1\rangle \,.
\eeq
\begin{theorem}
\textup{What is the ket-bra expression of the Hadamard gate?}
\end{theorem}

Above we have chosen to introduce the Hadamard gate in terms of its abstract action on the computational basis vectors, however, we could as well have chosen the matrix viewpoint. As you can easily check (do it!), in the computational basis the Hadamard gate has the following matrix representation,
\beq
H=\frac{1}{\sqrt{2}}
\begin{bmatrix}
1 & 1 \\
1 & -1
\end{bmatrix}
\,.
\eeq
\begin{theorem}\label{H2eq1}
\textup{Compute $H^2$. How do you interpret this result?}
\end{theorem}
\begin{theorem} 
\textup{The Pauli and Hadamard gates satisfy the relation $\sigma_a=\pm H\sigma_b H$. Find these relations for all the Pauli gates. Matrices $M$, such as the Hadamard gate, that satisfy $\sigma_a=\pm M\sigma_b M^{\dagger}$, are called \index{Clifford gate}\emph{Clifford gates}. Show that the Pauli gates are Clifford gates themselves.}
\end{theorem}

So far we have seen the Pauli matrices, rotations and the Hadamard gate. Let us introduce a couple of other useful single-qubit gates.

We know that in quantum mechanics two state vectors that differ by a global phase, actually represent the same quantum system. In the case of a single qubit, we can write this as $|q\rangle\sim e^{i\phi}|q\rangle$. However, if we add a relative phase between the components of a qubit, the two state vectors describe different quantum systems, $\alpha_0|0\rangle+\alpha_1|1\rangle~\nsim \alpha_0|0\rangle+e^{i\phi}\alpha_1|1\rangle$. We can add this relative phase factor $e^{i\phi}$ by letting our qubit enter the following gate,
\beq\label{relphasegate}
P(\phi)|0\rangle=|0\rangle \,, \qquad P(\phi)|1\rangle=e^{i\phi}|1\rangle \,;
\eeq
or, in matrix form,
\beq
P(\phi)=
\begin{bmatrix}
1 & 0 \\
0 & e^{i\phi}
\end{bmatrix}
\,.
\eeq
This unitary is known as the \index{Relative phase gate} \emph{relative phase gate}.
A special case occurs when $\phi=\pi/2$, 
\beq
S|0\rangle=|0\rangle \,, \qquad S|1\rangle=e^{i\pi/2}|1\rangle=i|1\rangle \,.
\eeq
This is the \index{S gate} \emph{S gate}.
Another useful case is when $\phi=\pi/4$,
\beq
T|0\rangle=|0\rangle \,, \qquad T|1\rangle=e^{i\pi/4}|1\rangle=\frac{1}{\sqrt{2}}(1+i)|1\rangle\,.
\eeq
No surprise, this is called the \index{T gate} \emph{T gate}, but sometimes it is also called the \index{$\pi/8$ gate}\emph{$\pi/8$ gate}.
In summary, $P(\phi=\pi/2)=S$ and $P(\phi=\pi/4)=T$.
\begin{theorem}
\textup{Prove that the $S$ gate is a Clifford gate.}
\end{theorem}
\begin{theorem}
\textup{Do you see why the $Z$ gate is also known as the \index{Phase flip gate}\emph{phase flip gate}?}
\end{theorem}
\begin{theorem}
\textup{Why do you think the gate $R=HSH$ is often called the \index{$\sqrt{\textup{NOT}}$ gate}$\sqrt{\textup{NOT}}$ gate?}
\end{theorem}
\begin{theorem}
\textup{Compute $P^m(\phi)$ for $m=2,3,4, \ldots$. Consider then the cases $\phi=\pi/2$ and $\phi=\pi/4$. How are these $P^m$'s related to the other unitaries?}
\end{theorem}
\begin{theorem}
\textup{In general, a relative phase gate is not Hermitian. What condition must a relative phase gate satisfy in order to be Hermitian?}
\end{theorem}

\subsection{Multiple Single-Qubit Gates}

Before explaining how a general unitary transformation acts on an $n$ qubit, let us first consider the simpler case of a gate that acts independently on the $n$ single qubits of an $n$-qubit product state,
\beq
U|q_n\rangle=U_1\otimes \ldots \otimes U_n\big(|q'\rangle\ldots|q'^n\rangle\big)=U_1|q'\rangle\ldots U_n|q'^n\rangle \,.
\eeq
As you see, these special transformations do not produce any entanglement between the single qubits of the incoming product state. 
\begin{figure}[H]
 \centering
\input{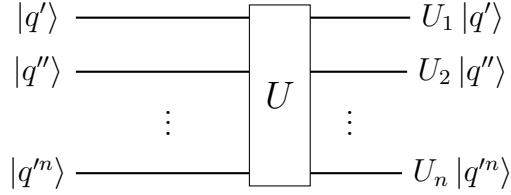}
 \caption{A non-entangling $n$-qubit gate.}
 \label{fig:galaxy}
\end{figure}
To be more precise, consider the action of $n$ independent Hadamard gates on the individual qubits of an $n$-product state,
\beq
H\otimes  \ldots \otimes H\big(|q'\rangle\ldots|q'^n\rangle\big)=H|q'\rangle\ldots H|q'^n\rangle \,.
\eeq
Let us start considering the easier cases. 

For a computational basis vector $|i\rangle\in \mathcal{H}_q$, a single \index{Hadamard gate}Hadamard gate acts as follows,
\begin{figure}[H]
 \centering
\input{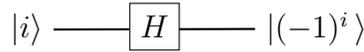}
 \caption{The Hadamard gate.}
 \label{fig:hadamardgate}
\end{figure}
\noindent Another useful way to write it is
\beq
H|i\rangle=\frac{1}{\sqrt{2}}|0\rangle+\frac{1}{\sqrt{2}}(-1)^i|1\rangle
=\frac{1}{\sqrt{2}} \sum_j (-1)^{ij}|j\rangle \,.
\eeq
Since $e^{i \pi}=-1$, a third common notation is
\beq
H|k\rangle=\frac{1}{\sqrt{2}}|0\rangle+\frac{1}{\sqrt{2}}e^{i\pi k}|1\rangle
=\frac{1}{\sqrt{2}} \sum_j e^{i\pi kj}|j\rangle \,.
\eeq
Note that in the last equation we used the letter $k$ instead of the usual $i$ to denote the computational basis vectors. We did this simply to avoid confusion with the imaginary $i$.

Thus, for a single qubit,
\begin{align}\label{Hon1qubit}
H|q\rangle&=H\sum_i\alpha_i|i\rangle=
\sum_i\alpha_i H|i\rangle
\nonumber\\[5pt]
&=\sum_i\alpha_i |(-1)^i\rangle=\frac{1}{\sqrt{2}} \sum_{i,j} (-1)^{ij}\alpha_i|j\rangle=\frac{1}{\sqrt{2}} \sum_{k,j}  e^{i\pi kj}\alpha_k|j\rangle \,.
\end{align}

\begin{theorem}
\textup{Use the index expressions above to prove that, as we already know from Exercise \ref{H2eq1}, $H(H|i\rangle)=|i\rangle$.}
\end{theorem}
Suppose now we have a product state $|i_1\rangle |i_2\rangle \in \mathcal{H}_{q_2}$ and we apply a Hadamard gate to each of the qubits,
\begin{figure}[H]
  \centering
  \input{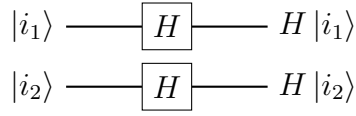}
 \caption{Two Hadamard gates in parallel.}
 \label{fig:cswap}
\end{figure}
\begin{align}
H|i_1\rangle H|i_2\rangle&=\frac{1}{\sqrt{2}}\bigg(|0\rangle+\frac{1}{\sqrt{2}}
(-1)^{i_1}|1\rangle\bigg)\frac{1}{\sqrt{2}}\bigg(|0\rangle+\frac{1}{\sqrt{2}}(-1)^{i_2}|1\rangle\bigg) \nonumber \\
&=\frac{1}{\sqrt{2^2}}\big(|00\rangle+(-1)^{i_2}|01\rangle+(-1)^{i_1} |10\rangle+(-1)^{i_1+i_2}|11\rangle \big) \nonumber \\
&=\frac{1}{\sqrt{2^2}} \sum_{j_1,j_2}(-1)^{i_1j_1+i_2j_2}|j_1\rangle|j_2\rangle \,.
\end{align}
We can rewrite the left hand side of this equation as follows,
\beq
H|i_1\rangle H|i_2\rangle=(H\otimes 1) (1\otimes H)|i_1\rangle|i_2\rangle=H\otimes H|i_1\rangle|i_2\rangle=H^{\otimes 2}|i_1\,i_2\rangle \,.
\eeq
The right hand side can also be written in a more compact and general form by using the notation $|x\rangle=|i_1~i_2\rangle$. Similarly, $|y\rangle=|j_1~j_2\rangle$. Putting these contributions together, we obtain
\beq\label{HHx}
H^{\otimes 2}|x\rangle=\frac{1}{\sqrt{2^2}}\sum_y(-1)^{x\cdot y}|y\rangle \,.
\eeq
Be aware that here $x$ denotes a binary string and not a decimal number as in equation \eqref{qubitindecnotation}. Moreover, $x\cdot y$ is a sort of dot product, $x\cdot y=i_1j_1+i_2j_2$, and not the multiplication of two decimal numbers. Finally, the sum over $y$ simply means
\beq
\sum_y=\sum_{j_1}\sum_{j_2}=\sum_{j_1,j_2}\,.
\eeq
You can easily generalize \eqref{HHx} to $n$ Hadamard gates acting independently on $n$ single qubits,
\beq\label{Hxnonx}
H^{\otimes n}|x\rangle=\frac{1}{\sqrt{2^n}}\sum_y (-1)^{x\cdot y}|y\rangle \,,
\eeq
where $x=i_1\ldots i_n$, $y=j_1\ldots j_n$ and $x\cdot y=i_1j_1+\ldots +i_nj_n$.
\begin{figure}[H]
 \centering
\input{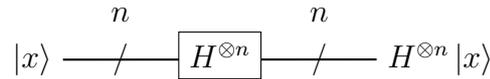}
 \caption{$n$ Hadamard gates in parallel acting on a computational basis vector of $\mathcal{H}_Q$.}
 \label{fig:galaxy}
\end{figure}
\begin{theorem} 
\textup{Explain how the general formula \eqref{Hxnonx} is obtained by considering three qubits, four qubits, etc.}
\end{theorem}
\begin{theorem}
\textup{Show that $H^{\otimes n}(H^{\otimes n}|x\rangle)=|x\rangle$.}
\end{theorem}
If we express the state vector of the $n$ qubit as a linear combination
\beq
|Q\rangle=\sum_x\alpha_x|x\rangle \,,
\eeq
the $n$ Hadamard gates will act according to
\begin{align}
H^{\otimes n}|Q\rangle=\sum_x\alpha_x H^{\otimes n}|x\rangle=\frac{1}{\sqrt{2^n}}\sum_{x,y}(-1)^{x\cdot y}\alpha_x|y\rangle \,.
\end{align}
Once again, remember that here, $x$ and $y$ are binary strings.

\subsection{Multi-Qubit Gates}

So far we have discussed quantum gates that act on single qubits,
the natural question now is: what about 2-qubit gates, 3-qubit gates, etc?
In principle, nothing prevents us from
conceiving quantum gates that act on $n$ qubits. In fact, the mathematical generalization is quite straightforward.
If $|Q\rangle$ is the state vector of an $n$ qubit, a general \index{Gate (quantum)}\emph{$n$-qubit gate} is a unitary transformation $U$ on $|Q\rangle$, $|Q\rangle \mapsto U|Q\rangle$. The only restriction on $U$ is that it must be unitary, $U^{-1}=U^{\dagger}$.
\begin{figure}[H]
 \centering
\input{images/multiqubitgate}
 \caption{A unitary transformation acting on an $n$ qubit.}
 \label{fig:multiqubitgate}
\end{figure}
Given that the computational basis vectors of $\mathcal{H}_Q=\mathcal{H}_{q'}\otimes \ldots\otimes\mathcal{H}_{q'^n} \cong \mathbb{C}^{2^n}$ are $|i_1\ldots i_n\rangle=|i_1\rangle\otimes \ldots \otimes  |i_n\rangle$, where 
$\{|i_r\rangle\}=\{|0\rangle,|1\rangle\}$ is the computational basis of $\mathcal{H}_{q'^r}$, $r=1,2,\ldots n$, every vector $|Q\rangle=\sum\alpha_{i_1\ldots i_n}|i_1\ldots i_n\rangle$ in  $\mathcal{H}_Q$ will have the following column vector representation,
\beq
|Q\rangle=\sum_{i_1,\ldots,i_n}\alpha_{i_1\ldots i_n}
\begin{bmatrix}
\delta_{i_10}\\
\delta_{i_11}
\end{bmatrix}
\otimes \ldots \otimes 
\begin{bmatrix}
\delta_{i_n0}\\
\delta_{i_n1}
\end{bmatrix}
\,,
\eeq
where $[\delta_{i_r0}~~\delta_{i_r1}]^T$ is the matrix representation of $|i_r\rangle$.

The unitary transformation $U$ will thus have a $2^n\times 2^n$ matrix representation,
\beq\label{2nx2nmatrixu}
U=
\begin{bmatrix}
U_{11} & \ldots & U_{12^n} \\
\ldots & \ldots & \ldots \\
U_{2^n1} & \ldots & U_{2^n2^n} 
\end{bmatrix}
\,.
\eeq
For example, the computational basis vectors of $\mathcal{H}_{q_2}$ are simply
\beq\label{q2cbmatrix}
|0\,0\rangle=
\begin{bmatrix}
1 \\ 0 \\ 0 \\ 0
\end{bmatrix}\,, \quad
|0\,1\rangle=
\begin{bmatrix}
0 \\ 1 \\ 0 \\ 0
\end{bmatrix}\,, \quad
|1\,0\rangle=
\begin{bmatrix}
0 \\ 0 \\ 1 \\ 0
\end{bmatrix}\,, \quad
|1\,1\rangle=
\begin{bmatrix}
0 \\ 0 \\ 0 \\ 1
\end{bmatrix}\,.
\eeq
 It follows that every 2-qubit state vector $|q_2\rangle=\sum\alpha_{ij}|i\,j\rangle$ will be represented by a column vector 
\beq
|q_2\rangle=
\alpha_{00}\begin{bmatrix}
1 \\ 0 \\ 0 \\ 0
\end{bmatrix}+
\alpha_{01}
\begin{bmatrix}
0 \\ 1 \\ 0 \\ 0
\end{bmatrix}+
\alpha_{10}
\begin{bmatrix}
0 \\ 0 \\ 1 \\ 0
\end{bmatrix}+
\alpha_{11}
\begin{bmatrix}
0 \\ 0 \\ 0 \\ 1
\end{bmatrix}=
\begin{bmatrix}
\alpha_{00} \\ \alpha_{01} \\ \alpha_{10} \\ \alpha_{11}
\end{bmatrix}
\,,
\eeq
and every unitary transformation on $|q_2\rangle$ will have the general $4 \times 4$ matrix form
\beq\label{U4x4}
U=
\begin{bmatrix}
U_{11} & U_{12} & U_{13} & U_{14} \\
U_{21} & U_{22} & U_{23} & U_{24} \\
U_{31} & U_{32} & U_{33} & U_{34} \\
U_{41} & U_{42} & U_{43} & U_{44}
\end{bmatrix}
\,,
\eeq
with $U_{rs}=U^*_{sr}$. 
\begin{theorem}
\textup{Can you find a way to rename the subscripts of the matrix elements $U_{rs}$ of \eqref{U4x4} so that the product $U|q_2\rangle$ has a tidy form in index notation?}
\end{theorem}

\vspace{10pt}
\begin{tcolorbox}[breakable, enhanced]
\vspace{5pt}
\begin{note} 
$\mathrm{\mathbf{Tensor~ product~ of ~operators.}}$
\end{note}
In \index{Tensor product}Box \ref{box:tensorproductspaces}, we recalled the definition of the tensor product of vectors as well as of entire Hilbert spaces. Now, we want to review how operators act on the individual Hilbert spaces of composite systems. 

~~Suppose two single qubits with Hilbert spaces $\mathcal{H}_q$ and $\mathcal{H}_{q'}$ and two operators acting on them,
\begin{align*}
A&\colon \mathcal{H}_q \to \mathcal{H}_q \,, \qquad ~\, |q\rangle \mapsto A|q\rangle \,,\\
B&\colon \mathcal{H}_{q'} \to \mathcal{H}_{q'} \,, \qquad |q'\rangle \mapsto B|q'\rangle \,.
\end{align*}
Let us say we form the 2-qubit system with Hilbert space $\mathcal{H}_ {q_2}= \mathcal{H}_q \otimes \mathcal{H}_{q'}$. We can associate to $A$ the operator $A \otimes 1\colon \mathcal{H}_ {q_2} \to \mathcal{H}_ {q_2}$, such that
\begin{equation*}
A\otimes 1|q_2\rangle=A\otimes 1 \Big(\sum_{i,j}\alpha_{ij}|i\,j\rangle\Big) = \sum_{i,j}\alpha_{ij}\big(A|i\rangle\big)\otimes 1|j\rangle=\sum_{i,j}\alpha_{ij}\big(A|i\rangle\big)|j\rangle \,.
\end{equation*}
A similar definition applies to the operator $B$. In general,
\beq
A\otimes B\Big(\sum_{i,j}\alpha_{ij}|i\,j\rangle\Big) = \sum_{i,j}\alpha_{ij}\big(A|i\rangle\big)\big(B|j\rangle\big)\,.
\eeq
\begin{theorem}
\textup{Show that a unitary transformation that entangles two single qubits cannot be expressed as the tensor product of two single-qubit gates.}
\end{theorem}
\begin{theorem}
\textup{Generalize everything said above for a Hilbert space that is the tensor product of $n$ single qubit spaces.}
\end{theorem}
Given two operators $A$ and $B$ and their respective matrix representations, to the tensor product $A \otimes B$ we associate the matrix
\beq
A\otimes B
=\begin{bmatrix}
    a_{11}&a_{12}\\
    a_{21}&a_{22}
\end{bmatrix}
\otimes
\begin{bmatrix}
    b_{11}&b_{12}\\
    b_{21}&b_{22}
\end{bmatrix}
=\begin{bmatrix}
    a_{11}B&a_{12}B\\
    a_{21}B&a_{22}B 
\end{bmatrix}
\,.
\eeq
The generalization to more than two operators and to higher order matrices is straightforward.

~~When there is no risk of confusion, we will drop the tensor product symbol $\otimes$ and simply write $AB$ for $A\otimes B$.
\begin{theorem}
\textup{Explain the choice \eqref{q2cbmatrix} for the basis vectors of $\mathcal{H}_{q_2}$.}
\end{theorem}
\vspace{5pt}
\end{tcolorbox}
\vspace{10pt}

One of the simplest $2^n\times 2^n$ unitary matrix transformations \eqref{2nx2nmatrixu} is the one formed by the tensor product of $n$ $2\times 2$ unitary matrices, 
\beq\label{UprodUs}
U=U_1\otimes \ldots \otimes U_n \,.
\eeq
This unitary acts on a product state $|Q\rangle=|q'\rangle\otimes \ldots \otimes |q'^{n}\rangle$ as follows,
\beq
U|Q\rangle=U_1\otimes \ldots \otimes U_n\big(|q'\rangle\otimes \ldots \otimes |q'^{n}\rangle\big)=U_1\big(|q'\rangle\ldots U_n|q'^{n}\rangle \,.
\eeq
Thus, the unitary \eqref{UprodUs} keeps the quantum state $|Q\rangle$ unentangled.
For instance, in the previous subsection we considered $U_1= \ldots =U_n=H$.

The advantage of a quantum computer over a classical one, though, is its ability to create and efficiently keep track of the superposition of all the possible states available to a quantum system. This includes, of course, entangled states. Thus, if we want to take full advantage of all the power of quantum mechanics, we need to introduce quantum gates that create entanglement. It can be proved that --- something we will not do here --- a single gate that produces entanglement, in addition to a complete set of single-qubit gates, is all we need to build any multi-qubit gate we want. The gate usually chosen is the so-called CNOT gate. We will first introduce it and then see how it enters into the production of other useful unitaries.

A quantum \index{Controlled gate} \emph{controlled gate} is a gate that operates on two qubits, one register by convention called the \index{Control qubit} \emph{control qubit} and the other the \index{Target qubit} \emph{target qubit}. While the
control qubit is a single qubit and it remains unchanged when passing through the gate, the
target qubit is in general an $n$ qubit and it gets modified depending on the value of the
control qubit. By definition, for $c\in \{0,1\}$, a controlled gate transforms
\beq
|c\rangle |Q_t\rangle \xmapsto{~~~} |c\rangle U(c)|Q_t\rangle \,,
\eeq
where $U(c)$ is a unitary on $|Q_t\rangle$ which action depends on the value of $c$.

The \index{Controlled-$U$ gate} \emph{controlled-U gate}
is defined as follows,
\beq
\textup{C}U|0\rangle|Q_t\rangle = |0\rangle|Q_t\rangle\,,\qquad
\textup{C}U|1\rangle|Q_t\rangle=|1\rangle U|Q_t\rangle \,.
\eeq
In ket-bra notation,
\beq\label{cuoperator}
\textup{C}U=|0\rangle\langle 0|\otimes 1 + |1\rangle\langle 1|\otimes U \,.
\eeq
Its circuit diagram is:
\begin{figure}[H]
  \centering
\input{images/controlledugate}
 \caption{The controlled-$U$ gate.}
 \label{fig:galaxy}
\end{figure}
\noindent Since there is nothing particular about the basis vector $|1\rangle$, we could as well have used the vector $|0\rangle$ to define a controlled gate. The latter is a \index{Controlled-$V$ gate} \emph{controlled}-$V$ \emph{gate},
\beq
\textup{C}V|0\rangle|Q_t\rangle = |0\rangle V|Q_t\rangle\,,\qquad
\textup{C}V|1\rangle|Q_t\rangle=|1\rangle |Q_t\rangle \,.
\eeq
As you can show,
\beq
\textup{C}V=|0\rangle\langle 0|\otimes V + |1\rangle\langle 1|\otimes 1 \,.
\eeq
The gate is commonly illustrated as follows,
\begin{figure}[H]
  \centering
\input{images/controlledvgate}
 \caption{The controlled-$V$ gate.}
 \label{fig:galaxy}
\end{figure}
\noindent We will almost exclusively deal with controlled-$U$ gates. 

Note that we can rewrite the definition of a controlled-$U$ gate as follows,
\beq
\textup{C}U|i\rangle|Q_t\rangle = |i\rangle U^i|Q_t\rangle \,,
\eeq
where 
\beq
U^i=\Bigg{\{}
\begin{matrix}
1~\mathrm{if}~i=0\\
U~\mathrm{if}~i=1 
\end{matrix}
\,.
\eeq
In particular, if we write the control qubit as $|c\rangle=\sum_{i}c_{i}|i\rangle$ and assume that the target qubit is a single qubit with state vector $|t\rangle=\sum_{j}t_{j}|j\rangle$, the transformation of a controlled-$U$ gate in index notation takes the general form
\beq\label{cugatec1qubit}
\textup{C}U(|c\rangle|t\rangle)=\sum_{i,j}c_{i}t_{j}|i\rangle U^i |j\rangle \,.
\eeq
\begin{theorem}
\textup{Prove that
\beq
\textup{C}V(|i\rangle|t\rangle)=|i\rangle V^{1-i}|t\rangle \,.
\eeq}
\end{theorem}
\noindent The matrix representation of a C$U$ gate on single qubits can easily be found:
\begin{align*}
\textup{C}U
\begin{bmatrix}
c_0t_0\\
c_0t_1\\
c_1t_0\\
c_1t_1
\end{bmatrix}
&=
c_0t_0
\begin{bmatrix}
1\\
0\\
0\\
0
\end{bmatrix}
+c_0t_1
\begin{bmatrix}
0\\
1\\
0\\
0
\end{bmatrix}
+c_1t_0
\begin{bmatrix}
0\\
0\\
\begin{bmatrix}
U&U\\
U&U
\end{bmatrix}
\begin{bmatrix}
1\\
0
\end{bmatrix}
\end{bmatrix}
+c_1t_1
\begin{bmatrix}
0\\
0\\
\begin{bmatrix}
U&U\\
U&U
\end{bmatrix}
\begin{bmatrix}
0\\
1
\end{bmatrix}
\end{bmatrix}\\[8pt]
&=\begin{bmatrix}
1 & 0 & 0 & 0\\
0 & 1 & 0 & 0\\
0 & 0 & 0 & 0\\
0 & 0 & 0 & 0
\end{bmatrix}
\begin{bmatrix}
c_0t_0\\
c_0t_1\\
0\\
0
\end{bmatrix}
+
\begin{bmatrix}
0 & 0 & 0 & 0\\
0 & 0 & 0 & 0\\
0 & 0 & U_{00} & U_{01}\\
0 & 0 & U_{10} & U_{11}
\end{bmatrix}
\begin{bmatrix}
0\\
0\\
c_1t_0\\
c_1t_1
\end{bmatrix} \\[8pt]
&=
\begin{bmatrix}
I & 0 \\
0 & U
\end{bmatrix}
\begin{bmatrix}
c_0t_0\\
c_0t_1\\
c_1t_0\\
c_1t_1
\end{bmatrix} \,.
\end{align*}
Thus, we have shown that a controlled-$U$ gate on single qubits has the following matrix representation,
\beq
\textup{C}U=\begin{bmatrix}
I & 0 \\
0 & U
\end{bmatrix} \,.
\eeq
\begin{theorem}
\textup{What is the matrix representation of a controlled-$V$ gate?}
\end{theorem}
For example, for a \index{Controlled-$X$ gate} \emph{controlled}-$X$ \emph{gate},
\beq
\textup{C}X\big(|c\rangle|t\rangle\big)
=
\textup{C}X\Big( 
\begin{bmatrix}
c_0 \\
c_1
\end{bmatrix}
\otimes 
\begin{bmatrix}
t_0 \\
t_1
\end{bmatrix}
\Big)=
\begin{bmatrix}
1 & 0 & 0 & 0\\
0 & 1 & 0 & 0\\
0 & 0 & 0 &1 \\
0 & 0 & 1 & 0
\end{bmatrix}
\begin{bmatrix}
c_0t_0\\
c_0t_1\\
c_1t_0\\
c_1t_1
\end{bmatrix}
=\begin{bmatrix}
c_0t_0\\
c_0t_1\\
c_1t_1\\
c_1t_0
\end{bmatrix}
 \,.
\eeq
\noindent If the control qubit is in the basis vector $|0\rangle=[1~0]^T$, we have
\begin{equation*}
\textup{C}X\big(|0\rangle|t\rangle\big)=
\begin{bmatrix}
1 & 0 & 0 & 0\\
0 & 1 & 0 & 0\\
0 & 0 & 0 &1 \\
0 & 0 & 1 & 0
\end{bmatrix}
\Big(
\begin{bmatrix}
1\\
0
\end{bmatrix}
\otimes
\begin{bmatrix}
t_0\\
t_1
\end{bmatrix}
\Big)=
\begin{bmatrix}
t_0\\
t_1\\
0\\
0
\end{bmatrix}
=\begin{bmatrix}
1 \\
0
\end{bmatrix}
\otimes 
\begin{bmatrix}
t_0 \\
t_1
\end{bmatrix}
=|0\rangle|t\rangle \,.
\end{equation*}
As expected, since any controlled-$U$ gate does nothing when the control qubit is in the state $|0\rangle$. If, on the other hand, $|c\rangle=|1\rangle=[0~1]^T$,
\begin{equation*}
\textup{C}X(|1\rangle|t\rangle)=
\begin{bmatrix}
1 & 0 & 0 & 0\\
0 & 1 & 0 & 0\\
0 & 0 & 0 &1 \\
0 & 0 & 1 & 0
\end{bmatrix}
\Big(
\begin{bmatrix}
0\\
1
\end{bmatrix}
\otimes
\begin{bmatrix}
t_0\\
t_1
\end{bmatrix}
\Big)=
\begin{bmatrix}
0\\
0\\
t_1\\
t_0
\end{bmatrix}
=\begin{bmatrix}
0 \\
1
\end{bmatrix}
\otimes 
\begin{bmatrix}
t_1 \\
t_0
\end{bmatrix}
=|1\rangle X|t\rangle \,.
\end{equation*}
\begin{theorem}
\textup{Show that a \index{Controlled-$Z$ gate} \emph{controlled}-$Z$ \emph{gate} transforms 
\beq
|0\rangle |t\rangle \xmapsto{~\textup{C}Z~} |0\rangle |t\rangle \,, \qquad
|1\rangle|t\rangle \xmapsto{~\textup{C}Z~} |1\rangle \sum_{j}(-1)^{j}t_{j}|j\rangle \,.
\eeq
What is the matrix corresponding to C$Z$?}
\end{theorem}
\begin{theorem}
\textup{A useful variant of the relative phase gate \eqref{relphasegate} is the \index{$R_l$ gate}$R_l$ \emph{gate} defined by
\beq
R_l|j\rangle=e^{\frac{2\pi i}{2^l}j}|j\rangle \,.
\eeq
Write its matrix representation. How would you define a controlled-$R_l$ gate? Write the corresponding matrix and draw the circuit diagram.}
\end{theorem}

The \emph{controlled-NOT} or \emph{CNOT gate}\index{CNOT gate}\index{Controlled-NOT gate} is another instance of controlled-$U$ gate on single qubits. For $i,j\in \{0,1\}$,
\beq
|i\rangle  |j\rangle \xmapsto{~\mathrm{CNOT}~} |i\rangle  |j\oplus i\rangle \,.
\eeq
The notation $i \oplus j$ is the standard way of denoting a binary sum: $i \oplus j=(i+j )\,\mathrm{mod}2$. For example, $0\oplus 0=0$, $0\oplus 1=1$, $1\oplus 0=1$ and $1\oplus 1=0$.
For the computational basis vectors, 
\begin{align*}
|0\rangle  |0\rangle &\xmapsto{~\mathrm{CNOT}~} |0\rangle  |0\oplus 0\rangle =|0\rangle  |0\rangle\,, \\
|0\rangle  |1\rangle &\xmapsto{~\mathrm{CNOT}~}  |0\rangle  |1\oplus 0\rangle = |0\rangle  |1\rangle\,,\\
|1\rangle  |0\rangle &\xmapsto{~\mathrm{CNOT}~}  |1\rangle  |0\oplus 1\rangle = |1\rangle  |1\rangle\,,\\
|1\rangle  |1\rangle &\xmapsto{~\mathrm{CNOT}~}  |1\rangle  |1\oplus 1\rangle = |1\rangle  |0\rangle\,.
\end{align*}
That is, 
\begin{align*}
|0\,0\rangle \xmapsto{~\mathrm{CNOT}~}  |0\,0\rangle \,,\quad |0\,1\rangle \xmapsto{~\mathrm{CNOT}~}  |0\,1\rangle \,, \quad
|1\,0\rangle \xmapsto{~\mathrm{CNOT}~}  |1\,1\rangle \,, \quad |1\,1\rangle \xmapsto{~\mathrm{CNOT}~}  |1\,0\rangle \,.
\end{align*}
From here, we read the matrix representation of the CNOT gate,
\beq
\mathrm{CNOT}=
\begin{bmatrix}
1 & 0 & 0 & 0\\
0 & 1 & 0 & 0\\
0 & 0 & 0 & 1\\
0 & 0 & 1 & 0
\end{bmatrix}
=\textup{C}X \,.
\eeq
We see that the CNOT gate is actually the same as the C$X$ gate. This is consistent with the fact that the $X$ gate flips the computational basis vectors $|0\rangle\leftrightarrow|1\rangle$ as well as the classical NOT gate flips the bits $0\leftrightarrow 1$. As we said, the CNOT gate is frequently used to create entanglement and build other unitary transformations.
\begin{figure}[H]
  \centering
\input{images/cnotcx}
 \caption{Circuit identity CNOT=C$X$.}
 \label{fig:galaxy}
\end{figure}
\begin{theorem}
\textup{Show that the CNOT gate is not the tensor product of two single-qubit gates. What is the physical meaning of this?}
\end{theorem}
\begin{theorem}
\textup{If $|i\rangle$ is a computational basis state vector, how would you write the transformed state $X|i\rangle$ using the $\oplus$ symbol?}
\end{theorem}

Another useful example of controlled-$U$ gate is the CSWAP gate. In this case the unitary $U$ is a 2-qubit gate known as the \index{SWAP gate}\emph{SWAP gate}, 
\beq
|u\rangle |b\rangle \xmapsto{~\textup{SWAP}~} |b\rangle|u\rangle \,.
\eeq
In components, the SWAP unitary interchanges $u_i\leftrightarrow b_i$. Its matrix representation can be obtained noting that
\beq
|u\rangle |b\rangle=
\begin{bmatrix}
u_0b_0 \\
u_0b_1 \\
u_1b_0 \\
u_1b_1 \\
\end{bmatrix}
\xmapsto{~\textup{SWAP}~} 
\textup{SWAP}
\begin{bmatrix}
u_0b_0 \\
u_0b_1 \\
u_1b_0 \\
u_1b_1 \\
\end{bmatrix}
=\begin{bmatrix}
b_0u_0 \\
b_0u_1 \\
b_1u_0 \\
b_1u_1 \\
\end{bmatrix}
=
\begin{bmatrix}
u_0b_0 \\
u_1b_0 \\
u_0b_1 \\
u_1b_1 \\
\end{bmatrix} \,.
\eeq
The matrix representation of the SWAP gate in the computational basis of $\mathcal{H}_{q_2}$ is then
\beq
\textup{SWAP}=
\begin{bmatrix}
1 & 0 & 0 & 0 \\
0 & 0 & 1 & 0 \\
0 & 1 & 0 & 0 \\
0 & 0 & 0 & 1 \\
\end{bmatrix}
\,.
\eeq
Graphically the SWAP gate is represented by the following diagram
\begin{figure}[H]
  \centering
\input{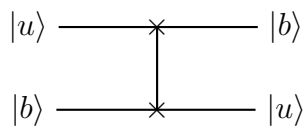}
 \caption{The SWAP gate.}
 \label{fig:galaxy}
\end{figure}
\begin{theorem}
\textup{
Show that the ket-bra expression for the SWAP gate is
\beq
\mathrm{SWAP}=\sum_{k,l}|k\, l\rangle\langle k\, l| \,.
\eeq
}
\end{theorem}
\begin{theorem}
\textup{
It is easy to check that the SWAP gate can be written as
\begin{align}
\mathrm{SWAP}=\frac{1}{2}\big(I\otimes I +X\otimes X +Y\otimes Y +Z\otimes Z\big)=\frac{1}{2}\sum_{A}\sigma_A\otimes \sigma_A \,,
\end{align}
where $A=I, X, Y, Z$. What is the physical reason for this?}
\end{theorem}
\begin{theorem}
\textup{Below is an illustration of the \index{Controlled-SWAP gate}\emph{controlled-SWAP gate} (\index{CSWAP gate}\emph{CSWAP}). What is the outgoing state $|\omega\rangle$?
\textup{\begin{figure}[H]
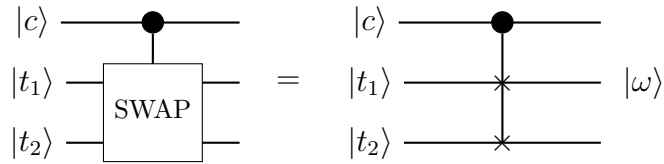

  \centering
  \include{images/cnotswap}
 \caption{The controlled-SWAP gate.}
 \label{fig:cswap}
\end{figure}}}
\end{theorem}
\noindent Now that we know how to create entangled states from product states using the CNOT gate, we would like to  known how to construct other multi-qubit gates using the CNOT gate. 

Since quantum gates are identified with unitary
transformations, then, according to the mathematical formalism of quantum mechanics, any gate will be the composition of certain unitary transformations (each of them, of course, corresponding to a particular gate),
\beq
|Q\rangle \mapsto U_1|Q\rangle \mapsto U_2 (U_1|Q\rangle) \mapsto \cdots 
\eeq
In terms of matrices, this means that any quantum gate will be equivalent to a product of matrices, each matrix corresponding to a gate in the circuit.
To illustrate how this works, consider the following circuit,
\begin{figure}[H]
  \centering
\input{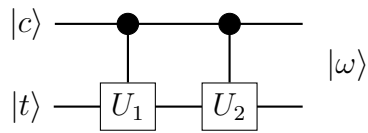}
 \caption{Two consecutive C$U$ gates.}
 \label{fig:galaxy}
\end{figure}
\begin{theorem}
\textup{Verify that each step below is correct:}
\end{theorem}\vspace{-20pt}
\begin{align*}
|c\rangle|t\rangle
&= c_0t_0|0\rangle|0\rangle+c_0t_1|0\rangle|1\rangle+c_1t_0|1\rangle|0\rangle+c_1t_1|1\rangle|1\rangle\\[5pt]
&\xmapsto {~\mathrm{C}U_1~}c_0t_0|0\rangle|0\rangle+c_0t_1|0\rangle|1\rangle+c_1t_0|1\rangle U_1|0\rangle+c_1t_1|1\rangle U_1|1\rangle\\[5pt]
&\xmapsto {~\mathrm{C}U_2~}c_0t_0|0\rangle|0\rangle+c_0t_1|0\rangle|1\rangle\\[5pt]
&~~~+c_1t_0|1\rangle \big(U_{1,00}U_2|0\rangle+ U_{1,10}U_2|1\rangle\big)+
c_1t_1|1\rangle \big(U_{1,01}U_2|0\rangle+ U_{1,11}U_2|1\rangle\big)\\[5pt]
&= c_0t_0|0\rangle|0\rangle+c_0t_1|0\rangle|1\rangle\\[5pt]
&~~~+\big[c_1t_0 \big(U_{1,00}U_{2,00}+ U_{1,10}U_{2,01}\big)+c_1t_1 \big(U_{1,01}U_{2,00}+ U_{1,11}U_{2,01} \big)\big] |1\rangle|0\rangle \\[5pt]
&~~~+ \big[c_1t_0 \big(U_{1,00}U_{2,10}+ U_{1,10}U_{2,11} \big)+c_1t_1 \big(U_{1,01}U_{2,10}+ U_{1,11}U_{2,11} \big)\big] |1\rangle|1\rangle \,,
\end{align*}
which is the product $\textup{C}U_2 \textup{C}U_1|c\rangle|t\rangle$.
\begin{theorem}
\textup{Compute the evolution of the incoming qubits as
they pass through the following gates,
\begin{figure}[H]
  \centering
\input{images/3gatesinsequence}
 \caption{}
 \label{fig:galaxy}
\end{figure}}
\end{theorem}
\begin{theorem}
\textup{Show the equivalence of the following circuits,
\begin{figure}[H]
  \centering
\input{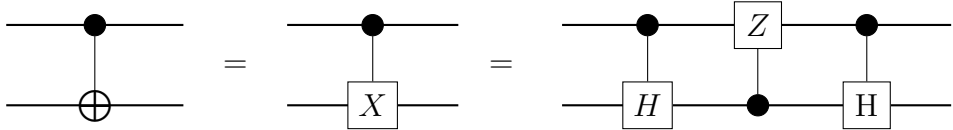}
 \caption{Circuit identity.}
 \label{fig:galaxy}
\end{figure}}
\end{theorem}
\begin{theorem}
\textup{What is the output state $|\omega\rangle$ of the circuit below? What if the C$V$ gate is followed by the C$U$ gate?
\begin{figure}[H]
  \centering
\input{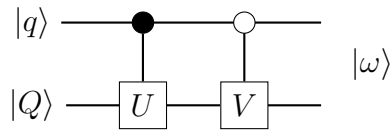}
 \caption{C$U$ gate followed by a C$V$ gate.}
 \label{fig:galaxy}
\end{figure}}
\end{theorem}
\begin{theorem}
\textup{Compare the outgoing states of the following circuits . Then, consider the special case $|t_i\rangle=|0\rangle$ for all $i=1,\ldots, N$.
\begin{figure}[H]
  \centering
\input{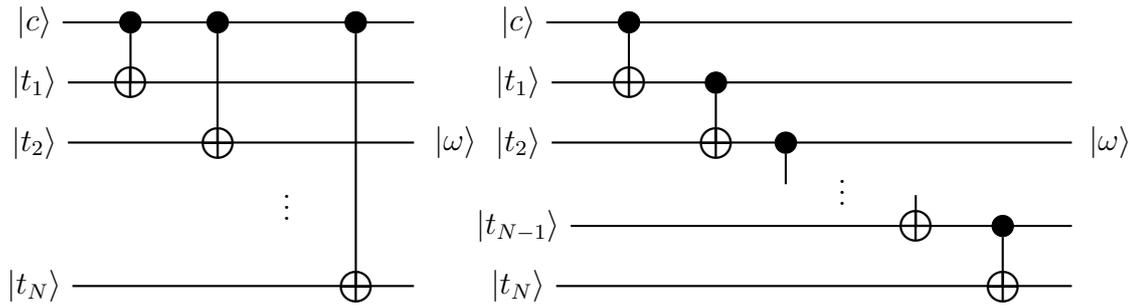}
 \caption{Circuits (a) and (b).}
 \label{fig:ncnotgates}
\end{figure}}
\end{theorem}
\begin{theorem}
\textup{What sequence of gates undo the action of the gates in Figure \ref{fig:ncnotgates}(b)?}
\end{theorem}
\begin{theorem}\label{exer:Zxnsimulation}
\textup{What is the outgoing state of the circuit below? Then, consider the case $U=R_z(\theta)$. 
\begin{figure}[H]
  \centering
\input{images/ncnotgates2}
 \caption{}
 \label{fig:galaxy}
\end{figure}}
\end{theorem}
If, as we said, any multi-qubit gate can be constructed using the CNOT gate and a set of universal single-qubit gates, we should be able to prove that the SWAP and CSWAP gates are concatenations of CNOT gates. The circuit that does it is shown below:
\begin{figure}[H]
  \centering
\input{images/swapcnot3}
 \caption{$\mathrm{CNOT}^3=\mathrm{SWAP}$.}
 \label{fig:CNOT3SWAP}
\end{figure}
\noindent Let us prove that it does exactly what we want:
\begin{align*}
|c\rangle|t\rangle=
\begin{bmatrix}
c_0\\
c_1
\end{bmatrix}
\otimes
\begin{bmatrix}
t_0\\
t_1
\end{bmatrix}&=
\begin{bmatrix}
c_0t_0\\
c_0t_1\\
c_1t_0\\
c_1t_1
\end{bmatrix}\nonumber \\[5pt]
&=c_0t_0|0\rangle|0\rangle+c_0t_1|0\rangle|1\rangle+c_1t_0|1\rangle|0\rangle+c_1t_1|1\rangle|1\rangle\\
&\xmapsto[~]{~\mathrm{CNOT}_2~}c_0t_0|0\rangle|0\rangle+c_0t_1|0\rangle|1\rangle+c_1t_0|1\rangle|1\rangle+c_1t_1|1\rangle|0\rangle\\
&\xmapsto[~]{~\mathrm{CNOT}{_1}~}c_0t_0|0\rangle|0\rangle+c_0t_1|1\rangle|1\rangle+c_1t_0|0\rangle|1\rangle+c_1t_1|1\rangle|0\rangle\\
&\xmapsto[~]{~\mathrm{CNOT}_2~}c_0t_0|0\rangle|0\rangle+c_0t_1|1\rangle|0\rangle+c_1t_0|0\rangle|1\rangle+c_1t_1|1\rangle|1\rangle\\
&=\begin{bmatrix}
c_0t_0\\
c_1t_0\\
c_0t_1\\
c_1t_1
\end{bmatrix}
=
\begin{bmatrix}
t_0\\
t_1
\end{bmatrix}
\otimes
\begin{bmatrix}
c_0\\
c_1
\end{bmatrix}
=|t\rangle|c\rangle \,.
\end{align*}
Thus, three consecutive CNOT gates acting as shown in Figure \ref{fig:CNOT3SWAP} are equivalent to a single SWAP gate. We can symbolically write this identity as $\mathrm{SWAP}=\mathrm{CNOT}_2\mathrm{CNOT}_1\mathrm{CNOT}_2$. Sometimes this identity is simply written $\mathrm{SWAP}=\mathrm{CNOT}^3$.\\

\noindent In index notation the proof is as follows,
\begin{align*}
|c\rangle|t\rangle=\sum_i c_i |i\rangle \sum_{j}t_{j}|j\rangle &=\sum_{i,j} c_i t_{j}|i\rangle|j\rangle \\
&\xmapsto[~]{~\mathrm{CNOT}_2~} \sum_{i,j} c_i t_{j}|i\rangle|i\oplus j\rangle \\
&\xmapsto[~]{~\mathrm{CNOT}_1~} \sum_{i,j} c_i t_{j}|i\oplus j \oplus i\rangle|i\oplus j\rangle \\
&\xmapsto[~]{~\mathrm{CNOT}_2~} \sum_{i,j} c_i t_{j}|i\oplus j \oplus i\rangle|i\oplus j\oplus i \oplus i\oplus j\rangle \,.
\end{align*}
But, since computational basis vectors satisfy $|i\oplus i\rangle=|0\rangle$, then
\begin{align*}
|c\rangle|t\rangle \xmapsto{~\mathrm{CNOT}_2~}\xmapsto[~]{~\mathrm{CNOT}_1~}\xmapsto[~]{~\mathrm{CNOT}_2~}&\sum_{i,j} c_i t_{j}|j\rangle|i\rangle=\sum_{j} t_{j}|j\rangle \sum_i c_i|i\rangle=|t\rangle|c\rangle \,.
\end{align*}
\begin{theorem}
\textup{Prove the circuit identity $\mathrm{SWAP}=\mathrm{CNOT}^3$ by using the ket-bra form \eqref{cuoperator} of the controlled-$U$ gates.}
\end{theorem}
\begin{theorem}
\textup{Show that the CSWAP gate can be implemented by the following equivalent sequence of gates:
\textup{\begin{figure}[H]
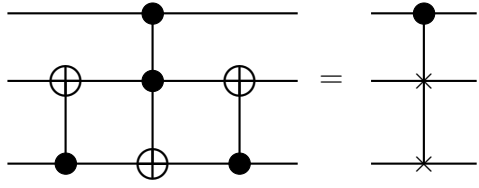

  \centering
  \vspace{10pt}
  \include{images/cswapnot3}
\vspace{-20pt}
 \caption{A sequence of gates equivalent to the CSWAP gate.}
 \label{fig:cswap}
\end{figure}}}
\end{theorem}
In the same spirit, we can construct quantum gates that operate on more than 2 qubits. One such gate is the \index{Controlled-controlled NOT gate}\emph{controlled-controlled NOT gate}, most commonly known as the \index{CCNOT gate} \emph{CCNOT} or the  \index{Toffoli gate}\emph{Toffoli gate}. It 
is a 3-qubit gate that transforms
\beq
|i\rangle |j\rangle |k\rangle \xmapsto{~\mathrm{TOFF}~}
|i\rangle |j\rangle |k\oplus ij\rangle \,.
\eeq
If we write the \index{Control qubit, first}\emph{first control qubit} as $c_1=\sum_i c_{1,i}|i\rangle$, the \index{Control qubit, second}\emph{second control qubit} as $c_2=\sum_i c_{2,j}|j\rangle$ and the target qubit as $t=\sum_{k} t_{k}|k\rangle$, the Toffoli gate transforms 
\beq
|c_1\rangle |c_2\rangle |t\rangle =\sum_{i,j,k}
c_{1,i}c_{2,j}t_{k}|i\rangle |j\rangle |k\rangle\xmapsto{~\mathrm{TOFF}~} \sum_{i,j,k}
c_{1,i}c_{2,j}t_{k}|i\rangle |j\rangle |k \oplus ij\rangle \,.
\eeq

\subsection{Measurement}

From the beginning of these notes I have assumed that
you are familiar with the crucial role played by the measurement
process in quantum mechanics. For example, in the pages above I took for granted that you knew that for a single
qubit $|q\rangle=\sum_i\alpha_i|i\rangle$, the probability of measuring
the state $|i\rangle$ is $|\alpha_i|^2$. Of course, this is simply because
\beq
P(|i\rangle)=|\langle i | q \rangle|^2=\Big|\langle i | \sum_j \alpha_j |j \rangle\Big|^2=
\Big|\sum_j \alpha_j \langle i |j \rangle\Big|^2=\Big|\sum_j\alpha_j \delta_{ij}\Big|^2=|\alpha_i|^2 \,.
\eeq
In terms of the projectors on the computational basis vectors, $P_i=|i\rangle\langle i|$, the formula for the probabilities is
\beq
P(|i\rangle)=|\langle i | q \rangle|^2=\langle i | q \rangle^*\langle i | q \rangle=\langle q | i \rangle\langle i | q \rangle =\langle q | P_i | q \rangle\,.
\eeq
In quantum computing, a measurement in the computational basis of a single qubit is depicted as follows,
\begin{figure}[H]
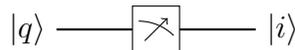

  \centering
  \include{images/measurement}
\vspace{-30pt}
 \caption{A measurement gate.}
 \label{fig:measurementgate}
\end{figure}
\begin{theorem}
\textup{Show that indeed the $P_i$'s are projectors, that is, $P_i^{\dagger}=P_i$ and $P_i^2=P_i$.}
\end{theorem} 
\begin{theorem}
\textup{If a single qubit $|q\rangle$ enters the sequence of gates $HP(\phi)H$, where $P(\phi)$ was defined in \eqref{relphasegate}, what is the probability of measuring $|0\rangle$ and $|1\rangle$? Consider then the case $|q\rangle=|0\rangle$. Draw the probabilities for $0\leq \phi \leq 2\pi$.}
\end{theorem}
\noindent Certainly, there is nothing new here for you. What you may not know, though, is what
happens to a 2 qubit when a measurement is performed on only one of the qubits. Let me recall it very quickly. 

Given a 2 qubit in a generic state $|q_2\rangle=\sum _{j,k}\alpha_{jk}|jk\rangle$,
we can, for example, ask about the probability of
finding the first qubit in the computational basis state $|i\rangle$. Because we do nothing to the second qubit, the probability
$P(|i\; \cdot\rangle)$ must take into account the two possibilities of the second qubit, that is,
\begin{align}
P(|i\; \cdot\rangle)&=P(|i\; 0\rangle) + P(|i\; 1\rangle)=|\langle i\,0|q_2\rangle|^2+|\langle i\,1|q_2\rangle|^2 \nonumber \\
&=|\alpha_{i0}|^2+|\alpha_{i1}|^2=\sum_{j}|\alpha_{ij}|^2 \,.
\end{align}
Similarly, the partial measurement of the second qubit comes with probabilities
\beq
P(|\cdot \,k\rangle)=\sum_{i}|\alpha_{ik}|^2 \,.
\eeq
\begin{theorem}
\textup{If $|i\rangle$ is the result of measuring the first qubit, what is the state vector of the second qubit?}
\end{theorem}
\begin{theorem}
\textup{Work explicitly the case of 3 qubits. Explore all possible measurements.}
\end{theorem}
To illustrate some interesting consequences of the measurement process in quantum computing, let us consider the following examples. But first, since we will need the controlled-$U$ gate to act on a general target qubit, rather than a single qubit as in \eqref{cugatec1qubit}, let us write it again in index notation,
\beq
|i\rangle|Q\rangle\xmapsto{~\textup{C}U~}|i\rangle U^i|Q\rangle \,. 
\eeq
That is, if the incoming product state  is $|q\rangle|Q\rangle$, we have that
\beq
|q\rangle|Q\rangle=\sum_i\alpha_i |i\rangle |Q\rangle \xmapsto {~\textup{C}U~} \sum_i\alpha_i |i\rangle U^i|Q\rangle \,.
\eeq
Without risk of confusion, we can also write this as 
\beq
\sum_i\alpha_i |i\rangle U^i|Q\rangle = 
\sum_i\alpha_i U^i|i\;Q\rangle \,,
\eeq
where it is understood that the $U^i$ in the right hand side  is $1\otimes U^i$.

Let us now examine the following circuit,
\begin{figure}[H]
  \centering
\input{images/generalgtest}
 \caption{}
 \label{fig:galaxy}
\end{figure}
\noindent In this example, $g$ is an arbitrary gate on
the single qubit $|q\rangle$ and $g^{-1}$ is its inverse. For instance, $g$ can be any of the Pauli unitaries or the Hadamard gate. The gate $U$, on the other hand, is an arbitrary unitary
transformation on the $n$ qubit $|Q\rangle$. As a special case, $|Q\rangle$ could be a single qubit.\\

\noindent Following the circuit, we have that
\beq
|q\,Q\rangle \xmapsto{~g~} \sum_{i,j} g_{ij}\alpha_i |i\,Q\rangle \xmapsto{~\mathrm{C}U~} \sum_{i,j} g_{ij}\alpha_j U^i|i\,Q\rangle \xmapsto{~g^{-1}~} \sum_{i,j,k} g^*_{ki}g_{kj}\alpha_j U^k|i\,Q\rangle \,.
\eeq
In full form, the output state vector $|\omega\rangle$ is
\begin{align}
|\omega\rangle &= \big[\alpha_0(g^*_{00}g_{00}+g^*_{10}g_{10}U)+
\alpha_1(g^*_{00}g_{01}+g^*_{10}g_{11}U) \big]|0\,Q\rangle \nonumber \\[3pt]
&~~~+ \big[\alpha_0(g^*_{01}g_{00}+g^*_{11}g_{10}U)+
\alpha_1(g^*_{01}g_{01}+g^*_{11}g_{11}U) \big]|1\,Q\rangle \,.
\end{align}
Of course, $g^*_{ij}=g_{ji}$ because $g$ is unitary.
\begin{theorem}
\textup{Check that the previous formulas are correct by using explicit matrix representations.}
\end{theorem}

Suppose now that $g$ is the Hadamard gate,
\begin{figure}[H]
  \centering
  \include{images/hadamardtest}
\vspace{-30pt}
 \caption{}
 \label{fig:HUH}
\end{figure}
\noindent In this case, the state vector $|\omega\rangle$ is
\begin{align}\label{HUHoutstate}
|\omega\rangle&=\frac{1}{2}\big[\alpha_0(1+U)+\alpha_1(1-U)\big]|0 \,Q\rangle+\frac{1}{2}\big[\alpha_0(1-U)+ \alpha_1(1+U)\big]|1 \,Q\rangle \nonumber \\[5pt]
&=\frac{1}{2}\sum_i \alpha_i\big( 1+(-1)^i U \big)|0 \, Q\rangle + \frac{1}{2}\sum_i \alpha_i\big( 1-(-1)^i U \big)|1 \, Q\rangle \,.
\end{align}
Once again, recall that we are using the shorthand notation $AB$ for the tensor product $A\otimes B$. Thus, by $(1\pm U)$ we really mean $(1\otimes 1 \pm 1\otimes U)$.\\

\noindent The probabilities of measuring the upper qubit in $|0\rangle$ and $|1\rangle$ are 
\begin{align*}
P(|0\,\cdot\rangle)=\frac{1}{4}\Big|\sum_i \alpha_i\big( 1+(-1)^i U \big)\Big|^2 \,,\qquad
P(|1\,\cdot\rangle)=\frac{1}{4}\Big|\sum_i \alpha_i\big( 1-(-1)^i U \big)\Big|^2 \,.
\end{align*}
If you prefer, we can write them more compactly as
\beq
P(|i\,\cdot\rangle)=\frac{1}{4}\Big|\sum_j \alpha_j\big( 1+(-1)^{i+j} U \big)\Big|^2 \,.
\eeq
\begin{theorem}
\textup{Prove that the sum of these two probabilities is equal to 1.}
\end{theorem}
\noindent In particular, if the control qubit $|q\rangle$ in \eqref{HUHoutstate} is prepared in the state $|0\rangle$, we get
\beq\label{hadamardteststate}
|\omega\rangle=\frac{1}{2}(1+U)|0 \, Q\rangle+\frac{1}{2}(1-U)|1\, Q\rangle \,.
\eeq 
\begin{theorem}
\textup{Show that 
\beq
P(|0\,\cdot\rangle)=\frac{1}{2}(1+\mathrm{Re}\langle Q|U|Q\rangle) \,.
\eeq
Find $P(|1\, \cdot\rangle)$ and show that the sum of the two probabilities is equal to 1.}
\end{theorem}

A similar set-up is at the core of the so called \index{Quantum phase estimation}\emph{quantum phase estimation algorithm}. Suppose that the unitary transformation $U$ acts as $U|Q\rangle=e^{i\theta}|Q\rangle$. In other words, assume that the state vector $|Q\rangle$ of the qubit  is an eigenvector of the unitary $U$. In this case, the outgoing state \eqref{hadamardteststate} becomes
\beq
|\omega\rangle=\frac{1}{2}(1+e^{i\theta})|0 \, Q\rangle+\frac{1}{2}(1-e^{i\theta})|1\, Q\rangle \,.
\eeq
As before, we are interested in the probabilities
\begin{align}
P(|0\,\cdot\rangle)&=\frac{1}{4}(1+e^{i\theta})(1+e^{-i\theta})=\cos^2(\theta/2) \,,\\
P(|1\,\cdot\rangle)&=\frac{1}{4}(1-e^{i\theta})(1-e^{-i\theta}) =\sin^2(\theta/2)\,.
\end{align}
As you see, there is a closed relationship between these probabilities and the phase angle. For example, if the phase angle is greater that 45$^{\circ}$, the probability of measuring the state $|1\rangle$ is greater than measuring $|0\rangle$. 
\begin{theorem}
\textup{If $U|Q\rangle=e^{i\theta}|Q\rangle$, what is the outgoing state in the following diagram? After this, do it for 3 and --- if you can --- generalize to an arbitrary number of incoming \index{Measuring qubit}\emph{measuring qubits} $|0\rangle$.}
\begin{figure}[H]
  \centering
\input{images/qft2}
 \caption{}
 \label{fig:galaxy}
\end{figure}
\end{theorem}
Another interesting case worth considering is when in the circuit shown in Figure \ref{fig:HUH}, the control qubit is in the state
$|q\rangle=(|0\rangle-i|1\rangle)/\sqrt{2}$, for which,
\beq
|\omega\rangle=\frac{(1-i)}{2\sqrt{2}}(1+iU)|0\,Q\rangle
+\frac{(1-i)}{2\sqrt{2}}(1-iU)|1\,Q\rangle\,.
\eeq 
\begin{theorem}
\textup{Draw the circuit diagram that implements the previous transformation. Find the probability $P(|0\,\cdot \rangle)$ and $P(|1\,\cdot \rangle)$.}
\end{theorem}
Suppose now that the incoming qubit $|Q\rangle$ in Figure \ref{fig:HUH} is a 2-qubit unentangled system, that is, suppose $|Q\rangle=|t_1\rangle|t_2\rangle$, and let $U$ be a SWAP gate. The circuit diagram becomes
\begin{figure}[H]
  \centering
\input{images/swaptest}
 \caption{}
 \label{fig:galaxy}
\end{figure}
\noindent The analysis of the circuit gives
\begin{align*}
|0\rangle|t_1\rangle|t_2\rangle &\xmapsto{~~H_c~~} \frac{1}{\sqrt{2}}\big(|0\rangle+|1\rangle \big)|t_1\;t_2\rangle\\
&\xmapsto{\mathrm{SWAP}} \frac{1}{\sqrt{2}}|0\rangle|t_1\;t_2\rangle + \frac{1}{\sqrt{2}}|1\rangle|t_2\;t_1\rangle\\
&\xmapsto{~~H_c~~} \frac{1}{2}\big(|0\rangle+|1\rangle \big)|t_1\;t_2\rangle + \frac{1}{2}\big(|0\rangle-|1\rangle \big)|t_2\;t_1\rangle\\
&~~~~~~~~=\frac{1}{2}|0\rangle\big(|t_1\;t_2\rangle+|t_2\;t_1\rangle\big)+\frac{1}{2}|1\rangle\big(|t_1\;t_2\rangle-|t_2\;t_1\rangle\big)=|\omega\rangle \,.
\end{align*}
If we measure the first qubit and leave alone the second and third qubits, the probabilities are
\beq\label{eq:prob3}
P(|i \;\cdot \; \cdot \rangle)=P\Big(\frac{1}{2}\big(|i\;t_1\;t_2\rangle+|i\;t_2\;t_1\rangle\big)\Big) \,.
\eeq
The $1/2$ in the right hand side is the normalization factor.
\begin{theorem}
\textup{Prove that, in fact, in the Hilbert space of outgoing states
\beq
1\otimes 1=1\otimes |t_1\;t_2\rangle\langle t_1\;t_2|+1\otimes |t_2\;t_1\rangle\langle t_2\;t_1| \;.
\eeq}
\end{theorem}
\noindent The explicit calculation of the probabilities \eqref{eq:prob3} is as follows, 
\begin{align}
P(|i\,\cdot \, \cdot\rangle)&=\frac{1}{2}\big(\langle i\; t_1\; t_2|+\langle i\;t_2\; t_1| \big)|\omega\rangle \nonumber \\
&=\frac{1}{2}\big(\langle i\; t_1\; t_2|+\langle i\;t_2\; t_1| \big)\frac{1}{2}\big(|i\; t_1\; t_2\rangle+(-1)^i| i\; t_2\; t_1\rangle\big) \nonumber\\
&=\frac{1}{4}\big(\langle t_1|t_1\rangle\langle t_2|t_2\rangle+(-1)^i\langle t_1|t_2\rangle\langle t_2|t_1\rangle \nonumber\\
&~~~+ \langle t_2|t_1\rangle\langle t_1|t_2\rangle 
+(-1)^i \langle t_2|t_2\rangle\langle t_1|t_1\rangle\big) \nonumber \\
&=\frac{1}{4}\big[\big(1+(-1)^i\big)+\big(1+ (-1)^i\big)\langle t_1|t_2\rangle\langle t_1|t_2\rangle^* \big] \nonumber\\
&=\frac{1}{4}\big(1+(-1)^i\big)\big(1+|\langle t_1|t_2\rangle|^2 \big) \,.
\end{align}
\vspace{-20pt}
\begin{theorem}
\textup{Show that $P(|0\,\cdot\,\cdot\rangle)+P(|1\,\cdot\,\cdot\rangle)=1$, regardless of the values of the incoming qubits $|t_1\rangle$ and $|t_2\rangle$.}
\end{theorem}
\noindent Notice that, if you prepare the two target qubits $|t_1\rangle$ and $|t_2\rangle$ such that they are perpendicular, $\langle t_1,t_2\rangle=0$, it follows that $P\big(|0\,\cdot\,\cdot\rangle\big)=1/2$ and $P\big(|1\,\cdot\,\cdot\rangle\big)=1/2$ as well. If, instead, they are prepared in the same state, $\langle t_1,t_2\rangle=1$, the probabilities are $P\big(|0\,\cdot\,\cdot\rangle\big)=1$ and $P\big(|1\,\cdot\,\cdot\rangle\big)=0$.

\section{Quantum Algorithms}

We often hear quantum computing experts and popular
science writers alike say that future quantum
computers will be much faster than standard computers. They will be so fast that, according to some, in a matter of
minutes or even seconds we will be able to solve problems
that would take billions of years (more than the age of the universe!) for the most powerful
classical supercomputers. Moreover, they say that there is good evidence
to think there are problems that, in principle,
a quantum computer will be able to solve but classical
computers will not, no matter how powerful they become or
how much time we give them to work on them.
All these claims seem to be unfounded
exaggerations, part of the contemporary hype around
quantum computers. However, there is something
that remains true: there is something in the
way a quantum computer processes information --- 
the superposition of quantum states --- that has the potential to 
make it faster than classical computers,
at least at solving certain problems.

Note that here we are not referring to the
physical realization of these devices, but to the
theoretical mode of computation. That is, on paper at least, quantum computers will be faster than classical ones thanks to their unique way of processing information
and not because of their implementation.
In other words, if we use the quantum circuit
model of computation instead of
the classical circuit model 
to design the solution to a problem, we may arrive
at a circuit that solves it in less time.

We have been careful to emphasize that quantum computers
will be faster than classical ones at solving some problems, but not all. In the circuit
model of computation, whether classical or quantum,
an \index{Algorithm}\emph{algorithm} is a circuit, that is, a specific
arrangements of gates, that given a certain
input, delivers the desired output. So, when
people, experts and non-experts, loosely say that
quantum computers will be much faster than
classical computers, what they really mean is that
we known some specific quantum algorithms that are faster than the classical algorithms created to solve the same problem.

After all this, you may be wondering, ``Ok, but what
exactly does ``faster" mean?" This is something that, as we will see in the next examples,
will depend on each particular problem.

\vspace{10pt}
\begin{tcolorbox}[breakable, enhanced]
\vspace{5pt}
\begin{note} 
$\mathrm{\mathbf{The~ probabilistic~model ~of~ computation.}}$
\end{note}
Another classical model of computation is the
probabilistic model. If we have a classical system and
there are various outcomes for an
experiment, we can use a probabilistic description of the system. For example, when you toss an unbiased coin in the air, you can describe the state of the system with a real two-dimensional vector
\beq
|C\rangle=\frac{1}{2}|H\rangle + \frac{1}{2}|T\rangle \,.
\eeq
The coefficients $1/2$ are the probabilities of observing head or tail when the coin stops. If the coin is biased, the state rector will take the more general form
\beq
|C\rangle=p_H|H\rangle + p_T|T\rangle \,,
\eeq
where $p_H,p_T\in [0,1]$ and $p_H+p_T=1$. The
probabilities $p_H$ and $p_T$ can be obtained by defining an inner product on $\mathbb{R}^2$ such that
\beq
\langle H|C\rangle = p_H \,, \qquad \langle T|C\rangle = p_T \,.
\eeq

~~No doubt, all this looks very similar to the mathematical
description of the single qubit \eqref{1qubit}. We can even write everything in column vectors and introduce matrix transformations on these vectors. It seems that the only difference between the two models is that the coefficients are real in one case and complex in the other. The two descriptions are so similar that many computer
scientists prefer to introduce the mathematics of
quantum mechanics by using the probabilistic model. Of course,
we already knew quantum mechanics, so we did not need
to do that. 

~~What we want to stress here, though, is that, despite the resemblance between the mathematics of the probabilistic and the quantum models of computation, there is a fundamental --- philosophical, if you wish --- difference between the two: the uncertainty in the probabilistic model is due to our limited knowledge of the system, whereas the uncertainty in the quantum model is intrinsic to nature. In principle, we can develop a probabilistic model of computation with complex coefficients (there is nothing wrong with that), but as long as it is based on a physical system which is classical, the superposition of states will not have the same meaning as the superposition of states occurring in the quantum world. In a classical system, whether deterministic or probabilistic, a measurement reveals the true state of the system before the measurement. In contrast, in quantum mechanics, a measurement fixes the state of the system.  
\vspace{5pt}
\end{tcolorbox}
\vspace{10pt}

\subsection{Deutsch's Algorithms}

We start with the
simplest and historically the first quantum algorithm ever conceived,
the algorithm proposed by David Deutsch in 1996 and then we 
discuss its generalization proposed a few months later by Deutsch himself and Richard Jozsa. The goal of these quantum algorithms is not their
real-life application, but
to prove that quantum algorithms, at least in principle, can solve computational problems faster than the fastest classical algorithm.\\

\noindent \index{Deutsch's algorithm}\emph{The Deutsch algorithm}\\

Suppose we are given a Boolean function $f\colon \{0,1\}\to \{0,1\}$ and we are told
that it is constant or balanced. However, we do not know which of the two is the case. By \index{Constant function}\emph{constant} we mean that $f(0)=f(1)$, whether because
\beq
f(0)=f(1)=0 \,,
\eeq
or 
\beq
f(0)=f(1)=1 \,,
\eeq
On the other hand, \index{Balanced function}\emph{balanced} means that $f(0)\neq f(1)$, that is,
\beq
f(0)=0\neq f(1)=1 \,,
\eeq
or
\beq
f(0)=1\neq f(1)=0 \,.
\eeq
To keep track of these two possibilities, we will indicate each of the previous
cases by $f_c$ and $f_b$, respectively. 

It seems clear that it is not enough to know the value
of the function at one single input, whether 0 or 1,
to determine if the function is constant or balanced;
we need to know the value of the function at both
0 and 1. If the function has to be evaluated at
two different values, computer scientists say that the
function has to be called ``twice" or ``two times".
In general, the more calls your algorithm makes to a
function, the more complex and slow it is. Conversely, the less calls you make to a function, the less complex and faster is your algorithm. This is
the principle of what is known as  \index{Query complexity}\emph{query complexity}. 

What Deutsch discovered is that we can find out
whether the function is constant or balanced by calling
the function only once.
The quantum circuit he conceived was the following,
\begin{figure}[H]
  \centering
  \input{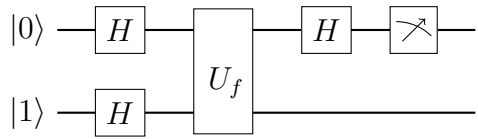}
 \caption{Circuit configuration for Deutsch's algorithm.}
 \label{fig:deutschalgorithm}
\end{figure}
\noindent The gate $U_f$, called an \index{Oracle}\emph{oracle} or more properly a \index{XOR oracle}\emph{XOR oracle}, transforms the
computational basis vectors according to
\beq\label{eq:oracle}
|i\rangle|j\rangle\xmapsto{~U_f~} |i\rangle  |j \oplus f(i) \rangle \,.
\eeq
Note that it is controlled gate. It is usually depicted as follows,
\begin{figure}[H]
  \centering
 \input{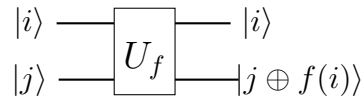}
 \caption{The oracle of the Deutsch algorithm.}
 \label{fig:deutschoracle}
\end{figure}
\noindent As we said, in the query complexity model we only care about the number of calls made by the algorithm to the function. The inherent complexity proper to the functioning of the oracle is ignored. This is why the oracle is often called a \index{Black box}\emph{black box}.
\begin{theorem}
\textup{Prove that $U_f$ is unitary.}
\end{theorem}
\begin{theorem}
\textup{What is the matrix representation of $U_f$?}
\end{theorem}
\noindent We have all the elements to analyze Deutsch's circuit in Figure \ref{fig:deutschalgorithm}:
\begin{align}
|0\,1\rangle\xmapsto{~H\otimes H~}&\,|+-\rangle \nonumber\\
&=\frac{1}{2}\big(|0\,0\rangle-|0\,1\rangle+|1\,0\rangle-|1\,1\rangle\big) \nonumber \\
\xmapsto{~~\,U_f~~\,}&\,\frac{1}{2}U_f\big(|0\,0\rangle-|0\,1\rangle+|1\,0\rangle-|1\,1\rangle\big) \nonumber\\
&=\frac{1}{2}\big(|0~0\oplus f(0)\rangle-|0~1\oplus f(0)\rangle+|1~0\oplus f(1)\rangle-|1~1\oplus f(1)\rangle\big) \,.
\end{align}
Let us first suppose that the function $f$ is constant. Substituting $f(1)$ by $f(0)$ and using $f_c$ instead of $f$,
\begin{align}
U_{fc}|+-\rangle&=\frac{1}{2}\big(|0~0\oplus f_c(0)\rangle-|0~1\oplus f_c(0)\rangle+|1~0\oplus f_c(0)\rangle-|1~1\oplus f_c(0)\rangle\big)\nonumber\\
&=\frac{1}{2}\big(|0\rangle+ |1\rangle\big) |f_c(0)\rangle
-\frac{1}{2}\big(|0\rangle+ |1\rangle\big)|1\oplus f_c(0)\rangle\nonumber\\
&=|+\rangle\frac{1}{\sqrt{2}}\big(|f_c(0)\rangle-|1\oplus f_c(0)\rangle\big) \,.
\end{align}
Consider now the case where $f$ is balanced. Since $f_b(1)=1-f_b(0)$, it follows that
\begin{align}
U_{fb}|+-\rangle&=\frac{1}{2}\big(|0~f_b(0)\rangle-|0~1\oplus f_b(0)\rangle+|1~1-f_b(0)\rangle-|1~1\oplus 1-f_b(0)\rangle\big)\nonumber\\
&=\frac{1}{2}|0\rangle\big(|f_b(0)\rangle-|1\oplus f_b(0)\rangle\big)+
\frac{1}{2}|1\rangle \big(|1-f_b(0)\rangle-|1\oplus 1-f_b(0)\rangle\big) \,.
\end{align}
When $f_b(0)=0$,
\begin{align}
U_{fb,0}|+-\rangle&=\frac{1}{2}|0\rangle  \big(|0\rangle-|1\rangle\big)+
\frac{1}{2}|1\rangle \big(|1\rangle-|1\oplus 1 \rangle\big)\nonumber\\
&=\frac{1}{\sqrt{2}}\big(|0\rangle-|1\rangle\big) \frac{1}{\sqrt{2}}\big(|0\rangle-|1\rangle\big)=|--\rangle \,,
\end{align}
and, when $f_b(0)=1$,
\begin{align}
U_{fb,1}|+-\rangle&=\frac{1}{2}|0\rangle  \big(|1\rangle-|1\oplus 1\rangle\big)+
\frac{1}{2}|1\rangle\big(|0\rangle-|1\oplus 1-1\rangle\big)\nonumber\\
&=-\frac{1}{\sqrt{2}}\big(|0\rangle-|1\rangle\big)\frac{1}{\sqrt{2}}\big(|0\rangle-|1\rangle\big)=-|--\rangle \,.
\end{align}
In summary, for $f$ constant,
\beq
U_{fc}|+-\rangle=\pm |+-\rangle \,,
\eeq
and, for $f$ balanced,
\beq
U_{fb}|+-\rangle=\pm |--\rangle \,.
\eeq
The last Hadamard gate in Figure \ref{fig:deutschalgorithm} gives: for $f$ constant,
\beq
U_{fc}|+-\rangle \xmapsto{~H\otimes 1~} \pm (H\otimes 1)|+-\rangle =\pm |0-\rangle \,,
\eeq
and, for $f$ balanced,
\beq
U_{fb}|+-\rangle  \xmapsto{~H\otimes 1~} \pm (H\otimes 1)|--\rangle =\pm |1-\rangle \,.
\eeq

Finally, we measure the state of the upper qubit. If the measurement gives the state $|0\rangle$, then we know with absolute certainty that the function is
constant. If, instead, we measure $|1\rangle$, then
the function is balanced. This completes the Deutsch algorithm. As stated, we can discover whether the function is constant or balanced by calling it just once.

For completeness' sake, let us present the Deutsch algorithm in a slightly more general form. Suppose that two qubits,
\beq
|u\rangle=\sum_i u_i |i\rangle \,, \qquad |b\rangle=\sum_j b_j |j\rangle \,,
\eeq
enter the oracle in Figure \ref{fig:deutschoracle}. The output is given by,
\begin{align}
|u\rangle|b\rangle=\sum_{i,j} u_i b_j |i\rangle |j\rangle  \xmapsto{~U_f~}& \sum_{i,j} u_i b_j |i\rangle |j \oplus f(i) \rangle\nonumber \\
&=u_0b_0 |0\rangle|0\oplus f(0)\rangle+u_0b_1 |0\rangle|1\oplus f(0)\rangle \nonumber \\[8pt]
&~~~+u_1b_0 |1\rangle|0\oplus f(1)\rangle+u_1b_1 |1\rangle|1\oplus f(1)\rangle \,.
\end{align}
When the function is constant, $f_c(0)=f_c(1)$, we group the first term
with the third and the second with the fourth,
\beq\label{ufc}
U_{fc}(|u\rangle|b\rangle)=\big(u_0b_0 |0\rangle+u_1b_0|1\rangle\big) |f_c(0)\rangle+\big(u_0b_1 |0\rangle+u_1b_1|1\rangle\big) |1\oplus f_c(0)\rangle \,.
\eeq
When $f$ is balanced, $f_b(0)\neq f_b(1)$, we group the first term with the fourth
and the second with the third,
\beq\label{ufb}
U_{fb}(|u\rangle|b\rangle)=\big(u_0b_0 |0\rangle+u_1b_1|1\rangle\big) |f_b(0)\rangle+\big(u_0b_1 |0\rangle+u_1b_0|1\rangle\big) |1\oplus f_b(0)\rangle \,.
\eeq
Note that equations \eqref{ufc} and \eqref{ufb} are telling us that the bottom incoming qubit cannot be 
in a state with $b_0= b_1$, if not we would not be able to identify whether $f$ is constant or balanced. So, for the algorithm to work, the first
condition is to set $b_0\neq b_1$. Now, $b_0$ and $b_1$ have to be chosen so that a single measurement of the upper
qubit will tell us if the function is constant or balanced. In general, of course, $|b_0|^2+|b_1|^2=1$; however, for simplicity we can choose $b_0=1/\sqrt{2}=-b_1$. That is, $|b\rangle=|-\rangle$. We then have that
\begin{align}
U_{fc}(|u\rangle|-\rangle)&=\frac{1}{\sqrt{2}}\big(u_0|0\rangle+u_1|1\rangle\big) |f_c(0)\rangle-\frac{1}{\sqrt{2}}\big(u_0|0\rangle+u_1|1\rangle\big) |1\oplus f_c(0)\rangle \nonumber \\
&=\frac{1}{\sqrt{2}}\big(u_0|0\rangle+u_1|1\rangle\big)\big(|f_c(0)\rangle- |1\oplus f_c(0)\rangle\big) \,,
\end{align}
and
\begin{align}
U_{fb}(|u\rangle|-\rangle)&=\frac{1}{\sqrt{2}}\big(u_0|0\rangle-u_1|1\rangle\big) |f_b(0)\rangle-\frac{1}{\sqrt{2}}\big(u_0|0\rangle-u_1|1\rangle\big) |1\oplus f_b(0)\rangle \nonumber \\
&=\frac{1}{\sqrt{2}}\big(u_0|0\rangle-u_1|1\rangle\big)\big(|f_b(0)\rangle- |1\oplus f_b(0)\rangle\big) \,.
\end{align}
Since we want a single measurement on the upper qubit to be able to unambiguously distinguish its state,
we need to choose $u_0$ and $u_1$ such that the two vectors $(u_0|0\rangle+u_1|1\rangle)/\sqrt{2}$ and $(u_0|0\rangle-u_1|1\rangle)/\sqrt{2}$ are
perpendicular. The condition is then,
\begin{align*}
\frac{1}{\sqrt{2}}\big( \langle0|u^*_0+ \langle1|u^*_1 \big)\frac{1}{\sqrt{2}}\big( u_0|0\rangle-u_1|1\rangle \big)&=\frac{1}{2}u^*_0u_0-\frac{1}{2}u^*_1u_1\\
&=\frac{1}{2}(1-u^*_1u_1-u^*_1u_1)=\frac{1}{2}-u^*_1u_1=0\\[5pt]
&=\frac{1}{2}(u^*_0u_0-1+u^*_0u_0)=u^*_0u_0-\frac{1}{2}=0 \,.
\end{align*}
This is enough to know whether $f$ is constant or
balance. For, example, as we did above, the usual choice is $u_0=u_1=1/\sqrt{2}$.\\

\noindent \index{Deutsch-Jozsa's algorithm}\emph{The Deutsch-Jozsa algorithm}\\

Suppose you are given a Boolean function $f\colon \{0,1\}^n\to \{0,1\}$
and you are told that it is constant or balanced. However, 
you do not know which of the two is the case. Again, as
for the Deutsch algorithm (for which $n= 1$), the
quantum circuit we will discuss below finds whether the
function is constant or balanced by calling the function
$f$ a fewer number of times than the classical
optimal solution. By \index{Constant function}\emph{constant}
we mean that $f$ takes the same value, 0 or 1,
for all the $x$'s in the domain $\{0,1\}^n$. By
\index{Balanced function}\emph{balanced} we mean that half
of the $x$'s in $\{0,1\}^n$ take the value 0 and the
other half the value 1. 
\begin{theorem}
\textup{Show that these definitions are consistent with
the ones given above for the Deutsch algorithm.}
\end{theorem} 
To easily generalize to the Boolean function $f\colon \{0,1\}^n\to \{0,1\}$,
let us start by considering $n=1$ and $n=2$. The case $n=1$ is just the Deutsch algorithm already discussed.
The evolution of the incoming product state 
$|0\rangle|1\rangle$ as it moves through the circuit  shown in Figure \ref{fig:deutschalgorithm} is
\begin{align}\label{wdeutschalg1q}
|0\,1\rangle \xmapsto{~H\otimes H~} & \,H|0\rangle \otimes H|1\rangle =\frac{1}{\sqrt{2}}\sum_i|i\rangle \frac{1}{\sqrt{2}}\sum_k(-1)^k|k\rangle \nonumber \\
&=\frac{1}{\sqrt{2^2}}\sum_{i,k}(-1)^k|i\;k\rangle \nonumber \\
\xmapsto{~~\,U_f\,~~} & \,\frac{1}{2}\sum_{i,k}(-1)^k|i\;k\oplus f(i)\rangle \nonumber \\
\xmapsto{~\,H\otimes 1\,~} & \,\frac{1}{2}\sum_{i,k}(-1)^k H|i\rangle |k\oplus f(i)\rangle \nonumber \\
&= \frac{1}{2}\sum_{i,k}(-1)^k \frac{1}{\sqrt{2}}\sum_j(-1)^{ij}|j\rangle |k\oplus f(i)\rangle \nonumber \\
&= \frac{1}{2}\sum_{i,j,k}(-1)^{k+ij} \frac{1}{\sqrt{2}}|j~k\oplus f(i)\rangle \,.
\end{align}
The following is a similar circuit, but
with three incoming single qubits instead of two, 
\begin{figure}[H]
  \centering
\input{images/deutschjozsaH2H}
 \caption{}
 \label{}
\end{figure}
\noindent The oracle in this case transforms
\beq
|i\rangle|j\rangle|k\rangle\xmapsto{U_f}|i\rangle|j\rangle|k\oplus f(i\;j)\rangle \,.
\eeq
The incoming state $|0\rangle|0\rangle|1\rangle$ evolves as follows,
\begin{align}\label{wdeutschalg2q}
|0\,0\,1\rangle \xmapsto{~H^{\otimes 2}\otimes H~} & \,H|0\rangle \otimes H|0\rangle \otimes H|1\rangle =\frac{1}{\sqrt{2}}\sum_{i_1}|i_1\rangle \frac{1}{\sqrt{2}}\sum_{i_2}|i_2\rangle \frac{1}{\sqrt{2}}\sum_k(-1)^k|k\rangle \nonumber \\
&=\frac{1}{\sqrt{2^3}}\sum_{i_1,i_2,k}(-1)^k|i_1\;i_2\;k\rangle \nonumber \\
\xmapsto{~~~\,U_f~~~\,} & \,\frac{1}{2^{3/2}}\sum_{i_1,i_2,k}(-1)^k|i_1\;i_2\;k\oplus f(i_1\; i_2)\rangle \nonumber \\
\xmapsto{\,~H^{\otimes 2}\otimes 1~} & \,\frac{1}{2^{3/2}}\sum_{i_1,i_2,k}(-1)^k H|i_1\rangle H|i_2\rangle |k\oplus f(i_1\; i_2)\rangle \nonumber \\
&= \frac{1}{2^{3/2}}\sum_{i_1,i_2,k}(-1)^k \frac{1}{\sqrt{2}}\sum_{j_1}(-1)^{i_1j_1}|j_1\rangle \frac{1}{\sqrt{2}}\sum_{j_2}(-1)^{i_2j_2}|j_2\rangle |k\oplus f(i_1\; i_2)\rangle \nonumber \\
&= \frac{1}{2^2}\sum_{i_1,i_2,j_1,j_2,k}(-1)^{k+i_1j_1+i_2j_2} \frac{1}{\sqrt{2}}|j_1\;j_2\;~k\oplus f(i_1\;i_2)\rangle \,.
\end{align}
In general, the control qubits form the state $|0\rangle^{\otimes n}$ and the oracle is
\begin{figure}[H]
 \centering
\input{images/generaloracle}
 \caption{Oracle of the Deutsch-Jozsa algorithm.}
 \label{fig:galaxy}
\end{figure}
\noindent or, in simplified form,
\begin{figure}[H]
 \centering
\input{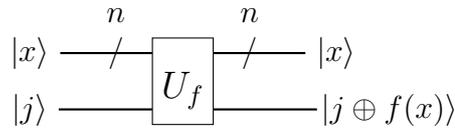}
 \caption{Simplified diagram for the oracle of the Deutsch-Jozsa algorithm.}
 \label{fig:galaxy}
\end{figure}
\noindent By induction, we see that the state of the system right before the measurements is given by,
\beq
|0\rangle^{\otimes n}\otimes |1\rangle \xmapsto{~H^{\otimes n}\otimes H~}\xmapsto{~U_f~}\xmapsto{~H^{\otimes n}\otimes 1~}|\omega\rangle_{f(x)}=\frac{1}{2^n}\sum_{x,y,k}(-1)^{k+x\cdot y} |y\rangle\frac{1}{\sqrt{2}}|k\oplus f(x)\rangle \,,
\eeq
where we are using the dot in $x\cdot y$ to indicate that $x$ and $y$ are in binary notation (not in decimal notation!). Recall the discussion concerning the notation used in equation \eqref{Hxnonx}. Writing explicitly the sum over $k$,
\beq
|\omega_{f(x)}\rangle=\frac{1}{2^n}\sum_{x,y}(-1)^{x\cdot y} |y\rangle\frac{1}{\sqrt{2}}\big(|f(x)\rangle-|1\oplus f(x)\rangle\big) \,.
\eeq
Note that $f(x)=0$ gives
\beq
|\omega\rangle_0=\frac{1}{2^n}\sum_{x,y}(-1)^{x\cdot y} |y\rangle|-\rangle \,,
\eeq
and for $f(x)=1$,
\beq
|\omega\rangle_1=\frac{1}{2^n}\sum_{x,y}(-1)^{x\cdot y} |y\rangle(-1)|-\rangle \,.
\eeq
Therefore,
\beq
|0\rangle^{\otimes n}\otimes |1\rangle\xmapsto{~~~}|\omega\rangle_{f(x)}
=\frac{1}{2^n}\sum_{x,y}(-1)^{f(x)+x\cdot y} |y\rangle|-\rangle\,.
\eeq
The probability of measuring all the upper qubits in the state $|0\rangle$ is
\begin{align}
P\big(|0\ldots 0 \;\cdot \rangle\big)&=\big|\langle 0\ldots 0|\frac{1}{2^n}\sum_{x,y}(-1)^{f(x)+x\cdot y}|y\rangle\big|^2 \nonumber \\
&=\frac{1}{2^{2n}}\Big|\sum_{i_1,\ldots,i_n}\sum_{j_1,\ldots,j_n}(-1)^{f(i_1\ldots i_n)+i_1j_1+\ldots+i_1j_n}\langle 0\ldots 0|j_1\ldots j_n\rangle\Big|^2 \nonumber \\
&=\frac{1}{2^{2n}}\Big|\sum_{i_1,\ldots,i_n}\sum_{j_1,\ldots,j_n}(-1)^{f(i_1\ldots i_n)+i_1j_1+\ldots+i_1j_n}\delta_{j_10}\ldots \delta_{j_n0}\Big|^2 \nonumber \\
&=\frac{1}{2^{2n}}\Big|\sum_{x}(-1)^{f(x)}\Big|^2 \,.
\end{align}
If $f$ is constant, we have two possibilities: whether $f(x)=0$, in which case 
\beq
P\big(|0\ldots 0 \;\cdot \rangle\big)=\frac{1}{2^{2n}}\big|2^n(-1)^0\big|^2=\frac{1}{2^{2n}}4^n=1 \,,
\eeq
or $f(x)=1$,
\beq
P\big(|0\ldots 0 \;\cdot \rangle\big)=\frac{1}{2^{2n}}\big|2^n(-1)^1\big|^2=\frac{1}{2^{2n}}4^n=1 \,.
\eeq
Thus, in both cases the probability of measuring the upper qubits in the state $|0\ldots 0\rangle$ is 1.

For $f$ balanced,
\beq
P\big(|0\ldots 0 \;\cdot \rangle\big)=\frac{1}{2^{2n}}\Big|\frac{2^n}{2}(-1)^0+\frac{2^n}{2}(-1)^1\Big|^2=\frac{1}{2^{2n}}\big|2^{n-1}-2^{n-1}\big|^2=0 \,.
\eeq
What this is saying is that in case the function $f$ is
balanced, it is impossible for all the upper qubits to be measured in the state $|0\rangle$; at least one of them is measured in $|1\rangle$. 

In conclusion, by just calling once the oracle and choosing
the appropriate states to measure, we can determine
whether the function $f$ is constant or balanced.
In the classical case, best case scenario we had to
call the function twice. 

\begin{theorem}
\textup{Apply the previous analysis to the results \eqref{wdeutschalg1q} and \eqref{wdeutschalg2q}.}
\end{theorem}

\subsection{Shor's Factoring Algorithm}

\index{Shor's algorithm}\emph{Shor's algorithm} is without doubt the most famous of all the quantum algorithms conceived so far. When it was invented in the mid-90s, it propelled the field of quantum
computing into a new era of development. However, despite its undeniable notoriety, historical importance and possible future application, here we will only give
a summary of the concepts it involves and its main attributes.
The motivation for this decision is twofold. First, as we said in the introduction, the
goal of the present notes is to sketch the main physical ideas and mathematical tools used in
quantum computing; alas, Shor's algorithm is too complex to be properly presented in a few pages.
Second, Shor's algorithm concerns a rather technical domain of quantum computing, that of secure transfer of information \index{Cryptography}(\emph{cryptography}), and we are instead more interested
in the physics of quantum computing.

We start with a rough definition of Shor's
algorithm to get an idea of the ingredients involved. Shor's algorithm is a quantum algorithm that solves the problem
of finding the prime factors of an integer number faster than any known classical algorithm. The
first thing we recognize is that some number theory must be at play here. In addition, the solution found by Shor exploits the
connection between prime factoring and
something we will describe below as period finding.
The latter uses a mathematical technique called
the quantum Fourier transform (See Box \ref{box:qft}).

Simply put, the prime factoring problem asks you to discover the two prime factors of a number that a priori is known to be the product of these numbers.
For example, you may be asked to find the prime factors of 15 or 21. Of course, in these simple cases you know that the prime factors are 3,5 and 3,7, respectively.
To check it, you simply multiply $3\times5$ and $3\times7$. However, it is not so easy to find the prime factors of a larger number such as 755,221. You can check that they
are 773 and 977. As you see, if I give you the two prime
factors, you can easily verify that they are in fact the
correct ones, however, to find them is not so easy.
As the number becomes larger and larger,
the problem of finding the prime factors becomes harder
and harder and eventually impossible to solve by classical computational methods. The difficulty of solving this problem
is at the heart of the modern
encoding process used to transfer secure data (the RSA cryptosystem).
Let us use the examples given above to see how the
prime factoring problem translates into the
period finding problem and how Shor's algorithm partially solves it.

Let us say we want to find the prime factors of 15. We
claim that the following ansatz will give us the prime factors,
\beq\label{1mod15}
x^2=1\,\mathrm{mod}15 \,.
\eeq
A solution is obviously $x= 4$. However, the equation says much more than that. In fact, note that
\begin{align*}
4^2=1\,\mathrm{mod}15&\Longrightarrow 4^2-1=0\,\mathrm{mod}15\\[5pt]
&\Longrightarrow (4+1)(4-1)=0\,\mathrm{mod}15\\[5pt]
&\Longrightarrow 5\cdot 3 =0\,\mathrm{mod}15 \,.
\end{align*}
So, the equation $x^2=1\,\mathrm{mod}15$ actually gives the prime factors of 15. A similar procedure applies to the number 21. We start with
\beq
x^2=1\,\mathrm{mod}21 \,,
\eeq
and find that
\begin{align*}
8^2=1\,\mathrm{mod}21&\Longrightarrow 8^2-1=0\,\mathrm{mod}21\\[5pt]
&\Longrightarrow (8+1)(8-1)=0\,\mathrm{mod}21\\[5pt]
&\Longrightarrow 3^2\cdot 7 =0\,\mathrm{mod}21 \,.
\end{align*}
By just a slight modification of the previous example,
we see that the equation $x^2=1\,\mathrm{mod}21$ indeed gives the
prime factors of 21.
\begin{theorem}
\textup{Apply this procedure to find the prime factors of 35.}
\end{theorem}
With the success of these examples at hand, we may be
tempted to generalize the formula and say that the two prime factors of any number $N=p_1p_2$ can be found by
solving  
\beq\label{1modN}
x^2=1\,\mathrm{mod}N \,.
\eeq
As before,
\begin{align*}
x^2=1\,\mathrm{mod}N&\Longrightarrow (x+1)(x-1)=0\,\mathrm{mod}N\\[5pt]
&\Longrightarrow p_1^{n_1} p_2^{n_2} =0\,\mathrm{mod}N \,.
\end{align*}
However, it seems that not all prime factors can be found by using the simple formula \eqref{1modN}. For example, the method does not apply to the number 77. Instead, we have that
\begin{align*}
9^2=4\,\mathrm{mod}77&\Longrightarrow (9+2)(9-2)=0\,\mathrm{mod}77\\[5pt]
&\Longrightarrow 11\cdot 7 =0\,\mathrm{mod}77 \,.
\end{align*}
\begin{theorem}
\textup{Use this procedure to find the prime factors
of 755,221.}
\end{theorem}
Thus, it seems that the problem of finding the prime factors is getting
more complicated: not only do we have to find $x$, but now also the number $c$ in front of $\mathrm{mod}N$. Perhaps we should modify the initial ansatz \eqref{1modN} as follows,
\beq\label{cmodN}
x^2=c\,\mathrm{mod}N \,.
\eeq
However, we do not need to do that. For example, consider again the prime
factors of 15. We saw that a solution of \eqref{1mod15} is 
\beq
2^4=1\,\mathrm{mod}15 \,,
\eeq
but another solution is 
\beq
2^8=1\,\mathrm{mod}15 \,.
\eeq
In fact, for any $k=0,1,2,\ldots$, the following are solutions,
\beq
2^{4k}=1\,\mathrm{mod}15 \,,\quad 2^{4k+1}=2\,\mathrm{mod}15 \,,\quad 2^{4k+2}=4\,\mathrm{mod}15 \,,\quad 2^{4k+3}=8\,\mathrm{mod}15 \,.
\eeq
Similarly, for 21 all the following are solutions
\begin{align*}
2^{6k}=1\,\mathrm{mod}21 \,,\quad 2^{6k+1}&=2\,\mathrm{mod}21 \,,\quad 2^{6k+2}=4\,\mathrm{mod}21 \\[5pt]
2^{6k+3}=8\,\mathrm{mod}21 \,,\quad 2^{6k+4}&=16\,\mathrm{mod}21 \,,\quad 2^{6k+5}=11\,\mathrm{mod}21 \,.
\end{align*}
This periodicity explains why the formula \eqref{cmodN} is also a solution to the prime factoring problem (in some special cases, of course). 
Thus, assuming that this procedure applies to the
integer number $N$, its prime factors will be given by the equation
\beq
a^{rk}=1\,\mathrm{mod}N \,,
\eeq
or, choosing $k=1$,
\beq\label{areq1modN}
a^{r}=1\,\mathrm{mod}N \,.
\eeq
The exponent $r$ is called the \index{Period}\emph{period}. Note that the period $r$ is the smallest non-trivial exponent $R$ for which $a^{R}=1\,\mathrm{mod}N$. Since we need to use the difference of
squares formula, we have that $r$ must be even,
\beq
a^r-1=0\,\mathrm{mod}N\Longrightarrow (a^{r/2}+1)(a^{r/2}-1)=0\,\mathrm{mod}N \,.
\eeq
\begin{theorem}
\textup{Show that the method does not apply if $N=21$ and $a=5$.}
\end{theorem}
\begin{theorem}
\textup{What is the period for $N=77$ and $a= 3$?}
\end{theorem}
In conclusion, here is the procedure: given a number $N$, we start by picking an integer $a$ and then we proceed to find the period $r$.
The prime factors of $N$ follows from equation \eqref{areq1modN} (of course, as long as $r$ is even).
What Shor's algorithm does is to determine the period $r$ faster than any classical algorithm invented so far.
\begin{theorem}
\textup{Show that $a^r=1\,\mathrm{mod}N$ is equivalent to
\beq
1=a^r\,\mathrm{mod}N \,.
\eeq}
\end{theorem}
\noindent Our goal then is to show how Shor's algorithm finds the period $r$ of the function
\beq
f(N,a,r)=a^r\,\mathrm{mod}N \,.
\eeq
Here, $a$ is an integer number coprime to $N$. That is, $a$ is an integer number
whose prime factors are not prime factors of $N$. 

As we said, we will not present Shor's algorithm in its most general form. Instead, let us see how it finds the prime factors of 15. The circuit is the following,
\begin{figure}[H]
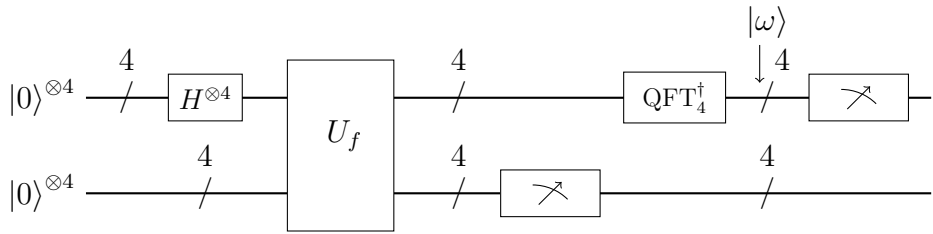

  \centering
  \include{images/shor15}
\vspace{-20pt}
 \caption{Circuit designed to find the prime factors of 15.}
 \label{fig:beta0cnot}
\end{figure}
The oracle $U_f$ is defined by,
\begin{align}
|x\rangle|0\rangle\xmapsto{~U_f~}&|x\rangle|0\oplus f(15,13,x)\rangle=|x\rangle|13^x\,\mathrm{mod}\rangle \,,
\end{align}
where everything is written in decimal notation.

The analysis of the circuit is as follows. First, we have the Hadamard gates that act on the four single qubits at the top of the diagram,
\begin{align}
|0\rangle^{\otimes 4}|0\rangle^{\otimes 4} \xmapsto{~H^{\otimes 4}\otimes 1^{\otimes 4}~}&
H^{\otimes 4}|0\rangle^{\otimes 4}|0\rangle^{\otimes 4}=\big(H|0\rangle\big)^{\otimes 4}|0\rangle^{\otimes 4}\nonumber\\[8pt]
&=\Big( \frac{1}{\sqrt{2}}\sum_i|i\rangle\Big)^{\otimes 4}|0\rangle^{\otimes 4}=\frac{1}{4}\sum_{i,j,k,l}|i\,j\,k\,l\rangle|0\rangle^{\otimes 4}
\nonumber\\[5pt]
&=\frac{1}{4}\sum_{x=0}^{15}|x\rangle|0\rangle\,.
\end{align}
Be aware that in the last step we changed from binary to decimal notation.

Then, there is the oracle
\begin{align}
\xmapsto{~U_f~}&\;\frac{1}{4}\sum_{x=0}^{15}|x\rangle|13^x\,\mathrm{mod}15\rangle\nonumber\\[5pt]
&=\frac{1}{4}\big(|0\rangle+|4\rangle+|8\rangle+|12\rangle\big)|1\rangle+\frac{1}{4}\big(|1\rangle+|5\rangle+|9\rangle+|13\rangle\big)|13\rangle\nonumber\\[5pt]
&+\frac{1}{4}\big(|2\rangle+|6\rangle+|10\rangle+|14\rangle\big)|3\rangle+\frac{1}{4}\big(|3\rangle+|7\rangle+|11\rangle+|15\rangle\big)|6\rangle\,.
\end{align}
Suppose now that the measurement of the bottom register gives $|6\rangle$. In this case, the state after the measurement is
\beq
\xmapsto{~M~} \frac{1}{2}\big(|3\rangle+|7\rangle+|11\rangle+|15\rangle\big)|6\rangle \,.
\eeq
If instead of $|6\rangle$, any of the other states came out, $(|1\rangle,|13\rangle$ or $|3\rangle)$, the analysis below would be similar. 

Then, we have the inverse quantum Fourier transform on the upper register,
\beq\label{shorw}
\xmapsto{~\mathrm{QFT}^{\dagger}_4~} \;\frac{1}{2}\big(\mathrm{QFT}^{\dagger}_4|3\rangle+\mathrm{QFT}^{\dagger}_4|7\rangle+\mathrm{QFT}^{\dagger}_4|11\rangle+\mathrm{QFT}^{\dagger}_4|15\rangle\big)=|\omega\rangle \,.
\eeq
We will sketch how to compute the first of these inverse QFT's, the
others are similar. Using the formula \eqref{qftinv},
\begin{align}
\mathrm{QFT}^{\dagger}_4|3\rangle&=\frac{1}{4}\sum_{y=0}^{15}e^{\frac{\pi i}{8}3y}|y\rangle\nonumber\\
&=\frac{1}{4}\big(|0\rangle+e^{\frac{\pi i}{8}3}|1\rangle+e^{\frac{\pi i}{8}6}|2\rangle+e^{\frac{\pi i}{8}9}|3\rangle\big)\nonumber\\
&+\frac{1}{4}\big(e^{\frac{\pi i}{8}12}|4\rangle+e^{\frac{\pi i}{8}15}|5\rangle+e^{\frac{\pi i}{8}18}|6\rangle+e^{\frac{\pi i}{8}21}|7\rangle\big)\nonumber\\
&+\frac{1}{4}\big(e^{\frac{\pi i}{8}24}|8\rangle+e^{\frac{\pi i}{8}27}|9\rangle+e^{\frac{\pi i}{8}30}|10\rangle+e^{\frac{\pi i}{8}33}|11\rangle\big)\nonumber\\
&+\frac{1}{4}\big(e^{\frac{\pi i}{8}36}|12\rangle+e^{\frac{\pi i}{8}39}|13\rangle+e^{\frac{\pi i}{8}42}|14\rangle+e^{\frac{\pi i}{8}45}|15\rangle\big)\,.
\end{align}
Computing all of them and substituting in \eqref{shorw} yields
\beq\label{wshor6}
|\omega\rangle=\frac{1}{2}|0\rangle+\frac{i}{2}|4\rangle-\frac{1}{2}|8\rangle-\frac{i}{2}|12\rangle \,.
\eeq
The corresponding probabilities are,
\beq
P\big(|0\rangle\big)=P\big(|4\rangle\big)=P\big(|8\rangle\big)=P\big(|12\rangle\big)=1/4 \,.
\eeq
\begin{theorem}
\textup{Do all the calculations that lead to \eqref{wshor6}.}
\end{theorem}
\noindent With this, we conclude the quantum analysis of Shor's algorithm. What remains is a classical post-processing, where the period $r=4$ is found. We will not provide the details here, but you can see it from the possible measurement outcomes $0, 4, 8, 12$.

\vspace{10pt}
\begin{tcolorbox}[breakable, enhanced]
\vspace{5pt}
\begin{note}\label{box:qft}
$\mathrm{\mathbf{The~ quantum~ Fourier ~transform.}}$
\end{note}
One of the mathematical tools used in Shor's algorithm --- but also in other quantum algorithms --- is the so called \index{Quantum Fourier transform, QFT}\emph{quantum Fourier transform} or \emph{QFT} for short. In few words, the
QFT is a change of basis from the computational basis to
the Fourier basis. 

~~In \eqref{qubitindecnotation} we saw how we can express any $n$-qubit state vector $|Q\rangle\in \mathcal{H}_Q\cong \mathbb{C}^{2^n}$ in binary as well as decimal notation,
\beq
|Q\rangle=\sum_{i_1,\ldots,i_n}\alpha_{i_1\ldots i_n}|i_1\,\ldots \,i_n\rangle
=\sum_{x=0}^{N-1}\alpha_x|x\rangle \,,
\eeq
where $N=2^n$.
However, the vectors $|x\rangle$ are not the only
possible vectors one can use to span $\mathcal{H}_Q$. In fact, an infinite amount of
alternative sets can be chosen. One such set is
the \index{Fourier basis}\emph{Fourier basis} with elements given by the following formula
\beq\label{eq:qft}
|x\rangle_{\mathrm{QFT}}=\frac{1}{\sqrt{N}}\sum_{y=0}^{N-1}e^{\frac{2\pi i}{N}xy}|y\rangle \,.
\eeq
The vector $|x\rangle_{\mathrm{QFT}}$ is the quantum Fourier transformed of $|x\rangle$, that is, $|x\rangle_{\mathrm{QFT}}=\mathrm{QFT}_n|x\rangle$. Be aware that here, $y$ and $x$ must be expressed in decimal notation.
\begin{theorem}
\textup{Show that $\mathrm{QFT}_1|i\rangle=|(-1)^i\rangle$, that is, $\mathrm{QFT}_1=H$.}
\end{theorem}

To show how the quantum Fourier transform works in more complex situations, let us show explicitly
how to obtain the Fourier basis vectors
of the 2-qubit Hilbert space $\mathcal{H}_{q_2}$, $|j\,k\rangle\mapsto \mathrm{QFT}_2|j\,k\rangle=|j\,k\rangle_{\mathrm{QFT}}$. 

The general formula \eqref{eq:qft} in this case is
\beq
\mathrm{QFT}_2|j\,k\rangle=\mathrm{QFT}_2|x=2j+k\rangle=\frac{1}{2}\sum_{y=0}^3 e^{\frac{\pi i}{2}xy}|y\rangle \,.
\eeq
The first two Fourier basis vectors are,
\begin{align*}
\mathrm{QFT}_2|0\,0\rangle&=\mathrm{QFT}_2|x=0\rangle =\frac{1}{2}\sum_{y=0}^3 e^{\frac{\pi i}{2}0y}|y\rangle\\
&=\frac{1}{2}\big(|0\rangle+|1\rangle+|2\rangle+|3\rangle\big) =\frac{1}{2}\big(|0\,0\rangle+|0\,1\rangle+|1\,0\rangle+|1\,1\rangle\big) \,,
\end{align*}
and
\begin{align*}
\mathrm{QFT}_2|0\,1\rangle&=\mathrm{QFT}_2|x=1\rangle =\frac{1}{2}\sum_{y=0}^3 e^{\frac{\pi i}{2}1y}|y\rangle\\
&=\frac{1}{2}\big(e^{\frac{\pi i}{2}0}|0\rangle+e^{\frac{\pi i}{2}1}|1\rangle+e^{\frac{\pi i}{2}2}|2\rangle+e^{\frac{\pi i}{2}3}|3\rangle\big) \\[5pt]
&=\frac{1}{2}\big(|0\,0\rangle+i|0\,1\rangle-|1\,0\rangle-i|1\,1\rangle\big) \,.
\end{align*}
\begin{theorem}
\textup{Find the Fourier  transformed of the other two basis vectors and show that the matrix representation of QFT$_2$ in the computational basis of $\mathcal{H}_{q_2}$ is
\beq
\mathrm{QFT}_2=
\frac{1}{2}
\begin{bmatrix}
1 & 1 & 1 & 1 \\
1 & i & -1 & -i \\
1 & -1 & 1 & -1 \\
1 & -i & -1 & i 
\end{bmatrix}
\,.
\eeq
}
\end{theorem}
\begin{theorem}
\textup{Find the matrix representation of  $\mathrm{QFT}_3$ and $\mathrm{QFT}_4$.}
\end{theorem}
\begin{theorem}
\textup{Show that the Fourier basis vectors of $\mathcal{H}_Q$ can also be written
\beq
\mathrm{QFT}_n|x\rangle=\frac{1}{\sqrt{N}}\bigotimes_{l=1}^n\sum_j e^{\frac{2\pi i}{2^l}xj}|j\rangle \,.
\eeq
Rewrite this expression in terms of $\omega_N=e^{2\pi i/N}$.
}
\end{theorem}
\begin{theorem}
\textup{What are the matrices of QFT$_2$, QFT$_3$ and QFT$_4$ in terms of $\omega_N=e^{2\pi i/N}$? Do you recognize any pattern? What about QFT$_n$?}
\end{theorem}
If the QFT takes us from a description of the qubit state vector in the computational basis to the Fourier basis, the \emph{inverse QFT}, denoted  $\mathrm{QFT}^{\dagger}_n=\mathrm{QFT}^{-1}_n$, does precisely the opposite,
\beq\label{qftinv}
\mathrm{QFT}^{\dagger}_n|x\rangle_{\mathrm{QFT}}=\frac{1}{\sqrt{N}}\sum_{y=0}^{N-1}e^{-\frac{2\pi i}{N}xy}|y\rangle \,.
\eeq

\vspace{5pt}
\end{tcolorbox}
\vspace{10pt}

\subsection{Superdense Coding and Teleportation}

Recall that, by definition, a separable or product state can always be written as the tensor
product of two state vectors. In contrast, an entangled state is non-separable (see equation \eqref{entangled2q}). By abuse of language, even though a 2 qubit is generally
entangled and its state vector is not the product of two single-qubit state vectors, in the literature the vectors of the first
and second Hilbert spaces are frequently called ``first"
and ``second" qubits, respectively.

Suppose now the following situation,
\begin{figure}[H]
  \centering
\input{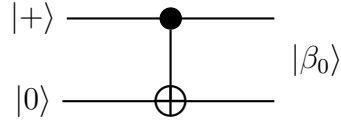}
\caption{Production of a Bell state using the CNOT gate.}
 \label{fig:beta0cnot}
\end{figure}
\noindent Initially, the two single-qubit states $|+\rangle$ and $|0\rangle$ are not entangled,
so the composite system is in the product state $|+\rangle|0\rangle$.
After the CNOT gate is applied, the outgoing state is
\begin{align*}
|+\rangle|0\rangle=\frac{1}{\sqrt{2}}\big(|0\,0\rangle+|1\,0\rangle\big) \xmapsto{~\mathrm{CNOT}~}\frac{1}{\sqrt{2}}\big(|0\,0\rangle+|1\,1\rangle\big)\equiv \beta_0\,.
\end{align*}
This is an entangled state in $\mathcal{H}_{q_2}$ and it is called a Bell state. Let us momentarily denote it by $\beta_0$. An alternative, but obviously equivalent form of creating $\beta_0$ is illustrated in the following diagram,
\begin{figure}[H]
  \centering
 \input{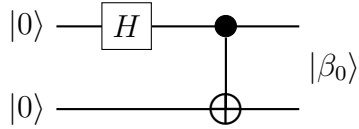}
 \caption{Alternative set-up to the circuit in Figure~\ref{fig:beta0cnot}.}
 \label{fig:galaxy}
\end{figure}
\noindent More generally, we can allow the incoming states to be in any of the computational basis state vectors, 
\begin{figure}[H]
  \centering
\input{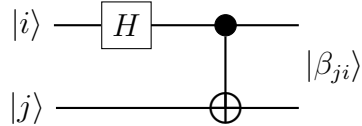}
 \caption{Set-up to create any Bell state.}
 \label{fig:galaxy}
\end{figure}
\noindent The possibilities are:
\begin{align*}
|0\rangle|0\rangle&\xmapsto{~H\otimes 1~}\frac{1}{\sqrt{2}}\big(|0\rangle+|1\rangle\big)|0\rangle \xmapsto{~\mathrm{CNOT}~}\frac{1}{\sqrt{2}}\big(|0\,0\rangle+|1\,1\rangle\big)\equiv  |\beta_0\rangle \,,\\
|0\rangle|1\rangle&\xmapsto{~H\otimes 1~}\frac{1}{\sqrt{2}}\big(|0\rangle+|1\rangle\big)|1\rangle \xmapsto{~\mathrm{CNOT}~}\frac{1}{\sqrt{2}}\big(|0\,1\rangle+|1\,0\rangle\big)\equiv |\beta_1\rangle \,,\\
|1\rangle|0\rangle&\xmapsto{~H\otimes 1~}\frac{1}{\sqrt{2}}\big(|0\rangle-|1\rangle\big)|1\rangle \xmapsto{~\mathrm{CNOT}~}\frac{1}{\sqrt{2}}\big(|0\,0\rangle-|1\,1\rangle\big)\equiv |\beta_2\rangle \,,\\
|1\rangle|1\rangle&\xmapsto{~H\otimes 1~}\frac{1}{\sqrt{2}}\big(|0\rangle-|1\rangle\big)|1\rangle \xmapsto{~\mathrm{CNOT}~}\frac{1}{\sqrt{2}}\big(|0\,1\rangle-|1\,0\rangle\big)\equiv |\beta_3\rangle \,.
\end{align*}
The four states $|\beta_0\rangle,|\beta_1\rangle,|\beta_2\rangle,|\beta_3\rangle\in \mathcal{H}_{q_2}$ are called
\index{Bell states}\emph{Bell states} or \index{EPR pair}\emph{EPR pairs}. Note that they are perpendicular, so they form a basis for $\mathcal{H}_{q_2}$. This basis is called the \index{Bell basis} \emph{Bell basis}.
Of course, we could also have obtained them in a quicker way by using index notation,
\begin{align}
|i\rangle|j\rangle&\xmapsto{~H\otimes 1~} H|i\rangle|j\rangle=\frac{1}{\sqrt{2}}
\big(|0\rangle+(-1)^i|1\rangle\big)|j\rangle \\
&\xmapsto{~\mathrm{CNOT}~} \frac{1}{\sqrt{2}}
\big(|0\rangle|j\rangle+(-1)^i|1\rangle|\bar{j}\rangle\big)=|\beta_{ji}\rangle \,.
\end{align}
Interchanging $i\leftrightarrow j$, the general Bell state vector becomes
\beq
|\beta_{ij}\rangle=\frac{1}{\sqrt{2}}
\big(|0\,i\rangle+(-1)^j|1\,\bar{i}\rangle\big) \,.
\eeq
Comparing with the states $|\beta_i\rangle$ defined above, we see that $|\beta_{00}\rangle=|\beta_0\rangle$, $|\beta_{01}\rangle=|\beta_1\rangle$, $|\beta_{10}\rangle=|\beta_2\rangle$ and $|\beta_{11}\rangle=|\beta_3\rangle$.
\begin{theorem}
\textup{Write the four computational basis vectors of $\mathcal{H}_{q_2}$ in terms of the Bell states.}
\end{theorem}

\index{Superdense coding}\emph{Superdense coding} is a quantum communication protocol
designed to communicate two classical bits of information
($b_1b_2=00,01,10$ or $11$) by sending only one single qubit.
That is, the code can be used to
communicate one of four classical pieces of information:
it can be four numbers, four colors, etc.
It works as follows. Imagine that there is a sender and a receiver, each with a physical qubit of a 2-qubit system forming a Bell state. The following sequence of unitary
transformations shows how the protocol operates.\\ 

\noindent At the sender's side:
\begin{align*}
|\beta_{00}\rangle=\frac{1}{\sqrt{2}}\big(|0\,0\rangle+|1\,1\rangle\big) &\xmapsto{~~~} \frac{1}{\sqrt{2}}\big(|0\oplus b_2 ~0\rangle+|1\oplus b_2~1\rangle\big) \nonumber \\
&\xmapsto{~~~}  \frac{1}{\sqrt{2}}\big((-1)^{b_1b_2}|b_2~0\rangle+(-1)^{b_1(b_2\oplus 1)}|1\oplus b_2~1\rangle\big)=|q\rangle \,.
\end{align*}
At the receiver's side:
\begin{align*}
|q\rangle&\xmapsto{~~~}  \frac{1}{\sqrt{2}}\big((-1)^{b_1b_2}|b_2~0\oplus b_2\rangle+(-1)^{b_1\bar{b_2}}|1\oplus b_2~1\oplus(1\oplus b_2)\rangle\big) \nonumber \\[5pt]
&~~~~=|(-1)^{b_1}\rangle|b_2\rangle \qquad \mathrm{(up~to~a~phase)}\nonumber \\[5pt]
&\xmapsto{~~~}  |b_1\rangle|b_2\rangle \nonumber \\[5pt]
&\xmapsto{~~~}b_1b_2\,.
\end{align*}
\begin{theorem}\label{ex:qteleportationdiagram}
\textup{First, determine each of transformations indicated above by arrows, then draw the circuit.}   
\end{theorem}
\begin{theorem}
\textup{Repeat the previous steps in case the preshared 2 qubit is a general Bell state $|\beta_{ij}\rangle$.} 
\end{theorem}
\index{Quantum teleportation}\emph{Quantum teleportation} is a communication protocol designed
to transfer the information of a single qubit through
a classical channel. The circuit diagram is similar to that of superdense coding, the difference being that the parts of the sender and the receiver are interchanged.
\begin{figure}[H]
  \centering
\input{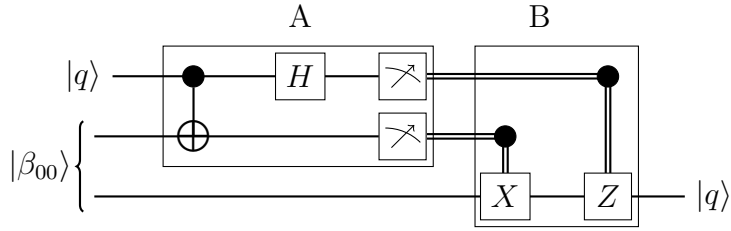}
  \caption{The quantum teleportation circuit.}
 \label{fig:qteleportation}
\end{figure}
\noindent Remember that the transfer of classical bits is represented graphically by a double line, while we use single lines for qubits.
\begin{theorem}
\textup{Write down the evolution of the initial state $|q\rangle|\beta_{00}\rangle$ at every step of the circuit and show
that the outgoing qubit is indeed $|q\rangle$.}
\end{theorem}
\begin{theorem}
\textup{Compare the circuit for quantum teleportation shown in Figure \ref{fig:qteleportation} with the circuit you drew in Exercise \ref{ex:qteleportationdiagram} for superdense coding.}
\end{theorem}

\subsection{Quantum Simulation}

The simulation of a quantum mechanical system by using a quantum computer was Feynman's seminal idea on quantum computers. He was convinced that quantum systems, such as common molecules, were so complex that the only way 
to predict their behaviour was through the
use of a device fully built according to the same physical principles as the system itself.
Despite Feynman's early vision and the effort made for more than twenty years in that direction, quantum simulation remains a challenging problem. It is not difficult to see why this is so.

Suppose, for example, that you have a quantum system
of $n$ interacting particles (let us say electrons),
each with two possible quantum states (the electrons can be up or down). A fully quantum mechanical
description of the system should keep track of the
quantum superposition of the $2^n$ possible configurations
of the system at every time $t$. If the number of particles is small and the interactions are simple enough, we may expect a classical computer to do the job. However, as soon as the
number of particles increases substantially, for instance to
$n=100$, the number of possible configurations to keep
track of becomes so large that the problem becomes intractable
for classical computers. For this, we need quantum computers.

As you know very well, the dynamics of a quantum system is described by the \index{Schr\"{o}dinger's equation}\emph{Schr\"{o}dinger equation},
\beq\label{schrodingereq}
\hat{H}|\psi(t)\rangle=i\frac{d}{dt}|\psi(t)\rangle \,,
\eeq
where $\hat{H}=\hat{H}^\dagger$ is the \index{Hamiltonian operator}\emph{Hamiltonian operator} and $|\psi(t)\rangle$ is the state of the system at some time $t$. That is, if $|\psi(t_0)\rangle$ is the state of the system at time $t_0$, the Schr\"{o}dinger equation tells you that, for time-independent Hamiltonians, there is an operator 
\beq
U(t,t_0)=e^{-i\hat{H}(t-t_0)} \,,
\eeq
called the \index{Time evolution operator}\emph{time evolution operator}, such that the initial state evolves in time according to 
\beq
|\psi(t_0)\rangle \mapsto |\psi(t)\rangle=U(t,t_0)|\psi(t_0)\rangle \,.
\eeq

\begin{theorem}
\textup{Show that $|\psi(t)\rangle=e^{-i\hat{H}(t-t_0)}|\psi(t_0)\rangle$ is, indeed, a solution to the Schr\"{o}dinger equation \eqref{schrodingereq}.}
\end{theorem}

The idea of \index{Quantum simulation}\emph{quantum simulation}, also known as \index{Hamiltonian simulation}\emph{Hamiltonian simulation}, consists of finding a
quantum circuit (built, of course, from elementary gates)
matching as accurately as possible the time-evolution operator of
the real physical system. Here we will only discuss the
simulation of the time-evolution operator and assume that we know
how to create an $n$-qubit state $|Q\rangle$ that reproduces the
initial state vector $|\psi(t_0)\rangle$ of the system, $|\psi(t_0)\rangle=|Q\rangle$.

To start with, suppose the simplest case of a two-level quantum system with Hamiltonian $\hat{H}=\hat{H}_{q_1}$. The matrix representation of the Hamiltonian in the computational basis is
\beq
\hat{H}_{q_1}=
\begin{bmatrix}
H_{00} & H_{01} \\
H_{10} & H_{11}
\end{bmatrix}
\,;
\eeq
where, because $\hat{H}$ is Hermitian, $H_{01}=H^*_{10}$. Now, since we know that any complex $2 \times 2$ matrix can be written as a linear
combination of the Pauli matrices and the identity matrix, then
\beq
\hat{H}_{q_1}=\sum_A h_A \, \sigma_A \,,
\eeq
where $A=I, X, Y, Z$ and, because of the hermicity of the Hamiltonian, $h_A\in \mathbb{R}$. \\

\noindent For the Hamiltonian of a 2-qubit quantum system, we can use an analogous result stating that any Hermitian $4 \times 4$ complex matrix can be written as
a real linear combination of the tensor product of Pauli matrices,
\beq
\hat{H}_{q_2}=\sum_{A,B} h_{AB} \,\sigma_A \otimes \sigma_B\,.
\eeq
\begin{theorem}
\textup{Write the matrices $\hat{H}_{q_1}$ and $\hat{H}_{q_2}$ as a linear combination of the Pauli matrices, displaying explicitly the coefficients $h_A$ and $h_{AB}$.}
\end{theorem}
\noindent Similarly, the most general Hamiltonian for a physical system corresponding to an $n$ qubit is of the form
\beq\label{nqubithamiltonian}
\hat{H}_{q_n}=\sum_{A_1,\ldots, A_n} h_{A_1 \ldots A_n} \, \sigma_{A_1} \otimes \ldots \otimes \sigma_{A_n} \,,
\eeq
where $A_1, \ldots , A_n=I, X, Y, Z$ and all the coefficients $h_{A_1 \ldots A_n}$ are real.

Let us now see some simple examples. 
Suppose we know that the Hamiltonian of a two-level system has the form of the Pauli operator $Z$, that is, $\hat{H}=\hat{H}_{q_1}=Z$. If $|q\rangle$ is associated to the initial state $|\psi(t_0)\rangle$, the evolution of the physical system will be described by $|q\rangle\mapsto U(t)|q\rangle=e^{-iZt}|q\rangle$. We now recall that the elementary gate $R_z(\theta)=e^{-iZ\theta/2}$, which implies that 
\beq
U(t)=e^{-iZt}=R_z(2t) \,.
\eeq
The quantum circuit that simulates the evolution of our physical system is then
\begin{figure}[H]
  \centering
\input{images/rz2t}
 \caption{}
 \label{fig:galaxy}
\end{figure}
\noindent The state $|\omega\rangle$ leaving the gate is assumed to perfectly match the final state $|\psi(t)\rangle$ of the real physical system we wanted to simulate.
\begin{theorem}
\textup{What if the Hamiltonian of the system is any of the other Pauli operators? For instance, for $\hat{H}=X$, show that the quantum circuit modelling the time-evolution operator is
\begin{figure}[H]
  \centering
\input{images/hrz2th}
 \caption{}
 \label{fig:galaxy}
\end{figure}}
\end{theorem}
To find a quantum circuit that simulates the evolution a 2 qubit is more difficult. Suppose, for simplicity, that $\hat{H}=\hat{H}_{q_2}=\sigma_A\otimes \sigma_A$, where $A=I,X,Y,Z$. The time-evolution operator is then $U(t)=e^{-i\sigma_A\otimes\sigma_A t}$. By Taylor expansion, 
\beq
U(t)=e^{-i\sigma_A\otimes\sigma_A t}=\cos(t) I-i\sin(t) \sigma_A\otimes\sigma_A \,.
\eeq
\begin{theorem}
\textup{Prove the previous formula.}
\end{theorem}
\begin{theorem}
\textup{What is the matrix representation of the operator $U(t)=e^{-i\sigma_A\otimes\sigma_A t}$?}
\end{theorem}
If, as we are assuming, $|\psi(t_0)\rangle=|q\rangle$, then the time evolution of the system will be given by,
\begin{align}
U(t)|q_2\rangle&=e^{-i\sigma_A\otimes\sigma_A t}\sum_{i,j}\alpha_{ij}|ij\rangle\nonumber\\
&=\cos(t) \sum_{i,j}\alpha_{ij}|ij\rangle -i\sin(t) \sum_{i,j}\alpha_{ij} \sigma_A|i\rangle\sigma_A|j\rangle \,.
\end{align}
\begin{theorem}
\textup{Show that the time evolution operator $U(t)=e^{-iZ\otimes Z t}$ can be implemented by the circuit
\begin{figure}[H]
  \centering
\input{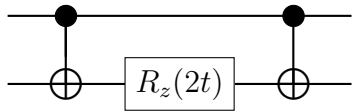}
 \caption{Circuit emulating $U(t)=e^{-iZ\otimes Z t}$.}
 \label{fig:galaxy}
\end{figure}}
\end{theorem}
Suppose now, more generally, a physical system modelled by an $n$ qubit evolving with $U(t)=e^{-i\sigma^{\otimes n}_A}t$. Taylor expanding as before, we get that
\beq
U(t)=e^{-i\sigma^{\otimes n}_A}=\cos(t) I-i\sin(t) \sigma_A^{\otimes n} \,.
\eeq
\vspace{-20pt}
\begin{theorem}
\textup{What is the quantum circuit corresponding to $U(t)=e^{-iZ^{\otimes n}}$? Compare your diagram with the circuit shown in Exercise \ref{exer:Zxnsimulation}.}
\end{theorem}

\vspace{10pt}
\begin{tcolorbox}[breakable, enhanced]
\vspace{5pt}
\begin{note} 
$\mathrm{\mathbf{The ~Pauli ~group.}}$
\end{note}
As you can easily verify, the product of two Pauli matrices is, up to a constant that can be $\pm 1$ or $\pm i$, another Pauli matrix.\index{Pauli group}
\begin{theorem}
\textup{Prove that the Pauli matrices form a group.}
\end{theorem}
The Pauli matrices and the $2\times 2$ identity matrix are said to form the single-qubit Pauli group $\mathcal{P}_1$. To remember that there are constants $\pm 1$ and $\pm i$ involved, the group is commonly denoted
\beq
\mathcal{P}_1=\{I, X, Y, Z;\pm 1, \pm i \} \,.
\eeq
We can be more economical and write $\mathcal{P}_1=\{ \sigma_A; \pm 1, \pm i \}$, where it is understood that $A=I, X, Y, Z$. 

~~The 2-qubit Pauli group $\mathcal{P}_2$, which acts on 2 qubits, is
\beq
\mathcal{P}_2=\{ \sigma_{A}\otimes \sigma_{B}; \pm 1, \pm i\} \,,
\eeq
where $A, B=I, X, Y, Z$. An arbitrary element of  $\mathcal{P}_2$ acts as follows,
\beq
(\sigma_A\otimes \sigma_B)|q_2\rangle =(\sigma_A\otimes \sigma_B)\Big(\sum_{i,j} \alpha_{ij}|i\rangle \otimes |j\rangle\Big)= \sum_{i,j} \alpha_{ij}\,\sigma_A|i\rangle \otimes \sigma_B|j\rangle
\eeq
In general, the Pauli group on $n$ qubits, also known as the $n$-qubit Pauli group for short, is denoted
\beq
\mathcal{P}_n=\{ \sigma_{A_1}\otimes \ldots \otimes \sigma_{A_n}; \pm 1, \pm i\} \,.
\eeq
The elements of a Pauli group are called \index{Pauli operators}\emph{Pauli operators}. Sometimes we simply write $\sigma_{A_1}\otimes \ldots \otimes \sigma_{A_n}=G_n$.
The following alternative notations can be used to denote a Pauli group,
\beq
\mathcal{P}_n =\{ \sigma_A; \pm 1, \pm i\}^{\otimes n}=\{ G_n; \pm 1, \pm i\}={\mathcal{P}_1}^{\otimes n } \,.
\eeq

~~It can be shown that, in general, any $2^n\times 2^n$ complex matrix $M$ can be written as a linear combination
\beq
M=\sum_{r=1}^{4^n} a_r G_r \,.
\eeq
\begin{theorem}
\textup{Show that any $4\times 4$ complex matrix is a linear combination of tensor products $\sigma_A\otimes \sigma_B$.}
\end{theorem}
For example, for the Hamiltonian of an $n$ qubit, instead of \eqref{nqubithamiltonian}, we can write
\beq
\hat{H}=\sum_{r=1}^{4^n} h_r G_r \,,
\eeq
where the hermicity of the Hamiltonian implies that $h^{\dagger}_r=h_r$.
\vspace{5pt}
\end{tcolorbox}
\vspace{10pt}

In the examples above, we have described the Hamiltonians of systems with very simple and somehow unrealistic behaviours. To tackle more interesting situations, we need more powerful methods. One of the simplest approaches is the so called \index{Product formula simulation}\emph{product formula simulation}.

First, we start by writing the total Hamiltonian as a sum of operators, 
\beq
\hat{H}=\sum_{l=1}^L \hat{H}_l \,,
\eeq
where each $\hat{H}_l$ acts, at most, on $k$ qubits. These individual terms are called \index{$k$-local Hamiltonians}$k$-\emph{local Hamiltonians}. 
We now divide the total time interval $t$ into $N$ subintervals, $\Delta t=t/N$ (for simplicity, we are taking $t_0=0$), and then use the \index{Product formula}\emph{product formula}
\beq\label{productformula}
e^{-i(A+B)t}=e^{-iAt}e^{-iBt}+O(\|[A,B]\|t^2/N) \,,
\eeq
to finally obtain 
\begin{align}\label{trotteraprox}
U(t)&=\lim_{N\to \infty}\big(U(t/N) \big)^N \nonumber \\[5pt]
&\approx \big(U(t/N) \big)^N \nonumber =\big( e^{-i\hat{H}t/N}\big)^N=\big( e^{-i\sum_{l=1}^L \hat{H}_lt/N}\big)^N \nonumber \\[5pt]
&=\big(e^{-i\hat{H}_1 t/N}\ldots e^{-i\hat{H}_L t/N}\big)^N
+O\Big(\sum_{l_1,l_2}^L\|[\hat{H}_{l_1},\hat{H}_{l_2}]\|t^2/N\Big) \,.
\end{align}
\begin{theorem}
\textup{Prove the product formula \eqref{productformula}.}
\end{theorem}
\noindent 
If we have enough reasons to neglect the higher-order terms in \eqref{trotteraprox}, the evolution of the initial state of the physical system will be given by
\beq
|\psi(t)\rangle \approx \bigg(\prod_{l=1}^L e^{-i\hat{H}_l t/N}\bigg)^N |\psi(t_0)\rangle \,.
\eeq
Thanks to this approximation, we do not need to find a quantum circuit for the entire time-evolution operator $U(t)$, but for the more manageable short-time operators
\beq
U_l(t/N)=e^{-i \hat{H}_l t/N} \,.
\eeq
Depending on the accuracy of the approximation, we then expect
\beq
|\psi(t)\rangle \approx \bigg(\prod_{l=1}^LU_l(t/N)\bigg)^N|q_n\rangle \,.
\eeq

\section{Quantum Error Correction}

Quantum computers are fragile objects, notably because their interactions with the environment,
for example, with external electromagnetic fields or tiny temperature changes, produce undesirable perturbations
that put at risk the performance of the device and ultimately our confidence in the computation. Knowing that these perturbations are unavoidable, from the very early days of quantum computer science, experts have been trying to build a theory to understand and have control over them. 

The interaction of a quantum system, in our case a qubit, with its environment, can symbolically be written as follows,
\beq\label{errorevolution}
|Q_0\rangle|e_0\rangle \xmapsto{~U~} \,\sum |Q_t\rangle|e_t\rangle \,.
\eeq
Here, $|Q_0\rangle$ and $|e_0\rangle$ are the initial states of the system and the environment, respectively. At this point, we are assuming that there is no entanglement between them. Since the quantum system is closed, even though it is a combination of two subsystems, according to the laws of quantum mechanics it evolves unitarily. As time passes, however, the mutual interaction produces a final state that is entangled. The state of the combined system at a later time is no longer a product state but an entangled state that we symbolically indicate in \eqref{errorevolution} with the summation symbol. 

The whole idea of \index{Quantum error correction, QEC}\emph{quantum error correction} (\emph{QEC}) is precisely to detect and correct the changes occurring in $|Q_0\rangle$ due to the interaction with the environment.
Only with such a theory can quantum computer scientists guarantee
that large-scale quantum computers will ever be useful.

\subsection{Entanglement with the Environment}

When a classical bit interacts with its environment, for example,
when it is transferred through a noisy channel (actually, all realistic channels are noisy to some extent), the
only effect the environment can have on the bit is to flip it. That is, if the bit sent is $b$, $\bar{b}$ may be received. This is the only type of error that must be taken into account
on a classical computing device. The way the
environment interacts with a qubit is more complex.
Moreover, the environment not only modifies the qubit, but in return it is affected by its interaction with the qubit.

Suppose that an instant before they start interacting, the qubit
is in its most general state $|q\rangle=\alpha_0|0\rangle+\alpha_1|1\rangle$ and the
environment is in the state $|e\rangle$. At this point, the composite system, single qubit plus
environment, is not an entangled state; that is, it is simply described by the tensor product $|q\rangle|e\rangle$.
Now, if we denote by $U$ the unitary transformation associated with the evolution of the interacting system, after a certain period of time we will have that
\begin{align}
U(|0\rangle|e\rangle)&=|0\rangle |e_{00}\rangle+|1\rangle |e_{01}\rangle \,, \\
U(|1\rangle |e\rangle)&=|0\rangle|e_{10}\rangle+|1\rangle|e_{11}\rangle \,;
\end{align}
or, in full form,
\begin{align}\label{1qealpha}
U(|q\rangle|e\rangle)&=U\big((\alpha_0|0\rangle +\alpha_1|1\rangle)|e\rangle\big)=\alpha_0 U(|0\rangle  |e\rangle + \alpha_1 U(|1\rangle |e\rangle \nonumber \\[8pt]
&=
\alpha_0 (|0\rangle |e_{00}\rangle + |1\rangle |e_{01}\rangle) +
\alpha_1 (|0\rangle |e_{10}\rangle + |1\rangle  |e_{11}\rangle) \nonumber \\[5pt]
&=\sum_{i,j}\alpha_i|j\rangle|e_{ij}\rangle \,.
\end{align}
Now, since any $2\times 2$ matrix acting on a single qubit can be written as a linear combination of the Pauli operators $I, X, Y, Z$,
we can rewrite this expression in a more convenient form,
\begin{align}\label{1qesigma}
U(|q\rangle|e\rangle)&=I|q\rangle|e_I\rangle+X|q\rangle|e_X\rangle+Y|q\rangle|e_Y\rangle+Z|q\rangle|e_Z\rangle \nonumber  \\
&=\sum_{A}\sigma_A|q\rangle|e_A\rangle \,,
\end{align}
where $A=I, X, Y, Z$ and, as usual, we use the shorthand notation $\sigma_A=\sigma_A\otimes I$. 

\begin{theorem}
\textup{How are the formulas \eqref{1qealpha} and \eqref{1qesigma} related?}
\end{theorem}

\noindent Suppose now that we have a 2 qubit $|q_2\rangle$ interacting with its environment $|e\rangle$. The entangled system is described by the state vector
\begin{align*}
U(|q_2\rangle|e\rangle)&=U\big((\alpha_{00}|0\,0\rangle +\alpha_{01} |0\,1\rangle+ \alpha_{10}|1\,0\rangle + \alpha_{11} |1\,1\rangle) |e\rangle \big)\\[5pt]
&=I\, I \,|q_2\rangle |e_{II}\rangle + I\, X\, |q_2\rangle |e_{IX}\rangle + I\, Y \,|q_2\rangle |e_{IY}\rangle + I\, Z \,|q_2\rangle |e_{IZ}\rangle \\[3pt]
&~~~+X\, I \,|q_2\rangle |e_{XI}\rangle + X\, X \,|q_2\rangle |e_{XX}\rangle + X\, Y \,|q_2\rangle |e_{XY}\rangle + X\, Z \,|q_2\rangle |e_{XZ}\rangle \\[3pt]
&~~~+Y\,I \,|q_2\rangle |e_{YI}\rangle + Y\, X \,|q_2\rangle |e_{YX}\rangle + Y\, Y \,|q_2\rangle |e_{YY}\rangle + Y\, Z \,|q_2\rangle |e_{YZ}\rangle \\[3pt]
&~~~+Z\, I \,|q_2\rangle |e_{ZI}\rangle + Z\, X \,|q_2\rangle |e_{ZX}\rangle + Z\, Y \,|q_2\rangle |e_{ZY}\rangle + Z\, Z\,|q_2\rangle |e_{ZZ}\rangle \,.
\end{align*}
Or, more simply,
\beq
U(|q_2\rangle|e\rangle)=\sum_{A,B}\sigma_A \,\sigma_B |q_2\rangle|e_{AB}\rangle \,,
\eeq
It is clear that for a 3 qubit,
\beq
U(|q_3\rangle|e\rangle)=\sum_{A,B,C}\sigma_A \, \sigma_B \, \sigma_C |q_3\rangle|e_{ABC}\rangle \,,
\eeq
and for an $n$ qubit,
\beq
U(|Q\rangle|e\rangle)=\sum_{A_1,\ldots,A_n}\sigma_{A_1}  \ldots\sigma_{A_n} |Q\rangle|e_{A_1\ldots A_n}\rangle \,.
\eeq
Since writing all the subscripts can easily become cumbersome, the following notation is usually used,
\beq
E_A=\sigma_{A_1} \ldots \sigma_{A_n} \,,
\eeq
where $A=1, \ldots, 4^n$. These new objects are referred as \index{Error operator}\emph{error operators}. Accordingly, the basis state vectors of the environment's Hilbert space are denoted
\beq
|e_A\rangle=|e_{A_1\ldots A_n}\rangle \,.
\eeq
Employing this new notation,
\beq
U(|Q\rangle|e\rangle)=\sum_{A}E_A |Q\rangle|e_A\rangle \,.
\eeq
The whole goal of QEC is to identify these $E_A$'s and reverse their action. For instance, the equation \eqref{1qesigma} is telling us that, due to its interaction with the environment, a single qubit can stay unaffected ($\sigma_A=I$) but at the same time it is prone to suffer from a bit flit ($\sigma_A=X$), a phase flip ($\sigma_A=Z$) and a combination of the two ($\sigma_A=Y$; remember that $Y=iXZ$). 

\vspace{10pt}
\begin{tcolorbox}[breakable, enhanced]
\begin{note} 
\vspace{5pt}
$\mathrm{\mathbf{Open~ quantum ~systems.}}$
\end{note}
In the previous sections we treated the qubits as almost perfectly
isolated quantum systems. We have only allowed them to
interact with the gates, which are also perfect quantum
objects in the sense that they act on the qubits as
unitary transformations. However, the truth is that quantum computers are not perfect quantum systems. For
example, the medium through which the qubits propagate to
go from one gate to the other can affect the
qubit we want to transmit. In order to built real quantum
computers, we need to deal with these undesirable situations.
The quantum mechanical subdiscipline dealing with this
type of phenomena is called ``open quantum systems".
Since this subject is not usually part of a conventional
quantum mechanics course, we will be very brief in our discussion. 

$~$ The idea, thus, is to provide a mathematical description  of the qubit where it no longer behaves as
an isolated quantum system with state vector evolving unitarily,
but as part of a larger quantum mechanical system
that includes other quantum objects affecting it. These external elements are generally called the \index{Environment}\emph{environment}. For example, in the transmission of a qubit, the
environment may be the medium through which it propagates.
The mathematical description of the qubit interacting
with its environment is given by the so called \emph{density operator} or 
\emph{density matrix formalism}. In it, a quantum system is not
described by a unit state vector $|\Psi\rangle$, but by a \index{Density operator (matrix)}\emph{density operator}, or \emph{density matrix}, defined by $\rho_{\Psi}=|\Psi\rangle\langle \Psi|$.
If the quantum system is a composite system, say a qubit and its environment, this approach allows us to understand the evolution of each of its interacting subsystems, in particular the qubit.

$~$ To begin with, let us assume that at some initial time the qubit and the
environment are not interacting (to be more precise, that they have never interacted). The composite system $|\Psi_0\rangle$ is
thus simply described by the product state $|Q_0\rangle|e_0\rangle$, where
$|Q_0\rangle$ and $|e_0\rangle$ are the qubit and the environment initial state vectors, respectively. After some time interacting, the composite
state evolves into an entangled state $|\Psi_t\rangle$,
\beq
|\Psi_0\rangle=|Q_0\,e_0\rangle\mapsto |\Psi_t\rangle=\sum_A E_A|Q_0\,e_A\rangle \,,
\eeq
where we have used the shorthand notation $E_A=E_A\otimes I$ and the $|e_A\rangle$'s form a basis for the Hilbert space of all possible final states of the environment. The $E_A$'s are the so called \index{Error operator}\emph{error operators}. The description of the evolution of the qubit in terms of the density operator formalism is as follows. Since the initial state of the qubit is $|Q_0\rangle$, the density operator is 
\beq
\rho_{Q_0}=|Q_0\rangle\langle Q_0| \,.
\eeq
Similarly, the density operator of the final composite system is
\begin{align}
\rho_{\Psi_t}=|\Psi_t\rangle\langle \Psi_t| &=\sum_{A,A'}E_A
|Q_0\,e_A\rangle\langle Q_0\, e_{A'}|E^{\dagger}_{A'}\nonumber\\
& =\sum_{A,A'}E_A |Q_0\rangle\langle Q_0|E^{\dagger}_{A'} \otimes |e_A\rangle\langle e_{A'}|\,.
\end{align}
The density operator corresponding to the final state of
the qubit is somehow contained within $\rho_{\Psi_t}$. It is called
the \index{Reduced density operator}\emph{reduced density operator} and it is given by 
\begin{align}
\rho_{Q_t}&=\mathrm{tr}_e \rho_{\Psi_t}=\mathrm{tr}_e|\Psi_t\rangle\langle\Psi_t|=\mathrm{tr}_e\sum_{A,A'}E_A |Q_0\rangle\langle Q_0|E^{\dagger}_{A'} \otimes |e_A\rangle\langle e_{A'}| \nonumber \\
&=\sum_{A,A'}E_A |Q_0\rangle\langle Q_0|E^{\dagger}_{A'} \mathrm{tr}_e|e_A\rangle\langle e_{A'}|=\sum_{A,A'}E_A |Q_0\rangle\langle Q_0|E^{\dagger}_{A'}\delta_{AA'} \nonumber\\
&= \sum_A E_A \rho_{Q_0}E^{\dagger}_A \,.
\end{align}
That is, if the initial state of the qubit is $\rho_{Q_0}=|Q_0\rangle\langle Q_0|$, after some time interacting with the environment, it will evolve into the state $\rho_{Q_t}=\sum_A E_A \rho_{Q_0}E^{\dagger}_A$.
\vspace{5pt}
\end{tcolorbox}
\vspace{10pt}

\subsection{Classical Error Correction}

Classical computers are also subject to undesirable perturbations arising from their interactions with the environment. Sometimes, for example,
we want to send a bit $b$ and it turns out that at the other
end of the wire a $\bar{b}$ is received. We review here
the \index{Classical repetition code}\emph{classical repetition code}, one of the multiple ways computer scientists have invented to protect classical
information from the destructive effects of the environment.
In the last pages, we will see how
a similar procedure can be applied to protect quantum
information. 

Suppose we want to communicate a bit $b$ through a
noisy channel and we know that there is a small probability
$p\ll 1$ for the bit of getting flipped to $\bar{b}$,
\beq
P(\bar{b})=p \,,\qquad P(b)=(1-p) \,.
\eeq
Since $p\ll 1$, it follows that $P(b)/P(\bar{b})\gg 1$. 

The repetition code instructs us to send multiple copies of $b$ if we want
to decrease the probability of receiving the wrong information.
For example, instead of one bit $b$, one can send three
copies of $b$, that is, we send $b\,b\,b$ rather than $b$. If
every single bit in the string $b\,b\,b$ can get flipped with
probability $p$, we can receive three, two, one or no $b$'s.
The corresponding probabilities are as follows,
\begin{align}
P(3b,0\bar{b})&=(1-p)^3 \,,\\[5pt]
P(2b,1\bar{b})&=3(1-p)^2 p \,,\\[5pt]
P(1b,2\bar{b})&=3(1-p) p^2 \,,\\[5pt]
P(0b,3\bar{b})&=p^3 \,.
\end{align}
From here we deduce that
\beq
\frac{P(3b,0\bar{b})}{P(2b,1\bar{b})}=\frac{(1-p)^3}{3(1-p)^2 p}=\frac{1}{3}\frac{P(b)}{P(\bar{b})} \,.
\eeq
That is, if we send three $b$'s instead of one,
the probability of receiving a string with one bit flipped is reduced by a third. Moreover, and this is what is really advantageous about
using the repetition code, the relative probability for
two bits to get flipped at the same time is
\beq
\frac{P(3b,0\bar{b})}{P(1b,2\bar{b})}=\frac{(1-p)^3}{3(1-p) p^2}=\frac{1}{3}\Big(\frac{P(b)}{P(\bar{b})}\Big)^2 \,.
\eeq
Therefore, the probability for two bits to get flipped simultaneously is very
small. It is even smaller for three bits,
\beq
\frac{P(3b,0\bar{b})}{P(0b,3\bar{b})}=\frac{(1-p)^3}{p^3}=\Big(\frac{P(b)}{P(\bar{b})}\Big)^3 \,.
\eeq
\begin{theorem}
\textup{To fix the ideas, substitute $p=10$ and $p=100$ in
the previous example.}
\end{theorem}
\begin{theorem}
\textup{Generalize this discussion to a classical
repetition code of $N$ bits. Consider both, an odd and
an even number of repetitions.}
\end{theorem} 
\noindent From the previous analysis, we arrive at the following conclusion: if we receive two or three $b$'s, is because the original bit string was $b\,b\,b$. In other words, we apply the \index{Majority rule}\emph{majority
rule}. Of course, we could also have received two or
three $b$'s when the original message was $\bar{b}\,\bar{b}\,\bar{b}$;
however, this is so unlikely that we simply ignore these possibilities. 

\begin{theorem}
\textup{For a repetition code of $N$ bits, what is the
maximum number of flips that can occur for the
code to give a correct result?}
\end{theorem}

\noindent Hence, we will assume that, if we send the bit string $b\,b\,b$, we can
receive $b_1\,b_2\,b_3$, where at most one of the $b_i$'s will be flipped, i.e., $b_1=\bar{b}$, $b_2=\bar{b}$ or $b_3=\bar{b}$.
Equivalently, we can say that out of the three initial
bits, at least two will remain unchanged: $b_1=b_2=b_3=b$, $b_1=b_2=b\neq b_3$, 
$b_1=b_3=b\neq b_2$ and $b_2=b_3=b\neq b_1$.
The question now is: how do we know if the bits have been corrupted or not? Of course, we can measure them to see if their values are $b$ or $\bar{b}$. But, we can also use the following alternative method that does not require a direct measurement of the bits. It only checks whether two bits have
the same or opposite values. This
procedure, generally called \index{Parity check}\emph{parity check}, works
as follows:
\begin{align*}
\mathrm{if}~b_1=b_2 \,,~b_3=b_1=b_2 \Rightarrow b_1\,b_2\,b_3 =b\,b\,b \,,\\[5pt]
\mathrm{if}~b_1=b_2 \,,~b_3\neq b_1=b_2 \Rightarrow b_1\,b_2\,b_3 =b\,b\,\bar{b} \,,\\[5pt]
\mathrm{if}~b_1=b_3 \,,~b_2\neq b_1=b_3 \Rightarrow b_1\,b_2\,b_3 =b\,\bar{b}\,b \,,\\[5pt]
\mathrm{if}~b_2=b_3 \,,~b_1\neq b_2=b_3 \Rightarrow b_1\,b_2\,b_3 =\bar{b}\,b\,b \,.
\end{align*}
After detecting which of the bits has been flipped --- if any --- we reverse it to its original value by applying to it a classical NOT gate.

\subsection{Generalities on QEC Codes}

Before the first quantum error-correcting codes were invented in the mid-nineties, it was thought that quantum computers were impossible to realize in practice due to the destructive nature of the interaction with the environment. Today, quantum error correcting-codes are known to exist and QEC is a well-established subfield of quantum computing. The general procedure we will follow here is summarized in the following steps: 
\begin{enumerate}
\item Starting with an $n$ qubit state vector $|Q\rangle$, we create an extended product state by simply adding $m$ \index{Ancillary qubits}\emph{ancillary qubits} (or \emph{ancillas}),
\beq
|Q\rangle \xmapsto{~~~} |Q\rangle|0\ldots 0\rangle=|Q\rangle_{anc} \,.
\eeq
The ancillary qubits are added because we want to use a quantum repetition code
inspired by the classical version discussed in the previous subsection.
\item We then encode the information contained in the original qubit $|Q\rangle$
in the extended state $|Q\rangle_{anc}$. This is done by acting with a unitary transformation $U_{enc}$ on the extended state created in Step 1,
\beq\label{qencoded}
|Q\rangle_{anc} \xmapsto{~~~} U_{enc}|Q\rangle_{anc}=|Q\rangle_{enc} \,.
\eeq
Of course, we need to find out the quantum circuit built from elementary gates that implements the unitary $U_{enc}$. For the quantum repetition code we will discuss here, this step is rather easy.
\item At this point, the error occurs:
\beq
|Q\rangle_{enc} \xmapsto{~~~} E|Q\rangle_{enc}=|Q\rangle_E \,.
\eeq
I said ``error", and not ``errors", because, as for the classical repetition code, we will assume that the probability for two errors to occur at the same time is negligible.
\item Here comes the difficult part. We must design a quantum circuit $R$ that detects and corrects the error,
\beq
|Q\rangle_E \xmapsto{~~~} R|Q\rangle_E=|Q\rangle_R \,.
\eeq
In fact, since we cannot measure directly the qubits without destroying the superposition of states, the error is detected indirectly; for example, as we will see, by parity check. 
\item We then decode the encoded state by undoing what the encoding operator did,
\beq
|Q\rangle_R \xmapsto{~~~} U_{dec}|Q\rangle_R=|Q\rangle_{anc} \,.
\eeq
Since $U_{dec}=U^{-1}_{enc}$, this is telling us that we need to built another circuit similar to the one corresponding to $U_{enc}$ but performing the inverse operation.
\item Finally, we get rid of the ancillary qubits and recover the original state vector $|Q\rangle$,
\beq
|Q\rangle_{anc} \xmapsto{~~~} |Q\rangle \,.
\eeq
\end{enumerate}

\subsection{Single Qubit Error Correction}

Given a single qubit with state vector $|q\rangle$ and two ancillary qubits prepared in the computational basis state $|0\rangle$, we let them pass through the following circuit,
\begin{figure}[H]
  \centering
\input{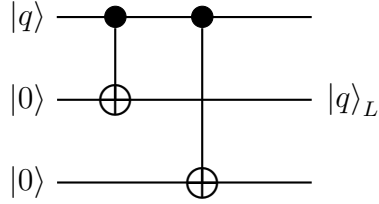}
 \caption{Production of the logical qubit.}
 \label{fig:logicqubit}
\end{figure}
\noindent The outgoing state is,
\beq\label{logicalqubit}
|q\rangle_{anc}=|q\rangle|0\rangle|0\rangle= \alpha_0|0\,0\,0\rangle+\alpha_1|1\,0\,0\rangle \xmapsto{~~~}
\alpha_0|0\,0\,0\rangle+\alpha_1|1\,1\,1\rangle\equiv|q\rangle_L \,.
\eeq
The state $|q\rangle_L$, corresponding to $|Q\rangle_{enc}$ in \eqref{qencoded}, is called the \index{Logical qubit}\emph{logical qubit} to distinguish it from the \index{Physical qubit}\emph{physical qubit} $|q\rangle$ we had originally. 

Note that, when we added the two ancillas $|0\rangle|0\rangle$ to the qubit $|q\rangle$, we extended the Hilbert space from two to eight dimensions,
\beq
|q\rangle \in \mathcal{H}_q\cong \mathbb{C}^2 \,,\qquad |q\rangle_{anc} \in \mathcal{H}_{q_{anc}}\cong \mathbb{C}^8 \,.
\eeq
Since it is in this extended Hilbert space that most of the error-correcting code we will discuss operates, it is worth mentioning some of its most relevant properties.

Two basis vectors of the Hilbert space $\mathcal{H}_{q_{anc}}$ are $|0\,0\,0\rangle$ and $|1\,1\,1\rangle$.
The other six basis vectors can be chosen to be, as usual, $|0\,0\,1\rangle$, $|0\,1\,0\rangle$, $|0\,1\,1\rangle$, $|1\,0\,0\rangle$, $|1\,0\,1\rangle$ and $|1\,1\,0\rangle$. Any vector $|\chi\rangle \in \mathcal{H}_{q_{anc}}$ will then be a linear combination of these basis vectors,
\begin{align}
|\chi\rangle&=\alpha_{000}|0\,0\,0\rangle+\alpha_{001}|0\,0\,1\rangle+\alpha_{010}|0\,1\,0\rangle +\alpha_{011}|0\,1\,1\rangle \nonumber \\[5pt]
&~~~+\alpha_{100}|1\,0\,0\rangle+\alpha_{101}|1\,0\,1\rangle+\alpha_{110}|1\,1\,0\rangle+\alpha_{111}|1\,1\,1\rangle \nonumber \\[5pt]
&=\big(\alpha_{000}|0\,0\,0\rangle+\alpha_{111}|1\,1\,1\rangle\big)+\big(\alpha_{100}|1\,0\,0\rangle +\alpha_{011}|0\,1\,1\rangle\big) \nonumber \\[5pt]
&~~~+\big(\alpha_{010}|0\,1\,0\rangle+\alpha_{101}|1\,0\,1\rangle\big)+\big(\alpha_{001}|0\,0\,1\rangle+\alpha_{110}|1\,1\,0\rangle\big) \,.
\end{align}
The vectors in parentheses are contained in four mutually orthogonal subspaces of $\mathcal{H}_{q_{anc}}$,
\begin{align}
\mathcal{F}_0 =\{|0\,0\,0\rangle,|1\,1\,1\rangle \} \,,\qquad
\mathcal{F}_1 =\{|1\,0\,0\rangle,|0\,1\,1\rangle \} \,,\nonumber \\
\mathcal{F}_2 =\{|0\,1\,0\rangle,|1\,0\,1\rangle \} \,,\qquad
\mathcal{F}_3 =\{|0\,0\,1\rangle,|1\,1\,0\rangle \} \,.
\end{align}
Note that we have chosen the basis vectors of the subspace $\mathcal{F}_1$ so that they correspond to the basis vectors of $\mathcal{F}_0$ with the first bit flipped, that is,
\beq
|1\,0\,0\rangle=X\,I\,I\,|0\,0\,0\rangle \,,\qquad
|0\,1\,1\rangle=X\,I\,I\,|1\,1\,1\rangle \,.
\eeq
Similar, of course, for the basis vectors of $\mathcal{F}_2$ and $\mathcal{F}_3$. 

\noindent It follows, then, that any vector in $\mathcal{H}_{q_{anc}}$ can be written as
\begin{align}
|\chi\rangle&=I\,I\,I\,\big(\alpha_{000}|0\,0\,0\rangle+\alpha_{111}|1\,1\,1\rangle\big) + X\,I\,I\,\big(\alpha_{100}|0\,0\,0\rangle+\alpha_{011}|1\,1\,1\rangle\big) \nonumber \\
&~~+I\,X\,I\,\big(\alpha_{010}|0\,0\,0\rangle+\alpha_{101}|1\,1\,1\rangle\big)+I\,I\,X\,\big(\alpha_{001}|0\,0\,0\rangle+\alpha_{110}|1\,1\,1\rangle\big) \,.
\end{align}

Now that we understand the basic geometry of the Hilbert space $\mathcal{H}_{q_{anc}}$, let us consider the effect of the environment. For a single qubit, we saw in \eqref{1qesigma} that,
\begin{equation}
|q\rangle \xmapsto{~~~} \sum_A \sigma_A |q\rangle \,.
\end{equation}
However, since we have encoded the information of the single qubit $|q\rangle$ in the logical qubit $|q\rangle_L$ given in \eqref{logicalqubit}, we have now to evaluate the effect of the environment on each physical qubit of $|q\rangle_L$. 

\noindent In general, several errors can simultaneously occur on each physical qubit, 
\beq\label{general1qerror}
|q\rangle_L=\sum_i \alpha_i |i\rangle|i\rangle|i\rangle \xmapsto{~~~} \sum_i \alpha_i \sum_A \sigma_A |i\rangle\sum_B \sigma_B |i\rangle\sum_C \sigma_C |i\rangle \,,
\eeq
where $\sigma_A, \sigma_B, \sigma_C=I, X, Y, Z$. But, since we want to consider at most one error per physical qubit, 
\begin{align*}
|q\rangle_L \xmapsto{~~~} \sum_i \alpha_i |i\,i\,i\rangle + \sum_i \alpha_i \sigma_a|i\rangle |i\rangle|i\rangle + \sum_i \alpha_i |i\rangle\sigma_b |i\rangle|i\rangle + \sum_i \alpha_i |i\rangle |i\rangle \sigma_c|i\rangle \,.
\end{align*}
where $\sigma_a, \sigma_b, \sigma_c=X, Y, Z$ and the first summation symbol takes into account the possibility that nothing happens to the qubits. This expression is still too general. In fact, it allows for errors of different nature and we are only interested in errors of the same type. Hence,
\begin{align*}
|q\rangle_L \xmapsto{~~~} \sum_i \alpha_i |i\,i\,i\rangle + \sum_i \alpha_i \sigma_a|i\rangle |i\rangle|i\rangle+ \sum_i \alpha_i |i\rangle\sigma_a |i\rangle|i\rangle + \sum_i \alpha_i |i\rangle |i\rangle \sigma_a|i\rangle \,.
\end{align*}
Finally, if we consider bit-flip errors, that is, $\sigma_a=X$, the corrupted qubit will be described by the following state vector
\begin{align}\label{X1qerror}
|q\rangle_L &\xmapsto{~~~} |q\rangle_E \nonumber \\[5pt]
&=\sum_i \alpha_i |i\,i\,i\rangle + \sum_i \alpha_i X|i\rangle |i\rangle|i\rangle + \sum_i \alpha_i |i\rangle X |i\rangle|i\rangle + \sum_i \alpha_i |i\rangle |i\rangle X|i\rangle \,.
\end{align}
More explicitly,
\begin{align}
|q\rangle_E&=\alpha_0|0\,0\,0\rangle+\alpha_1|1\,1\,1\rangle+\alpha_0|1\,0\,0\rangle+\alpha_1|0\,1\,1\rangle \nonumber \\[5pt]
&~~+\alpha_0|0\,1\,0\rangle+\alpha_1|1\,0\,1\rangle+\alpha_0|0\,0\,1\rangle+\alpha_1|1\,1\,0\rangle \,.
\end{align}
Remember that, actually, we do not know which physical qubit
of the logical qubit has been flipped. The goal is to identify and correct it.
The circuit that does this is the following:
\begin{figure}[H]
  \centering
\input{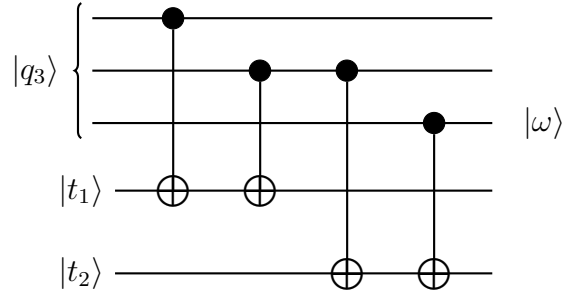}
 \caption{Three-qubit parity check.}
 \label{fig:shorparitycheck}
\end{figure}
\noindent As a matter of fact, the auxiliary qubits $|t_1\rangle$ and $|t_2\rangle$ introduced in Figure \ref{fig:shorparitycheck} need to be in the state $|0\rangle$. However, for
practice, let us first consider the most general case,
\begin{align*}
|q_3\rangle|t_1\rangle|t_2\rangle&=\sum_{i,j,k,l',m'} \alpha_{ijk} t_{1,l'}t_{2,m'}|i\,j\,k\rangle|l'\rangle|m'\rangle  \\
&\xmapsto{~\mathrm{CNOT}_{11}~}\sum_{i,j,k,l',m'} \alpha_{ijk} t_{1,l'}t_{2,m'}|i\,j\,k\rangle|l'\oplus i\rangle|m'\rangle  \\
&\xmapsto{~\mathrm{CNOT}_{21}~}\sum_{i,j,k,l',m'} \alpha_{ijk} t_{1,l'}t_{2,m'}|i\,j\,k\rangle|l'\oplus i\oplus j\rangle|m'\rangle  \\
&\xmapsto{~\mathrm{CNOT}_{22}~}\sum_{i,j,k,l',m'} \alpha_{ijk} t_{1,l'}t_{2,m'}|i\,j\,k\rangle|l'\oplus i\oplus j\rangle|m' \oplus j\rangle  \\
&\xmapsto{~\mathrm{CNOT}_{32}~}\sum_{i,j,k,l',m'} \alpha_{ijk} t_{1,l'}t_{2,m'}|i\,j\,k\rangle|l'\oplus i\oplus j\rangle|m' \oplus j \oplus k\rangle \,.
\end{align*}
Now, since we want $|t_1\rangle=|t_2\rangle=|0\rangle$, we substitute $t_{1,0}=t_{2,0}=0$ and 
$t_{1,1}=t_{2,1}=1$ in the previous result, giving
\begin{align*}
|q_3\rangle|0\rangle|0\rangle&\mapsto \sum_{i,j,k} \alpha_{ijk} |i\,j\,k\rangle|1\oplus i\oplus j\rangle|1 \oplus j \oplus k\rangle  \\
&=\alpha_{000} |0\,0\,0\rangle|1\oplus 0\oplus 0\rangle|1 \oplus 0 \oplus 0\rangle + \alpha_{001} |0\,0\,1\rangle|1\oplus 0\oplus 0\rangle|1 \oplus 0 \oplus 1\rangle  \\[5pt]
&~~~+\alpha_{010} |0\,1\,0\rangle|1\oplus 0\oplus 1\rangle|1 \oplus 1 \oplus 0\rangle + \alpha_{011} |0\,1\,1\rangle|1\oplus 0\oplus 1\rangle|1 \oplus 1 \oplus 0\rangle  \\[5pt]
&~~~+\alpha_{100} |1\,0\,0\rangle|1\oplus 1\oplus 0\rangle|1 \oplus 0 \oplus 0\rangle + \alpha_{101} |1\,0\,1\rangle|1\oplus 1\oplus 0\rangle|1 \oplus 0 \oplus 1\rangle  \\[5pt]
&~~~+\alpha_{110} |1\,1\,0\rangle|1\oplus 1\oplus 0\rangle|1 \oplus 1 \oplus 0\rangle + \alpha_{111} |1\,1\,1\rangle|1\oplus 1\oplus 1\rangle|1 \oplus 1 \oplus 1\rangle  \\[5pt]
&=\big(\alpha_{010} |0\,1\,0\rangle + \alpha_{101} |1\,0\,1\rangle \big) |0\,0\rangle + \big(\alpha_{100} |1\,0\,0\rangle + \alpha_{011} |0\,1\,1\rangle \big) |0\,1\rangle  \\[5pt]
&~~~+\big(\alpha_{001} |0\,0\,1\rangle + \alpha_{110} |1\,1\,0\rangle \big) |1\,0\rangle + \big(\alpha_{000} |0\,0\,0\rangle + \alpha_{111} |1\,1\,1\rangle \big) |1\,1\rangle  \\[5pt]
&=I\,X\,I \big(\alpha_{010} |0\,0\,0\rangle + \alpha_{101} |1\,1\,1\rangle \big) |0\,0\rangle + X\,I\,I \big(\alpha_{100} |0\,0\,0\rangle + \alpha_{011} |1\,1\,1\rangle \big) |0\,1\rangle  \\[5pt]
&~~~+I\,I\,X \big(\alpha_{001} |0\,0\,0\rangle + \alpha_{110} |1\,1\,1\rangle \big) |1\,0\rangle + I\,I\,I \big(\alpha_{000} |0\,0\,0\rangle + \alpha_{111} |1\,1\,1\rangle \big) |1\,1\rangle \,.  
\end{align*}
Finally, since the arbitrary qubit $|q_3\rangle$ used above is indeed $|q\rangle_E$ given explicitly in \eqref{X1qerror}, we must take
\begin{align*}
\alpha_{010}=\alpha_{100}=\alpha_{001}=\alpha_{000}=\alpha_{0} \,, \\
\alpha_{101}=\alpha_{011}=\alpha_{110}=\alpha_{111}=\alpha_{1} \,.
\end{align*}
After substituting, we get
\begin{align}
|q\rangle_E|0\,0\rangle& \xmapsto{~~}|q\rangle_R \\[5pt]
&=I\,X\, I\, |q\rangle_L|0\,0\rangle + X\,I\, I\, |q\rangle_L|0\,1\rangle 
+I\,I\, X\, |q\rangle_L|1\,0\rangle + I\,I\, I\, |q\rangle_L|1\,1\rangle \nonumber \,.
\end{align}
We now measure the auxiliary qubits and do the following:
\beq
\begin{matrix}\label{fliperrorcorrection}
\mathrm{if~the~measurement~gives~}|0\rangle|0\rangle,\hspace{-8pt} & \mathrm{\,we~apply~}I\,X\,I \,,\\
\mathrm{if~the~measurement~gives~}|0\rangle|1\rangle, \hspace{-8pt} & \mathrm{\,we~apply~}X\,I\,I \,,\\
\mathrm{if~the~measurement~gives~}|1\rangle|0\rangle, \hspace{-8pt} & \mathrm{\,we~apply~}I\,I\,X \,,\\
\mathrm{if~the~measurement~gives~}|1\rangle|1\rangle, \hspace{-8pt} & \mathrm{we~apply~}I\,I\,I \,.
\end{matrix}
\eeq
Regardless of the measurement outcome, the procedure \eqref{fliperrorcorrection} will always result in the state vector $|q\rangle_L$. We finally get rid of the ancillary qubits by using the following circuit,
\begin{figure}[H]
  \centering
\input{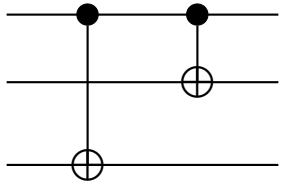}
 \caption{Decoding gate.}
 \label{fig:galaxy}
\end{figure}
\noindent Indeed, the outgoing state is,
\beq
|q\rangle_L=\alpha_0 |0\,0\,0\rangle + \alpha_1 |1\,1\,1\rangle \xmapsto{~~~} \alpha_0 |0\rangle + \alpha_1 |1\rangle =|q\rangle \,.
\eeq
We have provided a complete description of the
bit flip error-correcting code. However, as equation \eqref{general1qerror} shows, many other errors can occur to the logical qubit $|q\rangle_L$. The treatment of the general case will be the subject of future notes.

\begin{figure}[H]
  \centering
\input{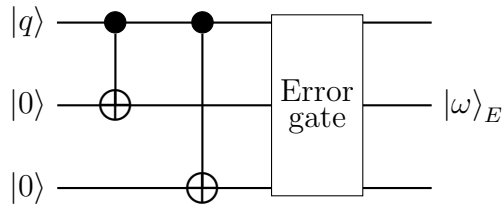}
 \caption{The error gate.}
 \label{fig:galaxy}
\end{figure}

\section{Bibliography}

If you think you need additional supporting material, maybe
because an idea or calculation in my notes is not clear enough, consult the book by Kaye \emph{et al}. It is a bit more elementary and it is very well-written. A textbook at
the same level as these notes is Nielsen \& Chuang, which
is the classic reference on the subject. In addition to these books, you may find useful the free online resources I list below. In particular, the notes by Preskill are worth studying, especially because they were written from the viewpoint of a theoretical physicist and the video lectures can be found online. Finally, I highly recommend that you watch the online lectures by Nathan Wiebe.

\printindex

\begin{thebibliography}{99}

\bibitem{aaronson} S.~Aaronson, ``Introduction to Quantum Information Science: Lecture Notes".
\bibitem{ekert} A.~Ekert, ``Introduction to Quantum Computation".
\bibitem{girvin} S.~Girvin, ``Introduction to Quantum Error Correction and Fault Tolerance".
\bibitem{jozsa} R.~Jozsa, ``Quantum Information and Computation".
\bibitem{kaye} P.~Kaye, R.~Laflamme \& M.~Mosca, \emph{An Introduction to Quantum Computing}.
\bibitem{knill} E.~Knill et al., ``Introduction to Quantum Information Processing".
\bibitem{lloyd} S.~Lloyd, ``Quantum Information Science".
\bibitem{mosca} M.~Mosca, ``Quantum Algorithms".
\bibitem{nielsenchuang} M.~Nielsen \& I.~Chuang, \emph{Quantum Computation and Quantum Information}.
\bibitem{preskill} J.~Preskill, ``Quantum Computation: Lecture Notes".
\bibitem{steane} A.~Steane, ``Quantum Computing".
\bibitem{dewolf} R.~de Wolf, ``Quantum Computing: Lecture Notes".
\end{thebibliography}
\end{document}